\documentclass[fleqn,usenatbib]{mnras}

\usepackage{newtxtext,newtxmath}

\usepackage[T1]{fontenc}

\DeclareRobustCommand{\VAN}[3]{#2}
\let\VANthebibliography\thebibliography
\def\thebibliography{\DeclareRobustCommand{\VAN}[3]{##3}\VANthebibliography}

\usepackage{graphicx}	
\usepackage{amsmath}	
\usepackage{gensymb}
\usepackage{subfigure}

\newcommand{\msun}{$\mathrm{M_{\odot}}$}
\newcommand{\rsun}{$\mathrm{R_{\odot}}$}
\newcommand{\Ni}{$^{56}$Ni\ }
\newcommand{\Co}{$^{56}$Co\ }

\title[SN 2018hna]{SN~2018hna: Adding a Piece to the Puzzles of the Explosion of Blue Supergiants}
\author[Xiang et al.]{
	Danfeng Xiang,$^{1}$\thanks{E-mail: xiangdf@mail.tsinghua.edu.cn}
	Xiaofeng Wang,$^{1,2,3}$\thanks{E-mail: wang\_xf@mail.tsinghua.edu.cn}
	Xinghan Zhang,$^{1}$
	Hanna Sai,$^{1}$
	Jujia Zhang,$^{4}$
	Thomas G. Brink,$^{5}$ \newauthor
	Alexei V. Filippenko,$^{5}$
	Jun Mo,$^{1}$
	Tianmeng Zhang,$^{6,7}$
	Zhihao Chen,$^{1}$
	Luc Dessart,$^{8}$
	Zhitong Li,$^{9,7}$
	\newauthor
	Shengyu Yan,$^{1}$
	Sergei I. Blinnikov,$^{10,11}$
	Liming Rui,$^{1}$
	E. Baron$^{12}$
	and J.~M.~DerKacy$^{13,12}$
	\\
	$^{1}$Department of Physics and Tsinghua Center for Astrophysics (THCA), Tsinghua University, Haidian District, Beijing 100084, China\\
	$^{2}$Beijing Planetarium, Beijing Academy of Sciences and Technology, Beijing 100044, China\\
	$^{3}$Purple Mountain Observatory, Chinese Academy of Sciences, Nanjing 210023, China\\
	$^{4}$Yunnan Observatories, Chinese Academy of Sciences, Kunming 650216, China\\
	$^{5}$Department of Astronomy, University of California, Berkeley, CA 94720-3411, USA\\
	$^{6}$Key Laboratory of Space Astronomy and Technology, National Astronomical Observatories, Chinese Academy of Sciences,  Beijing 100101, China\\
	$^{7}$School of Astronomy and Space Science, University of Chinese Academy of Sciences, Beijing 101408, China\\
	$^{8}$Institut d'Astrophysique de Paris, CNRS-Sorbonne Universit\'{e}, 98 bis boulevard Arago, F-75014 Paris, France\\
	$^{9}$Key Laboratory of Optical Astronomy, National Astronomical Observatories, Chinese Academy of Sciences, Beijing 100101, China\\
	$^{10}$National Research Center ``Kurchatov Institute'', 123182 Moscow, Russia\\
	$^{11}$Kavli IPMU (WPI), UTIAS, The University of Tokyo, Kashiwa, Chiba 277-8583, Japan\\
	$^{12}$Homer L. Dodge Department of Physics and Astronomy, University of Oklahoma, 440 W. Brooks, Norman, OK 73019-2061, USA\\
	$^{13}$Department of Physics, Virginia Tech, 850 West Campus Drive, Blacksburg VA, 24061, USA\\
}
\date{Accepted XXX. Received YYY; in original form ZZZ}

\pubyear{20xx}

\begin{document}
	\label{firstpage}
	\pagerange{\pageref{firstpage}--\pageref{lastpage}}
	\maketitle
	
\begin{abstract}	
	We present extensive optical/ultraviolet observations and modelling analysis for the nearby SN 1987A-like peculiar Type II supernova (SN) 2018hna. Both photometry and spectroscopy covered phases extending to $>$500 days after the explosion, making it one of the best-observed SN~II of this subtype. SN~2018hna is obviously bluer than SN 1987A during the photospheric phase, suggesting higher photospheric temperature, which may account for weaker \ion{Ba}{ii}~$\mathrm{\lambda}$6142 lines in its spectra. Analysis of early-time temperature evolution suggests a radius of $\sim$45~\rsun\ for the progenitor of SN~2018hna, consistent with a blue supergiant (BSG). By fitting the bolometric light curve with hydrodynamical models, we find that SN~2018hna has an ejecta mass of $\sim$(13.7--17.7)~\msun, a kinetic energy of $\sim$ (1.0--1.2) $\times 10^{51}$~erg, and a \Ni mass of about 0.05~\msun. Moreover, based on standard stellar evolution and the oxygen mass (0.44--0.73~\msun) deduced from nebular [\ion{O}{i}] lines, the progenitor of SN 2018hna is expected to have an initial main-sequence mass $<$16~\msun. In principle, such a relatively low-mass star cannot end as a BSG just before core-collapse, except some unique mechanisms are involved, such as rapid rotation, restricted semiconvection, etc. On the other hand, binary scenario may be more favourable, like in the case of SN~1987A. While the much lower oxygen mass inferred for SN~2018hna may imply that its progenitor system also had much lower initial masses than that of SN~1987A. 
	
\end{abstract}
	
	
	
\section{Introduction}\label{sec:intro}
	Stars in the Universe may end their lives in different ways after experiencing long-term evolution,  and they will eventually ``die'' in silence or spectacular explosions. 
	Core-collapse supernovae (SNe) are the violent deaths of massive stars that have initial masses $\gtrsim~8$~\msun. 
	Some of these stars keep most of their hydrogen envelopes prior to explosion, so the spectra of these SNe show strong hydrogen lines with P~Cygni profiles. 
	These H-rich SNe are classified as SNe~II, while H-poor ones are classified as SNe~I \citep[e.g.,][]{1997ARA&A..35..309F}. 
	The progenitors of SNe~II are believed to be red supergiants (RSGs) with large low-density hydrogen envelopes. 
	Unlike H-poor SNe (SNe~Ia and Ibc), recombination of H in the ejecta plays an important role during the photospheric phase of SNe~II.
	
	At the beginning of the explosion, the shock breaks out from the stellar surface, and then the shocked envelope cools down by emitting light in ultraviolet (UV) and optical bands. 
	During this very early phase, the light curves rise to a peak quickly, usually in a few days. 
	Since the temperature of the H-rich ejecta is high, almost all H atoms are ionised at early phases. 
	As time goes by, the expanding ejecta cool down and the H$^{+}$ ions start to recombine with electrons to form neutral H, releasing photons. 
	According to theoretical analysis, during the recombination phase the photospheric temperature is roughly the H-recombination temperature ($\sim 5500$~K), and the size of the photosphere remains almost constant. 
	As a result, the SN luminosity remains unchanged for $\sim 100$ days, and this period is called a ``plateau'' in light curves of SN~IIP. 
	Some other H-rich SNe do not have plateaus; their light curves decline linearly (in magnitudes) and thus are called Type IIL. 
	It has been confirmed that the progenitor stars of SNe~IIP are consistent with red supergiants \cite[RSGs; see the review by][]{2015PASA...32...16S}.
	
	While most H-rich SNe can be classified as Type IIP or IIL (but see \citealt{2014ApJ...786...67A} and the review by \citealt{2017hsn..book..195G} for arguments for a continuous distribution of SNe~II), there are some other peculiar subtypes that show different properties. 
	The most well-known one is the SN 1987A-like subtype, named after its prototype SN~1987A, which exploded in the Large Magellanic Cloud (LMC) and was the closest SN discovered over the past few centuries. 
	Extensive and detailed observations of SN~1987A show that it had the spectral characteristics of SNe~II (i.e., the persistent H lines), but it took more than 80 days to reach its primary light-curve peak after a short and rapid rise \citep{1987MNRAS.227P..39M,1987MNRAS.229P..15C,1988MNRAS.231P..75C,1988AJ.....96.1864S,1988MNRAS.234P...5W,1995ApJS...99..223P}.
	There is no plateau in the light curves of SN~1987A-like SNe~II (hereafter ``87A-like SNe"); instead, they rise to their peaks slowly. 
	After a rise of $\sim 80$ days, these SNe start to decline rapidly, followed by a linear decay which also appears in SNe~IIP/IIL. 
	It is now clear that the distinctive behaviour of 87A-like SN light curves is the result of smaller, denser, and hotter progenitors -- blue supergiants (BSGs). 
	In comparison with RSGs, the envelopes of BSGs have steeper density profiles and are nonconvective. 
	Some studies also show that 87A-like SNe ejected more mass and synthesised more \Ni in the explosion \citep{2016A&A...588A...5T}, which can explain the peculiar light-curve evolution of this subclass. 
	
	Moreover, the spectra of 87A-like SNe show prominent \ion{Ba}{ii} absorption lines in addition to features typically seen in typical SNe~II \citep{1995A&A...303..118M}.
	\ion{Ba}{ii} lines also appear in normal SNe~II, but later ($>40$ days), and they tend to emerge in dimmer events \citep{2017ApJ...850...89G}.
	There remains the issue of why SN~1987A displayed stronger \ion{Ba}{ii} lines than other SNe~II, but the enhanced He and $s$-process elements in the circumstellar material suggested that these apparent enhancements are caused by real abundance effects \citep{1988PASA....7..434H,1992A&A...258..399M}. 
	Recent studies have made some attempts to interpret the spectral signatures of peculiar SNe~II \citep[e.g.,] {2005A&A...437..667D,2008MNRAS.383...57D,2010MNRAS.405.2141D,2018A&A...619A..30D,2019A&A...622A..70D}. 
	The connection of $s$-process, He burning, and binary interaction can probably shed light on this issue.
	
	SN~1987A was well studied in many aspects, including the analysis of its pre-explosion star. 
	The progenitor of SN~1987A, Sk$-$69\degree202, was identified as a compact BSG with radius $R \approx$ 45~\rsun\ and temperature $T_{\mathrm{eff}} \approx 16,000$~K \citep{1988ApJ...324..466W}. 
	While stellar-evolution theory predicts that massive stars with initial masses lower than 30~\msun\ explode at the RSG stage ($R>100$~\rsun), SN~1987A was an exception. 
	New hypotheses were proposed to explain BSGs as progenitors of H-rich SNe, such as low metallicity, new treatment of convection, dredge-up of He layers, rapid rotation, binary interaction, and star mergers \cite[see the review by][]{1992PASP..104..717P}. 
	Furthermore, the triple-ring structure around SN~1987A found by \cite{1990ApJ...362L..13W} imposes additional constraints on the evolution of the precursor, favouring a binary evolution scenario, in which the progenitor of SN~1987A first evolved to RSG phase and then went blue again after obtaining matter from the companion by either accretion or merging.
	Nevertheless, the evolutionary mechanism of the progenitors of other 87A-like SNe remains unclear.
	
	Thus far, only a few well-observed SNe~II have been identified as 87A-like objects: SN~1998A \citep{2005MNRAS.360..950P}, SN~2006au and SN~2006V \citep{2012A&A...537A.140T}, SN~2009E \citep{2012A&A...537A.141P}, and recently SN~2018hna \citep{2019ApJ...882L..15S}. Other samples are presented by \cite{2011MNRAS.415..372K}, \cite{2016A&A...588A...5T}, \cite{2016MNRAS.460.3447T}, and \cite{2016ApJ...831..205K}. 
	All of the above SNe have long rise times and H-rich spectra. 
	Among this sample, SN~2018hna is the best observed after SN~1987A. This peculiar SN II exploded in a nearby irregular galaxy UGC 7534, and it was discovered by Koichi Itagaki on 22.82 October 2018 \footnote{\url{https://www.wis-tns.org/object/2018hna}} (MJD 58413.82; UT dates are used throughout this paper). \cite{2019ApJ...882L..15S} studied the observed properties of SN~2018hna, suggesting that it had a progenitor with an initial mass of 14--20~\msun\ and an ejected \Ni mass of $\sim 0.087$~\msun\ according to their hydrodynamical modelling of the multiband light curves. 
	Polarimetric observations of SN~2018hna were published by \cite{2021MNRAS.503..312M} and \cite{2021NatAs...5..544T} in optical and near-infrared (NIR) bands, respectively.
	SN~2018hna showed significant polarization in the near-infrared (NIR) bands at 182 days, with a typical value of $p \approx 2$\%, but dust formation was not responsible for this large polarization. 
	In the optical, it had $p < 1$\% in RINGO3 $b^*$ and $g^*$ bands before $t \approx 115$ days and 1.29\% in $r^*$ at day 102. 
	By comparing with SN~1987A, \cite{2021NatAs...5..544T} concluded that the polarization behaviour of SN~2018hna is similar to that of SN~1987A, so their ejecta share a similar underlying geometry observed at a similar inclination angle. 
	
	In this paper, we present extensive optical (and UV) observations SN~2018hna, including light curves and spectra covering pre-maximum times to very late nebular phases. 
	These datasets are used to constrain the explosion parameters and determine the properties of the progenitor star. 
	Furthermore, along with SN~2018hna, we collect some well-observed 87A-like SNe in history, to discuss the common characteristics of this particular family and the differences among the individuals.
	The paper is structured as follows. 
	Our photometric and spectroscopic observations are presented in Sec.~\ref{sec:obs}. 
	Then, in Sec.~\ref{sec:obs-prop-lc} we discuss the properties of the multi-band and bolometric light curves of SN~2018hna compared to other 87A-like SNe, followed by comparisons of spectral features in Sec.~\ref{sec:obs-prop-spec}.
	Next, in Sec.~\ref{sec:bol-prop}, we estimate and analyse the explosion parameters using both semi-analytical and hydrodynamic methods. 
	In Sec.~\ref{sec:progenitor-analysis}, we explore the properties and possible evolutionary pathways of the progenitor star of SN~2018hna through the characteristics of its nebula phase spectra. 
	Finally, we summarise our conclusions in Sec.~\ref{sec:summary}. 
	
\section{Data}\label{sec:obs}
	Our photometric and spectroscopic observations of SN 2018hna were obtained by ground-based instruments, including the 80~cm Tsinghua-NAOC Telescope at Xinglong Observatory \citep[hereafter TNT;][]{2008ApJ...675..626W, 2012RAA....12.1585H}, the 2.4~m telescope at Lijiang Observatory (hereafter LJT), the 2.16~m telescope at Xinglong Observatory (hereafter XLT), the ARC 3.5~m telescope at Apache Point Observatory (hereafter ARC), and the 10~m Keck-I telescope on Maunakea, Hawai`i. 
	The Zwicky Transient Facility \cite[ZTF;][]{2019PASP..131a8002B} Bright Transient Survey \citep[BTS;][]{2020ApJ...904...35P} monitored SN~2018hna in the $gr$ bands until very late phases, and the data are publicly available (ZTF ID: ZTF18acbwaxk)\footnote{\url{lasair.roe.ac.uk/object/ZTF18acbwaxk/}}. 
	We also include early-time UV--optical photometry obtained by the Ultraviolet/Optical Telescope \citep[UVOT;][]{2005SSRv..120...95R, 2004ApJ...611.1005G} onboard the \textit{Neil Gehrels Swift Observatory} \citep{2004ApJ...611.1005G}.
	Details of these observations are presented below.
	
\subsection{Optical photometry}\label{sec:obs-phot}
	Photometric observations by TNT spanned the phases from 4 Dec. 2018 to 3 Mar. 2019, lasting for over 200 days with an average cadence of $<4$ days. 
	The filters used by TNT were the generic Johnson $BV$ and Sloan $gri$ bands. 
	The reference images were obtained in all corresponding bands after the SN faded away. 
	
\begin{figure}
		\includegraphics[width=1.0\columnwidth]{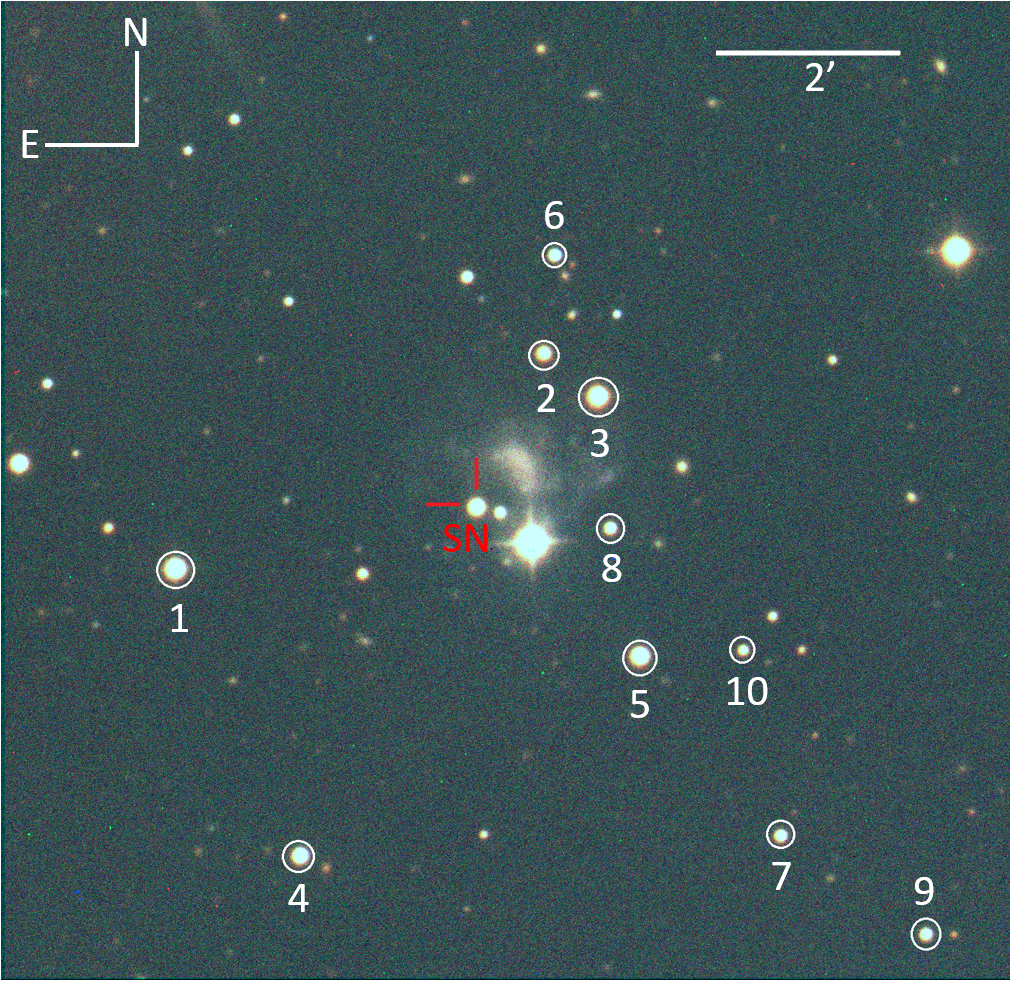}
		\caption{Pseudo-colour finder chart of SN~2018hna and its host galaxy UGC 7534, produced from the $BVr$-band images obtained by the Tsinghua-NAOC 0.8-m telescope on 4th January, 2018. The SN is located southeast of the centre of UGC 7534. The reference stars used for flux calibration are marked by white circles. North is up and east is left. }
		\label{fig:finderchart}
	\end{figure}

	All of the $BVgri$ images are preprocessed using standard \texttt{IRAF}\footnote{IRAF is distributed by the National Optical Astronomy Observatories, which are operated by the Association of Universities for Research in Astronomy, Inc., under cooperative agreement with the National Science Foundation (NSF).} routines, which includes corrections for bias, flat field, and removal of cosmic rays. 
	To remove contamination from the host galaxy, we applied a template-subtraction technique based on the \texttt{hotpants} package\footnote{\url{https://github.com/acbecker/hotpants}}. 
	For SN~2018hna and the reference stars, the instrumental magnitudes were measured using standard-point-spread-function (PSF) photometry. 
	These were then converted to standard Johnson $BV$ and Sloan $gri$ magnitudes using the zero points and colour terms of the filters, which are obtained by comparing the instrumental and standard magnitudes of the reference stars in field of the SN (see Fig.~\ref{fig:finderchart}). 
	Standard magnitudes of the reference stars  in $BV$-bands are obtained from the AAVSO Photometric All-Sky Survey (APASS)\footnote{\url{www.aavso.org/apass}}, and magnitudes in $gri$-bands are obtained from the Sloan Digital Sky Survey (SDSS; \citealp{2007AJ....134..973I}). 
	Photometric results of the reference stars and SN~2018hna are listed in Appendix~\ref{sec:appendixA}.

	\begin{figure*}
		\includegraphics[width=1.9\columnwidth]{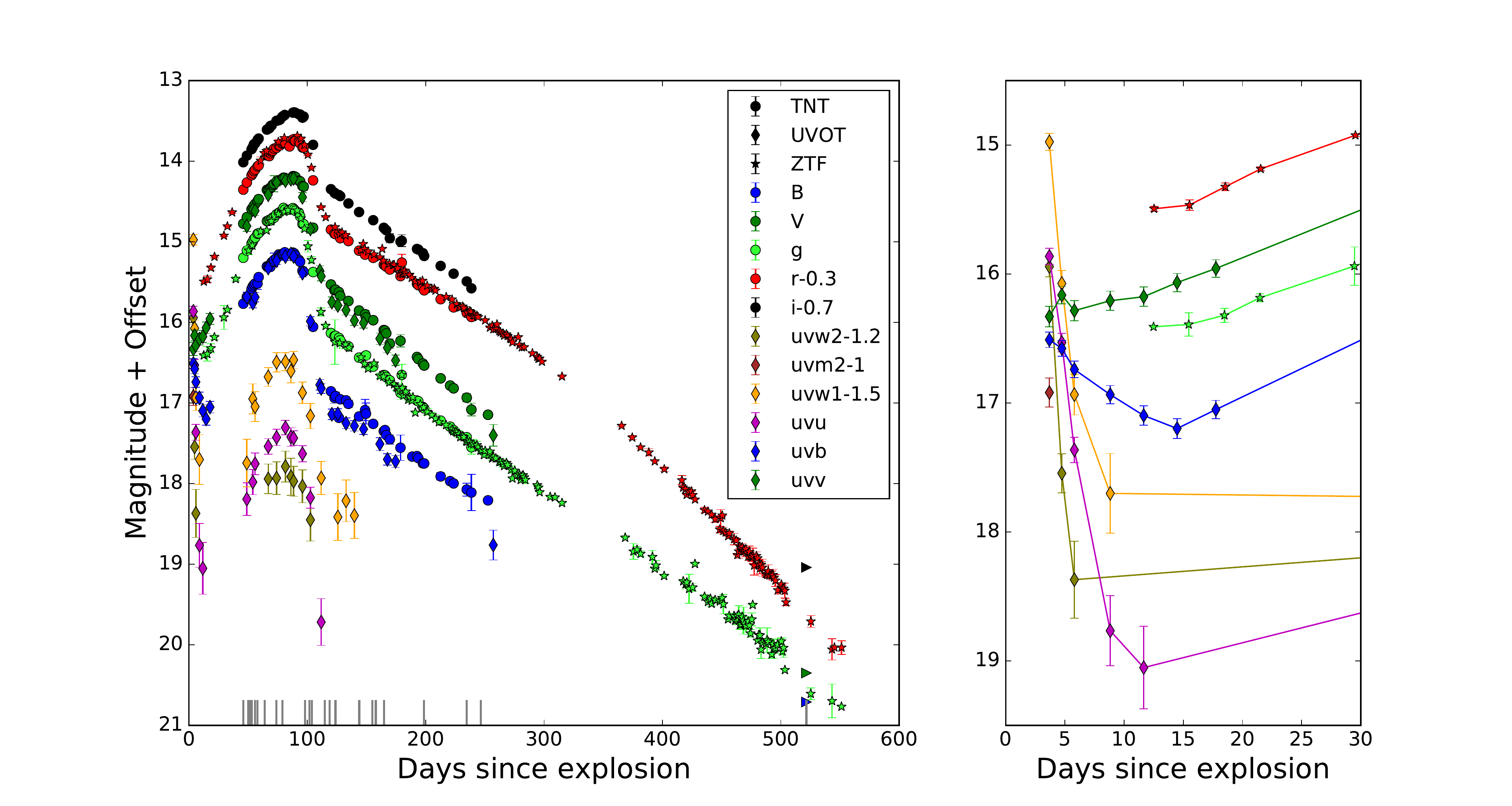}
		\caption{UV--optical light curves of SN~2018hna. The grey ticks at the bottom mark the epochs of spectroscopic observations. Magnitudes in $BVi$ bands inferred from the Keck spectrum at day 522 are presented by the coloured triangles. The right panel shows a close-up view of the early phases. }
		\label{fig:lc}
	\end{figure*}
	
	Our light curves together with the public ZTF and UVOT data are shown in Fig.~\ref{fig:lc}. 
	Light curves from different sources agree well with each other, except for the UVOT $uvb$ and TNT $B$ observations in the postmaximum phases. 
	We notice these differences and evaluate that they have no impact on our analysis.
	We also made comparisons with light curves from \cite{2019ApJ...882L..15S} and found consistency of both datasets.
	We include complementary data from \cite{2019ApJ...882L..15S} when necessary.
	
\subsection{Optical spectroscopy}\label{sec:obs-spec}
\begin{figure*}
		\includegraphics[width=2.0\columnwidth]{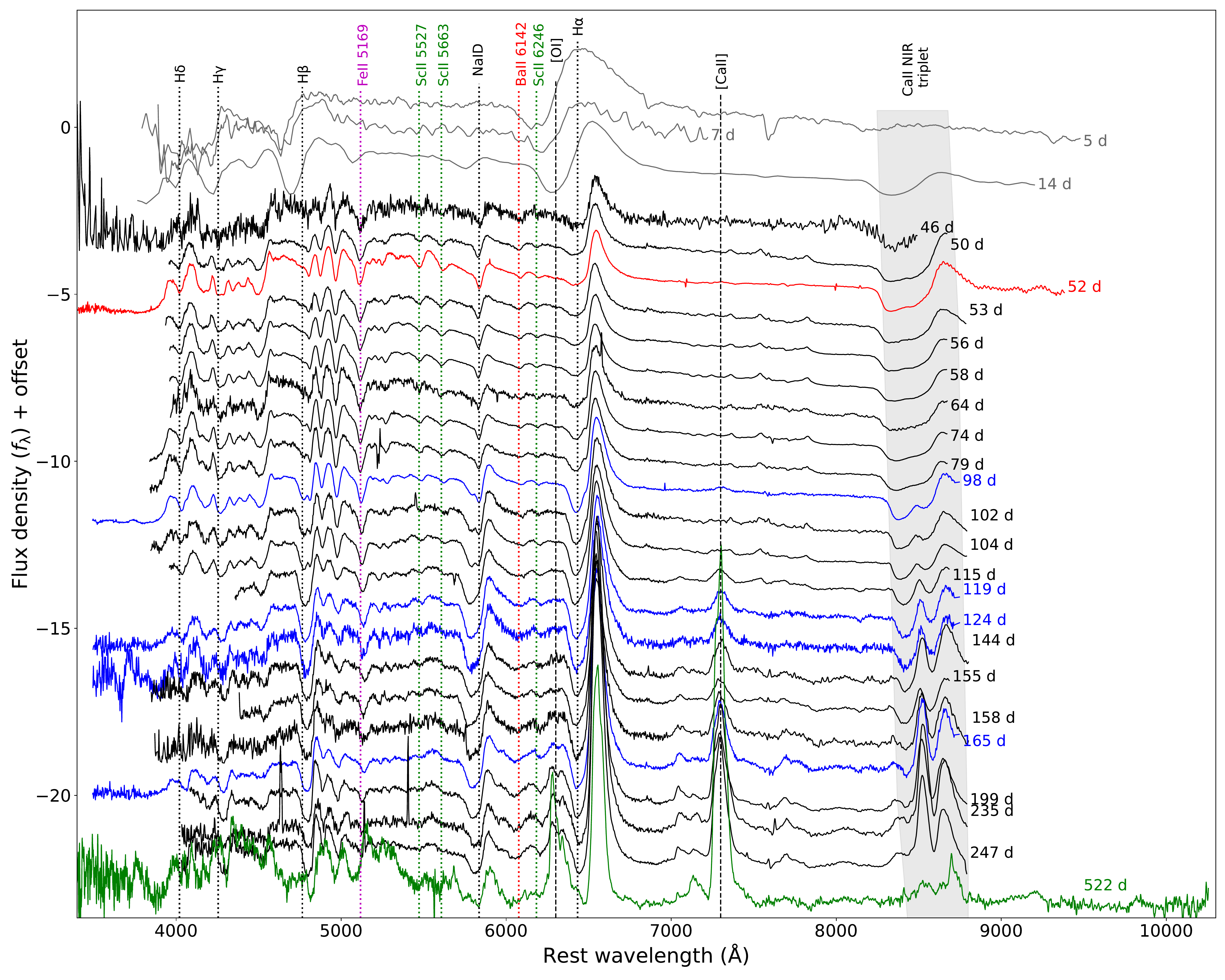}
		\caption{Spectral sequence of SN~2018hna. Different colours represent different instruments: black--XLT, red--ARC, blue--LJT, green--Keck-I, grey--TNS. Numbers to the right of each spectrum display phases since explosion (MJD 58411). The dominant spectral features are marked by dotted vertical lines.}
		\label{fig:spectra}
	\end{figure*}

	A sequence of optical spectra was obtained by the 2.16~m telescope at Xinglong Observatory of NAOC, the 2.4~m telescope at Lijiang Observatory, and the 3.5~m telescope at Apache Point Observatory. 
	A late-time spectrum was taken by the 10~m Keck-I telescope with the Low-Resolution Imaging Spectrometer (LRIS; \citealp{1995PASP..107..375O}) on 24 March 2020. 
	The spectroscopy of Keck-I used a slit with width of 1.0\arcsec\ and the slit was placed along the parallactic angle. 
	The resultant spectrum has a resolution of $\sim$4~\AA\ in the blue side (<5600\AA) and $\sim$6.9~\AA\ in the red side (>5600~\AA). All our spectra cover the phases from $\sim 46$ to 522 days after the explosion of SN~2018hna. 
	Three pieces of optical spectra are available on the Transient Name Server (TNS). Details of the TNS spectra are listed in Table~\ref{tab:TNS-spec}. 
	Observation details are listed in Table~\ref{tab:spec-log} and the entire spectral sequence is displayed in Fig.~\ref{fig:spectra}. 
	
	The spectral evolution of SN~2018hna is characterised by typical features of SNe~II, showing strong P~Cygni lines of the hydrogen Balmer series as well as absorption lines of \ion{Fe}{ii}, \ion{Na}{i}, \ion{Ca}{ii}, \ion{Sc}{ii}, and \ion{Ba}{ii}.
	In the nebular phase, the spectra are dominated by emission lines of H, \ion{Na}{i}, and \ion{Ca}{ii}, along with forbidden lines of [\ion{Fe}{ii}], [\ion{Ca}{ii}], and [\ion{O}{i}]; all of these are commonly seen in late-time spectra of SNe~II.
	
\subsection{Distance and extinction}
	The host galaxy of SN~2018hna, UGC 7534, is an irregular dwarf galaxy, which was classified as an IBm-type galaxy according to the third reference catagloue of bright galaxies (RC3.9, \citealp{1991rc3..book.....D}). 
	The metallicity of UGC 7534 was found to be subsolar ($Z\sim0.3$~Z$\odot$, \citealp{2019ApJ...882L..15S}).
	The helliocentric velocity of UGC~7534 gives a redshift of 0.002408 \citep{1999PASP..111..438F}. 
	The distance to this galaxy was estimated to be 10.50~Mpc through the Tully-Fisher relation \citep{2013AJ....145..101K}. 
	Given a measurement error of 25\% \citep{2013AJ....145..101K}, we apply a distance of 10.50$\pm$2.63~Mpc, and a distance modulus of $\mu = 30.11 \pm 0.54$~mag.
	
	Since there is no significant \ion{Na}{I}~D absorption in optical spectra of SN~2018hna at the redshift of the host, we adopt a Galactic reddening of $E(B-V) = 0.01$~mag \citep{2011ApJ...737..103S} and neglect extinction in the host galaxy. 
	Following \cite{2019ApJ...882L..15S}, we adopt the first-light date as MJD~=~58411 in this paper.
	
\section{Light Curves and Colour Evolution}\label{sec:obs-prop-lc}
	In this section, we compare the multi-band light curves and colour evolution of SN~2018hna to other 87A-like SNe.
	As shown in Fig.~\ref{fig:lc}, the light curves of SN~2018hna are similar to those of typical 87A-like SNe~II, showing a long rise to peak. 
	Light curves in the UV and $B$ bands exhibit early bumps before rising to the main peaks, which is due to cooling of the shocked H-rich envelope, common in SNe~II. 
	Then the light curves rise slowly to peaks at $\sim 80$~days after explosion, followed by a rapid decline in $\sim 30$ days. 
	After that, the light curves decline linearly with a rate of $\sim 1$~mag (100~days)$^{-1}$ in all optical bands except in $r$, where a slower decline is observed before $\sim 250$ days and a faster decline is seen after that. 
	This phenomenon is also observed in SN~1987A, and is most likely due to the emission of hydrogen, as the transmission curve of the $r$-band filter peaks near the wavelength of H$\alpha$. 
	We use the Keck-I spectrum and the observed magnitudes in $g$ and $r$ as flux calibrators to determine the $Bvi$ magnitudes on day 522.
	
\begin{figure}
		\includegraphics[width=0.95\columnwidth]{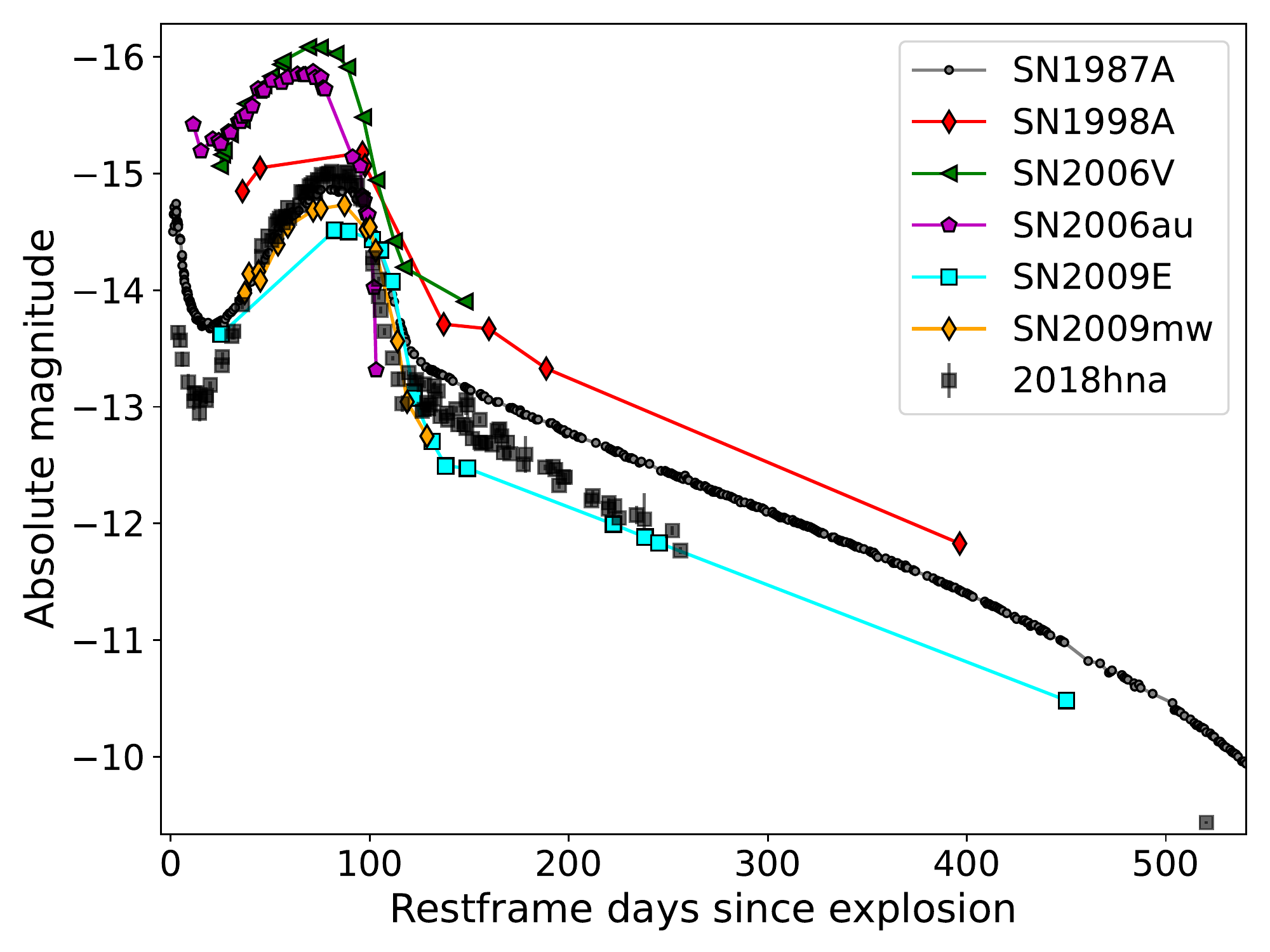}
		\includegraphics[width=0.95\columnwidth]{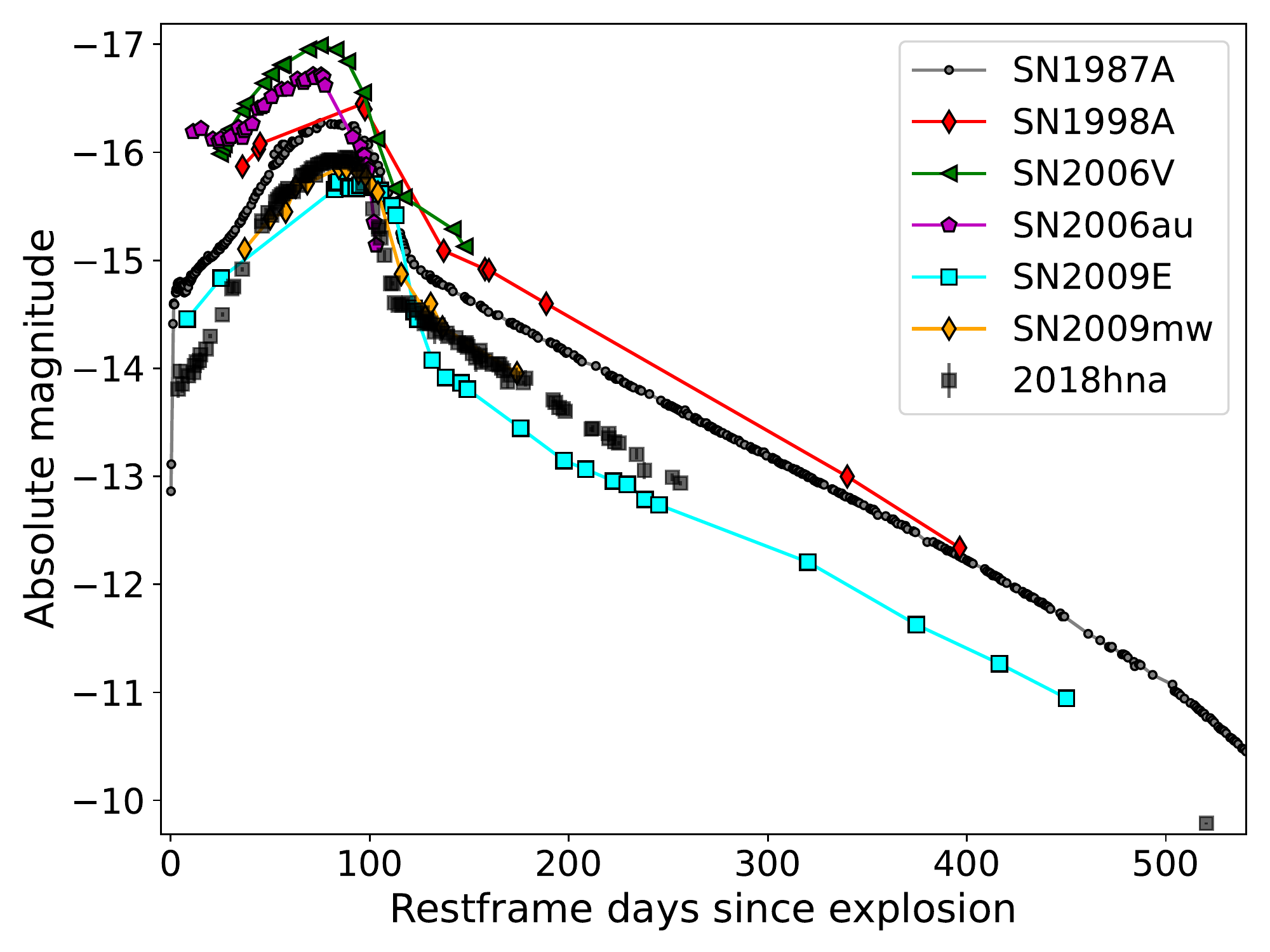}
		\includegraphics[width=0.95\columnwidth]{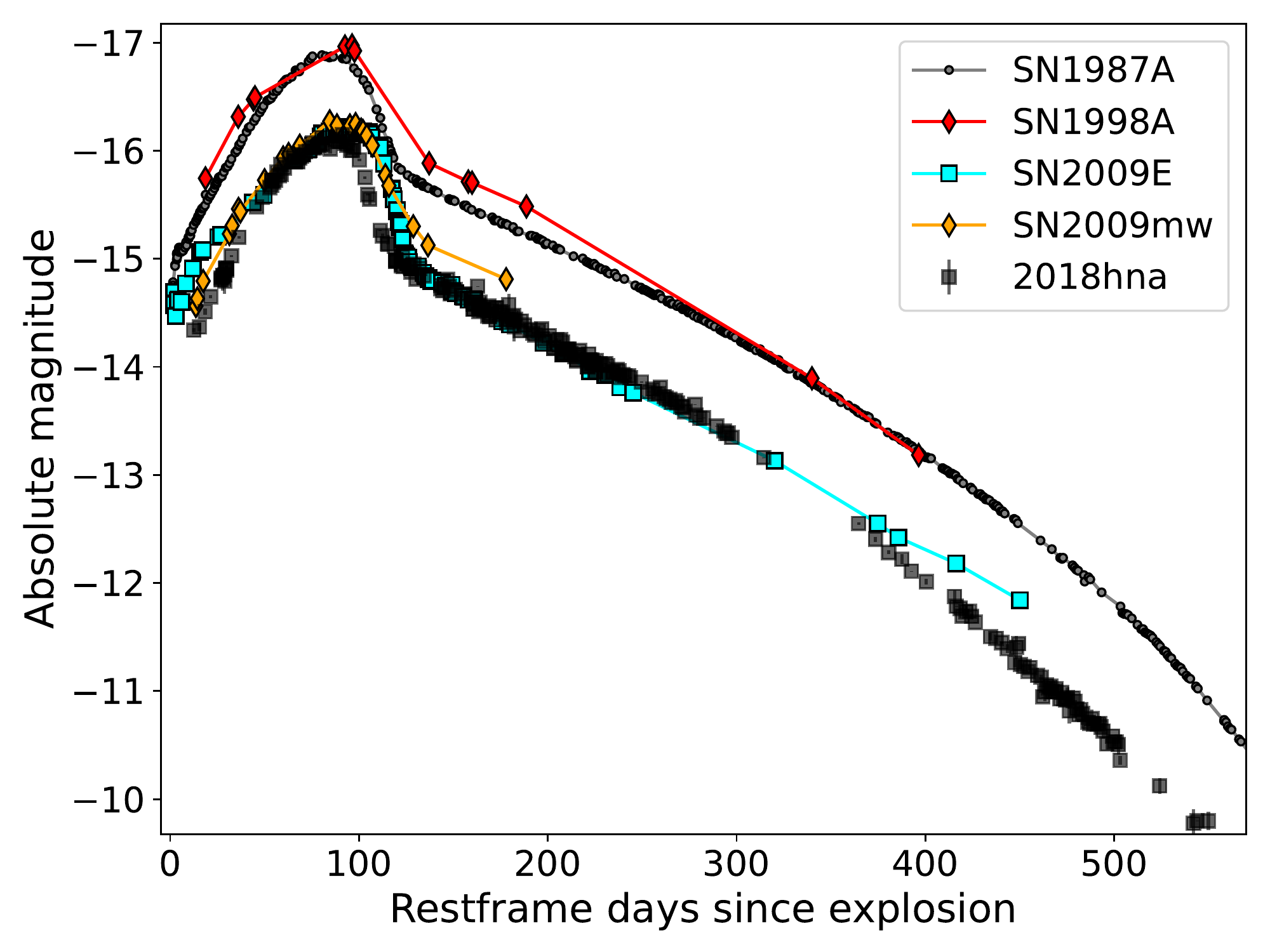}
		\caption{Light curves of SN~2018hna compared with those of other 87A-like SNe in $B$ (\textit{top}), $V$ (\textit{middle}), and $r$ (\textit{bottom}). The $R$-band light curve of SN~1987A is plotted instead of $r$. }
		\label{fig:lc_cmp}
	\end{figure}

	In Fig.~\ref{fig:lc_cmp}, we compare the $BVr$ light curves of SN~2018hna with those of other 87A-like SNe.
	The peak absolute magnitudes of SN~2018hna in these bands are fainter than the bright SN~2006V and SN~2006au, but comparable to SN~1987A and SN~2009mw. 
	In $B$-band, although SN~2018hna is similar to SN~1987A around the peak, its shock-cooling-phase emission and the radioactive tail are much fainter than those of the latter, while in $V$-band it is fainter than SN~1987A at all phases. 
	In $r$-band, the overall light-curve evolution of SN~2018hna becomes quite similar to that of SN~2009E and SN~2009mw, although SN~2009E shows an early-time bump.
	
\begin{figure}
		\includegraphics[width=0.95\columnwidth]{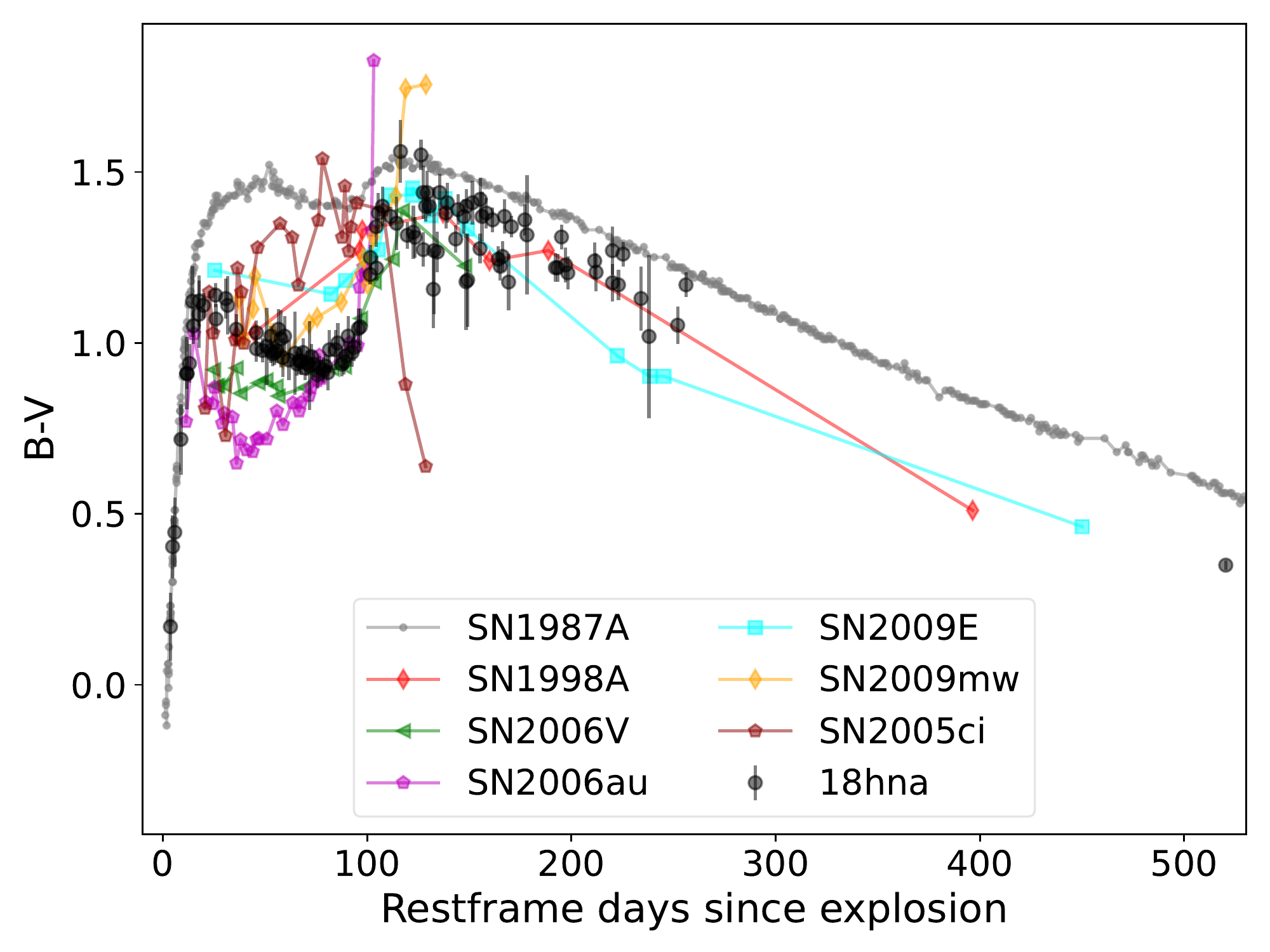}
		\includegraphics[width=0.95\columnwidth]{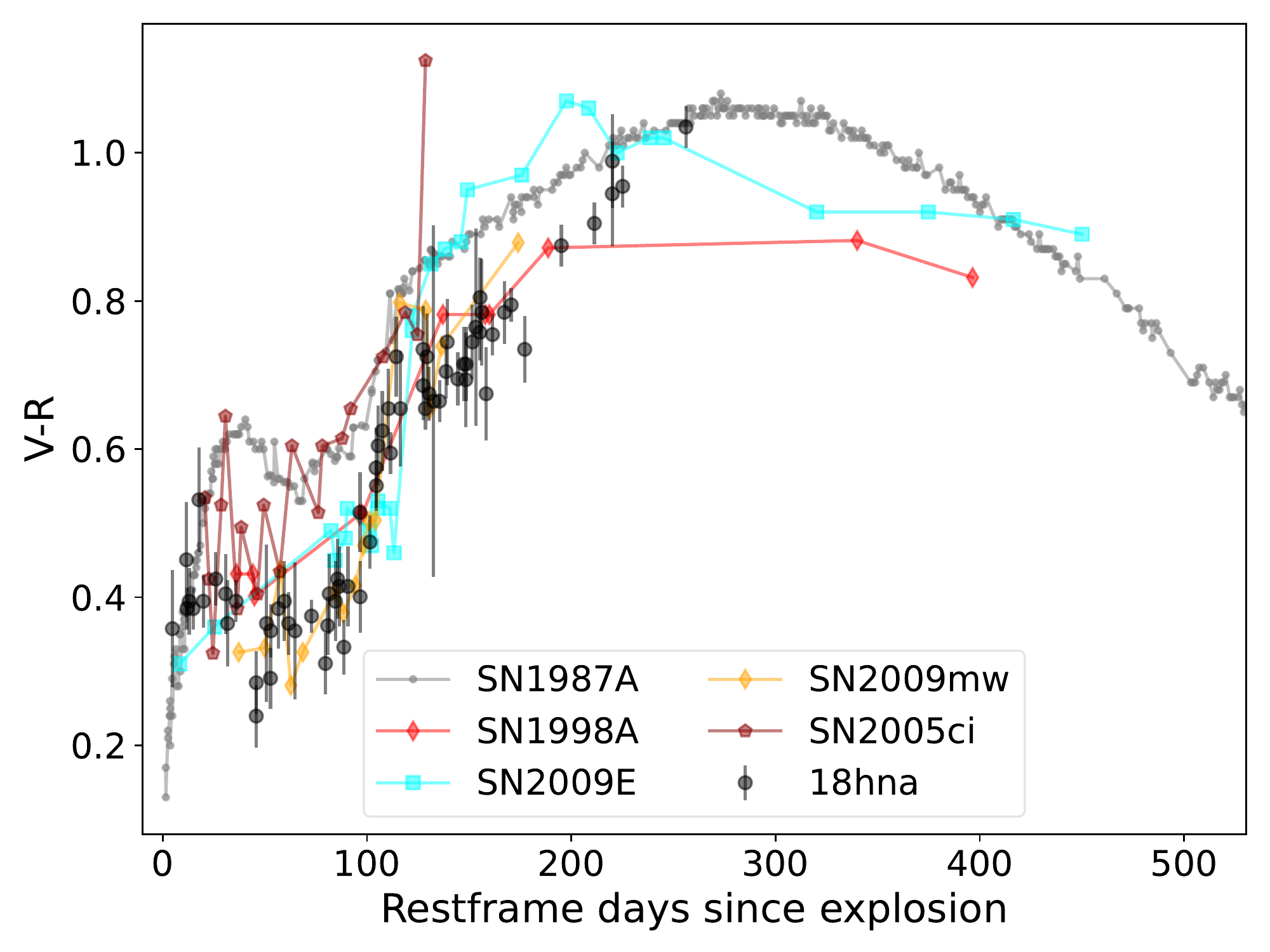}
		\includegraphics[width=0.95\columnwidth]{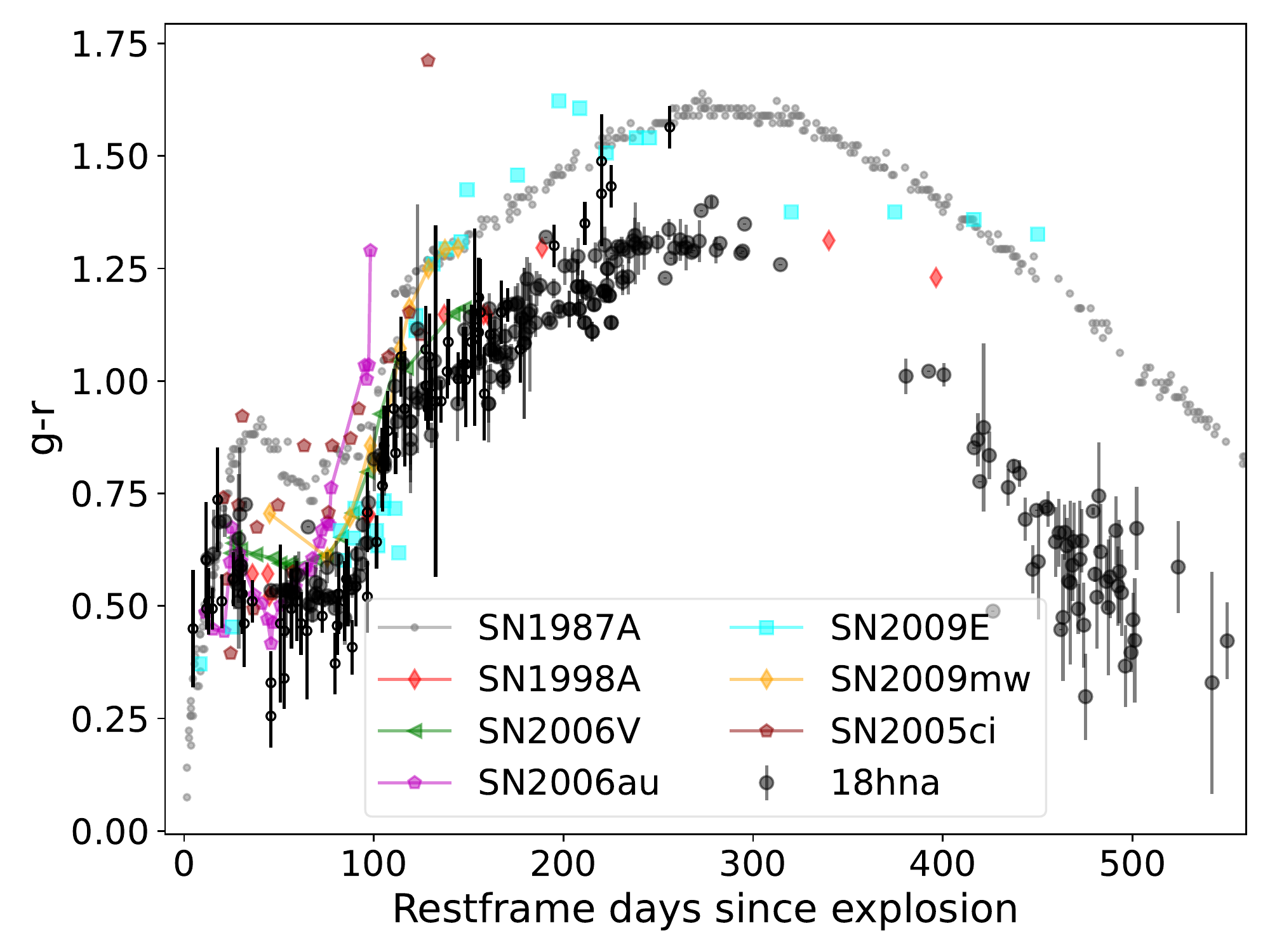}
		\caption{Colour curves of SN~2018hna compared with those of other 87A-like SNe.}
		\label{fig:lc_cmp_color}
	\end{figure}

	The colour evolution of SN~2018hna, compared with other 87A-like SNe is shown in Fig.~\ref{fig:lc_cmp_color}. 
	The $V-R$ colours of SNe are transformed to $g-r$ using the conversion formula from Table 3 of \cite{2006A&A...460..339J}.
	The $B-V$ colour curves of 87A-like SNe all show two peaks. 
	They share similar evolution before reaching the first peak, although the first-peak time varies. 
	One can see that the earlier the first peak occurs, the shorter duration it takes the colour curves to turn red again. 
	This might be also related to the luminosities of the SNe. 
	For example, the most luminous SN~2006V and SN~2006au have a faster turnoff time, while the remaining relatively faint events like SNe~1987A, 2009E, and 2018hna show slower colour-turn even though their colour differences can be as large as 0.5~mag.
	Since the early $B-V$ colours are tied to the photospheric temperature, the correlation of colour evolution with peak luminosity may be a result of different progenitor properties such as stellar radius and the temperature of the shocked envelope. 
	Among the slow events, SN~1987A shows unique behaviour in $B-V$: it is slightly brighter than SN~2018hna but its $B-V$ turn is slower than that of all others. 
	This might imply that the specific shock energy (i.e., $\sim E/M$) in SN~1987A is lower than in other objects, which can also be confirmed by its low expansion velocities (see Fig.~\ref{fig:vel-comp}).
	After the second peak, the $B-V$ colours of SN~2018hna and all comparison SNe~II exhibit similar linear declines. 
	The dispersion of $B - V$ colour evolution reaches a minimum at $\sim 100$ days, at which time recombination ends and all SNe share a similar effective temperature. 
	After that, SN~2018hna enters the fast-declining phase. 
	At $t \approx 100$ days, the $V - R$ and $g - r$ colours of different 87A-like SNe also tend to become more consistent, and they reach the reddest peak values at much later phase ($t \approx 300$ days). 
	After that, both the $V - R$ and $g - r$ colours evolve blueward in a linear fashion.  
	
\section{Spectroscopic Evolution}\label{sec:obs-prop-spec}
	As shown in Fig.~\ref{fig:spectra}, the spectral evolution of SN~2018hna is similar to that of a typical SN~II, and we do not detect any significant narrow emission lines which probe interaction \citep[e.g.,][]{2001MNRAS.326.1448C}. In the photospheric phase, the spectra are dominated by Balmer lines with P~Cygni profiles, along with absorption lines of \ion{Na}{i}~D, \ion{Fe}{ii}, \ion{Ca}{ii}, \ion{Sc}{ii}, and the characteristic \ion{Ba}{ii} lines of 87A-like SNe.
	At later phases, H absorption lines vanish as the ejecta expand and become geometrically diluted, and forbidden lines of [\ion{O}{i}] and [\ion{Ca}{ii}] emerge.
	Below we discuss the spectral properties of SN~2018hna, such as line velocities, line strength, in comparison with other 87A-like SNe II.

\subsection{Line velocities}\label{sec:velocity}
\begin{figure}
		\includegraphics[width=1.0\columnwidth]{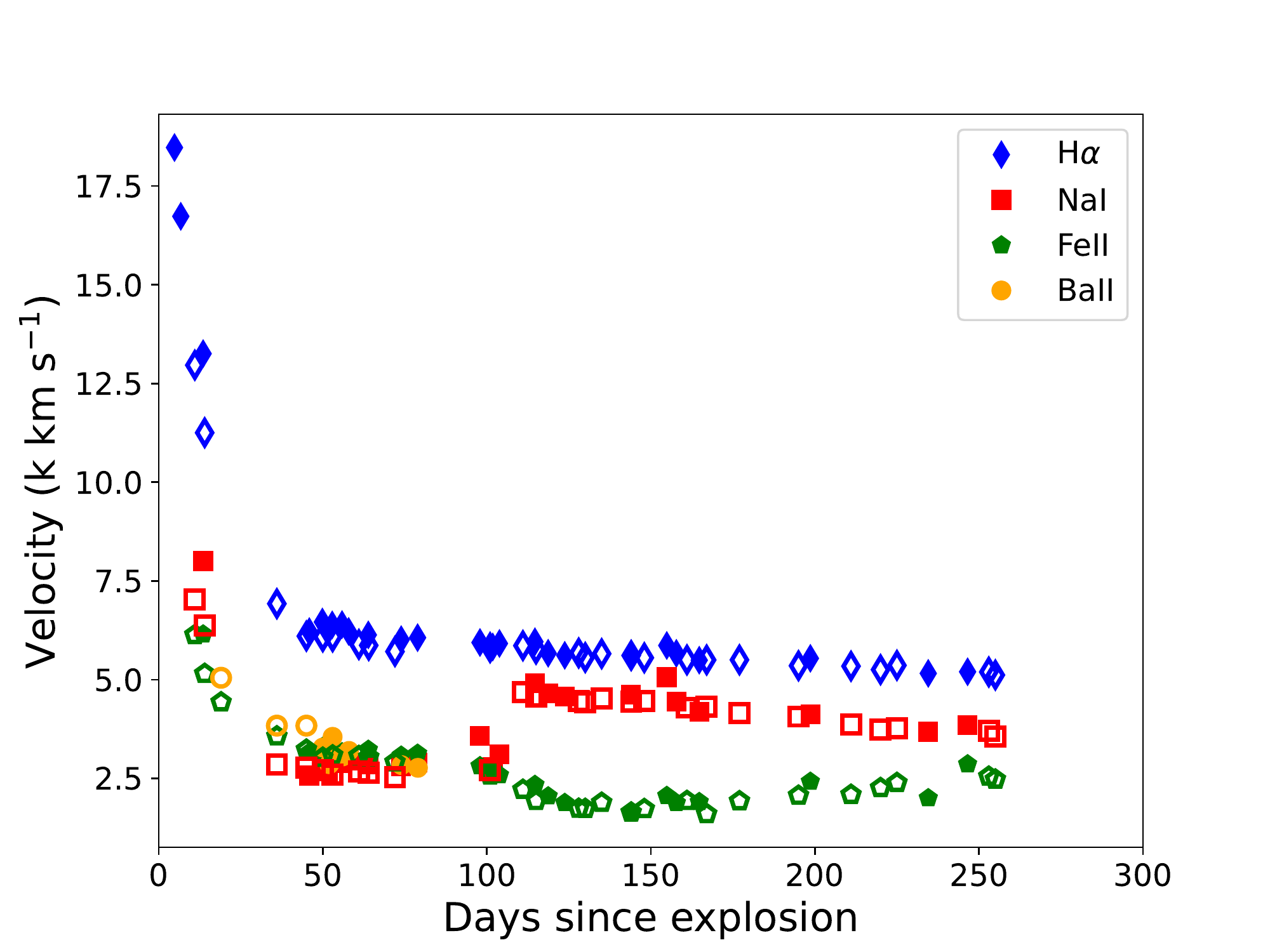}
		\caption{Expansion velocities of SN~2018hna inferred from H${\alpha}$, \ion{Na}{i}~D, \ion{Fe}{ii}~$\mathrm{\lambda}$5169, and \ion{Ba}{ii}~$\mathrm{\lambda}$6142 absorption in the spectra. Open data points represent measurements from data available in the literature \citep{2019ApJ...882L..15S}.}
		\label{fig:vel}
	\end{figure}

\begin{figure}
		\includegraphics[width=1.0\columnwidth]{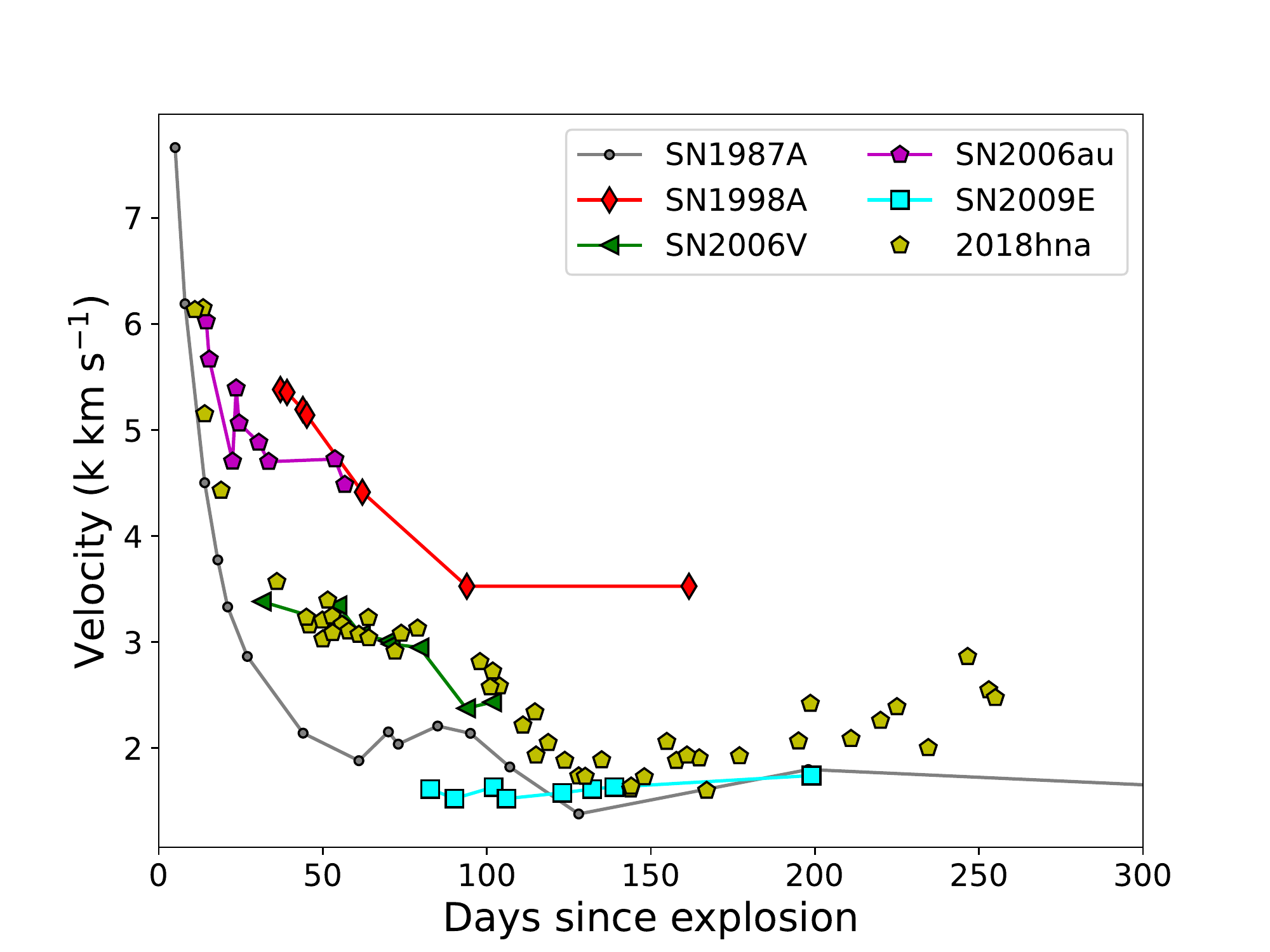}
		\caption{\ion{Fe}{ii} line velocities of SN~2018hna in comparison with several other 87A-like SNe. }
		\label{fig:vel-comp}
	\end{figure}

	We measured the expansion velocities of several absorption lines, as shown in Fig.~\ref{fig:vel}. 
	The line velocities are measured from the blueshifted minima of the corresponding absorption troughs.
	Spectral data from \cite{2019ApJ...882L..15S} are also included in our analysis. 
	Note that the velocity inferred from \ion{Na}{i}~D absorption shows a steep rise at $\sim 105$~days, probably a measurement artifact since the absorption can be contaminated by other species such as \ion{He}{i}.
	In Fig.~\ref{fig:vel-comp}, the velocity of \ion{Fe}{ii}~$\lambda$5169, which is often used to trace the SN photosphere, is measured for SN~2018hna and compared with those of several 87A-type objects. 
	As it can be seen, the 87A-like SNe II show a large diversity in Fe II velocity, with SN~2018hna lying between SN~1987A and SN~1998A but close to SN~2006V. 
	No strong correlation seems to exist between the expansion velocity and peak luminosity for this peculiar subgroup of SNe II when inspecting the light curves shown in Fig.~\ref{fig:lc_cmp}, although this correlation has been proposed for normal SNe IIP\citep[e.g.,][]{2002ApJ...566L..63H}. 
	Among our comparison sample, SN~2006au and SN~2006V have comparable peak luminosities, while their velocities could differ by $\sim$1500 km s$^{-1}$.
	
\subsection{Spectral evolution}\label{sec:comp-spec-phot}
\begin{figure}
		\centering
		\includegraphics[width=1.0\columnwidth]{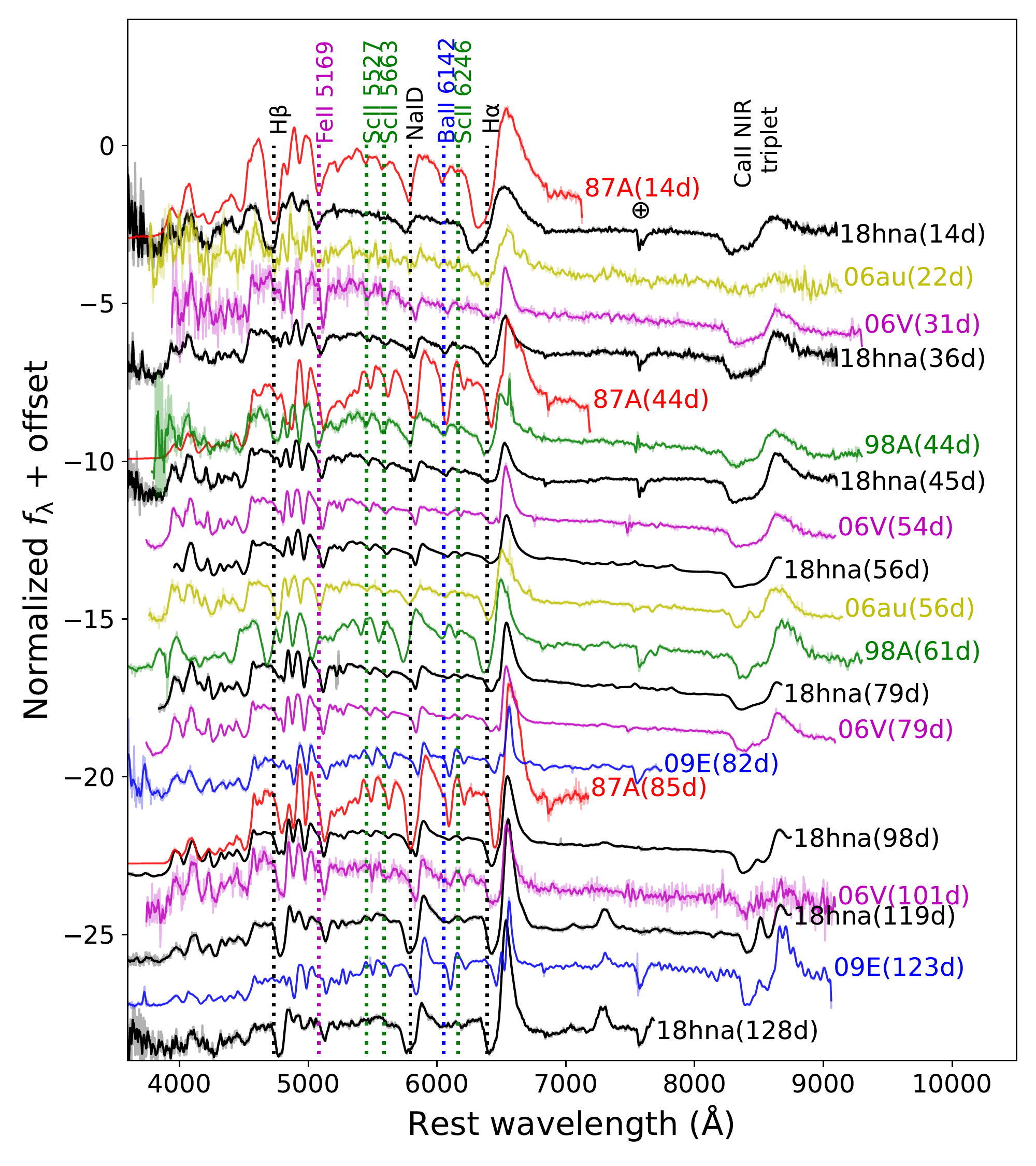}
		\caption{Spectra of SN~2018hna compared with those of other 87A-like SNe at photospheric phases. All spectra are corrected for extinction using the Milky Way law \citep{1999PASP..111...63F}. Numbers in brackets are the corresponding phases after explosion. All spectra are normalised by dividing by their average fluxes, and also shifted vertically for clarity.}
		\label{fig:spec-comp-phot}
	\end{figure}

	Fig.~\ref{fig:spec-comp-phot} shows the spectral evolution of SN~2018hna during the photospheric phase, together with that of the comparison sample of 87A-like SNe II. 
	SN~2018hna is found to show close resemblances to SN~2006V, except that the latter has slightly narrower line profiles. 
	Compared to SN~1987A, the metal lines such as \ion{Na}{i}, \ion{Sc}{ii}, \ion{Fe}{ii} and the characteristic \ion{Ba}{ii} are quite weak in SN~2018hna, SN~2006V and SN~2006au at photospheric phase. 
	Note that SN~1998A has metal lines with strength lying between the above three SNe and SN 1987A. 
	In the spectra of 87A-like SNe, the strength of \ion{Ba}{ii} lines is generally proportional to that of other metal lines; while SN~2009E is an exception, which has weak metal lines as SN~2018hna but it shows stronger \ion{Ba}{ii}~$\mathrm{\lambda}$6142. 
	This indicates that, in addition to metal abundance, there are other factors that may affect the formation of \ion{Ba}{ii} absorption feature. 
	To figure out the possible factors affecting the intensity of spectral lines in SN spectra, we try to compare observations with theoretical models.
	
	\cite{2019A&A...622A..70D} computed a set of nonlocal thermodynamic equilibrium (NLTE) time-dependent radiative transfer models based on the explosion of a BSG which originally evolved from a star having an initial mass of 15~\msun\ and extremely low metallicity and subsequently exploded with a thermal bomb. 
	Strong mixing is imposed, which erases the original low metallicity of the progenitor (i.e., the metal-rich inner layers pollute the outer H-rich layers). 
	These models are characterized with a set of different initial conditions such as ejecta mass, kinetic energy, \Ni mass and element abundance. 
	We caution that the models of \cite{2019A&A...622A..70D}, which are based on limited ejecta parameters, are not expected to match all of the observational properties. 
	But comparing the spectra of SN~2018hna with those predicted by models can give a clue on how different physical conditions can influence the spectra. 
	As an example, we perform a detailed analysis with one of these models below.
	
\begin{figure}
		\includegraphics[width=0.9\columnwidth]{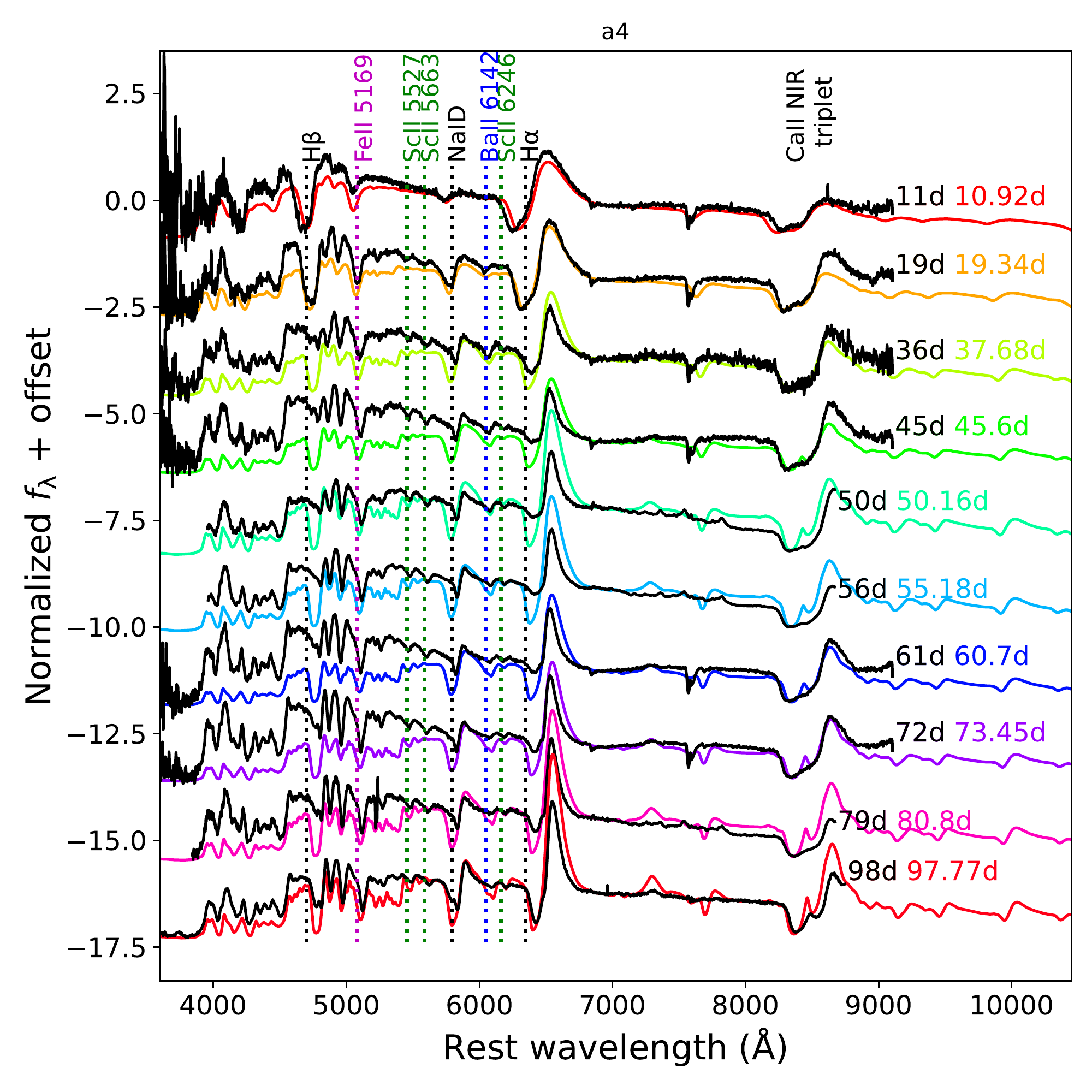}
		\includegraphics[width=0.9\columnwidth]{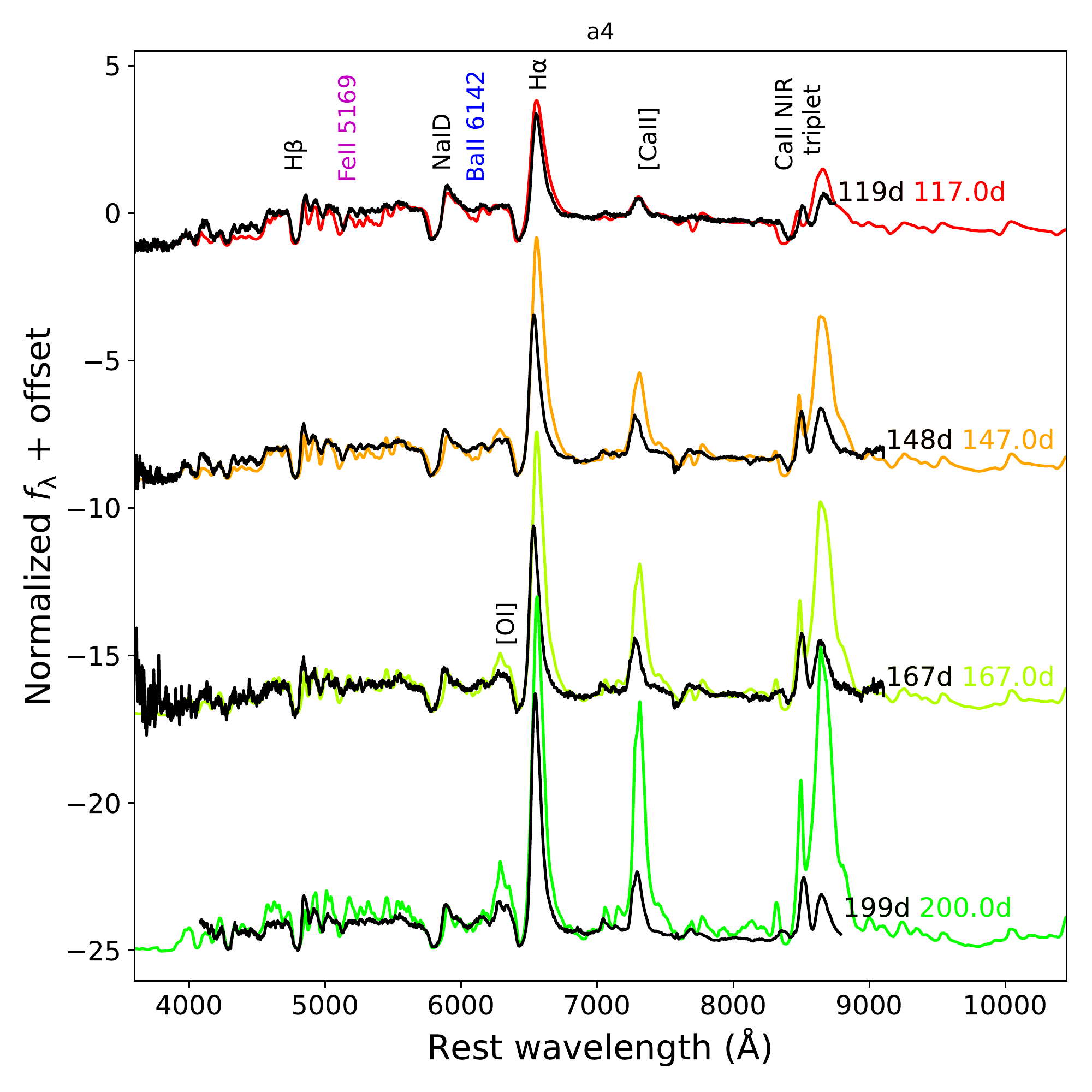}
		\caption{Comparison of the spectra of SN~2018hna in the nebular phases ($<100$ days; upper) and early nebular phases (100--200 days; lower) with model a4. Model spectra are plotted with coloured lines, and the phase of each spectrum is indicated by the text of the corresponding colour. Each model spectrum is normalised to match the flux within 6780--7000~\AA\ of the overplotted observed spectrum. }
		\label{fig:comp-spec-LDmodel}
	\end{figure}

	Model a4 from \cite{2019A&A...622A..70D} has an ejecta mass of 13.22~\msun, a kinetic energy of 1.26$\times10^{51}$~erg, a \Ni mass of 0.084~\msun, and a mixing length of 4~\msun. 
	Comparison of the spectra of SN~2018hna with model a4 is shown in Fig.~\ref{fig:comp-spec-LDmodel}.
	In the early phases (i.e, at t $<30$ days), model a4 has similar line velocities but weaker \ion{Ba}{ii} and \ion{Fe}{ii} lines in comparison with the observations. 
	Subsequently, \ion{Ba}{ii} lines in the models become stronger than the observations, but other metal lines are still weaker until day 119. 
	The metal lines match better with the observations in model a4ni (with enhanced \Ni by a factor of $\sim 1.5$ compared to model a4), despite that they have weak Ca emission after transition to the nebular phase; less amount of \Ni might reduce the difference. 
	Since the models have strong mixing thus high metal abundance in the ejecta, the weak metal lines in SN~2018hna are likely due to lower metallicity in the ejecta, hence smaller mixing length in the model.
	
	Discrepancy between the models and observations are large at intermediate phases (30--110 days after the explosion), diminishing around day 120. 
	This is likely due to the growing asymmetry in the SN seen from the polarimetry \cite{2021MNRAS.503..312M}. 
	On the other hand, it can also be a result of faster receding of the photosphere due to less mixing or clumping in the ejecta \citep[e.g.,][]{2018A&A...619A..30D}. 
	The weak absorption component in H$\mathrm{\alpha}$ and the vanishing of H$\mathrm{\beta}$ between days 19 and 36 can also be a result of interaction \citep[e.g.,][]{2022A&A...660L...9D}. 
	After day 119, when the photosphere has receded to the inner H-poor region, the emission lines of \ion{Ca}{ii} and [\ion{Ca}{ii}] in the models grow faster, implying less calcium in the inner ejecta.
	
\subsection{\ion{Ba}{ii} line strength in 87A-like SNe}\label{sec:BaII-line}
\begin{figure}
		\includegraphics[width=1.0\columnwidth]{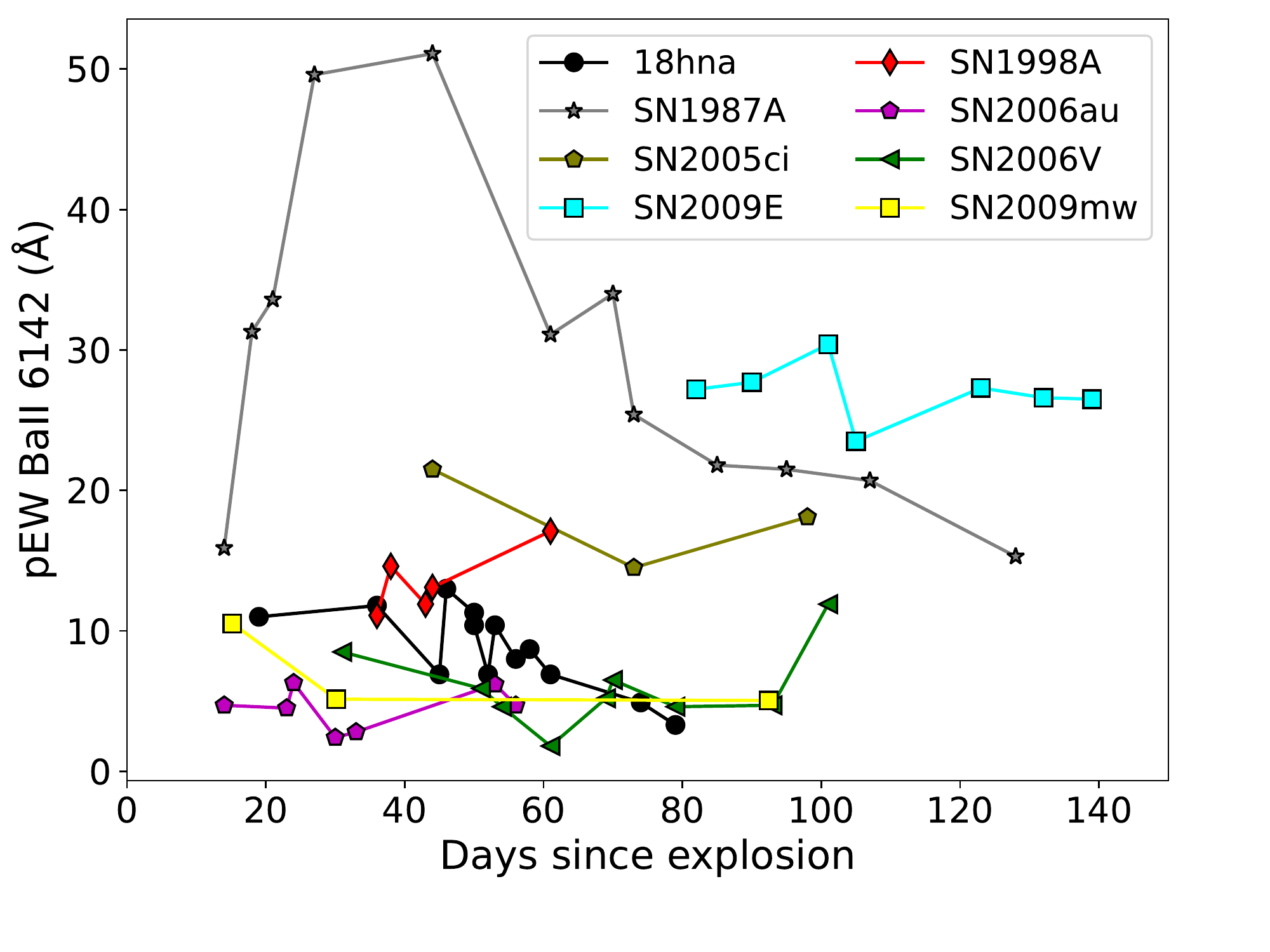}
		\includegraphics[width=1.0\columnwidth]{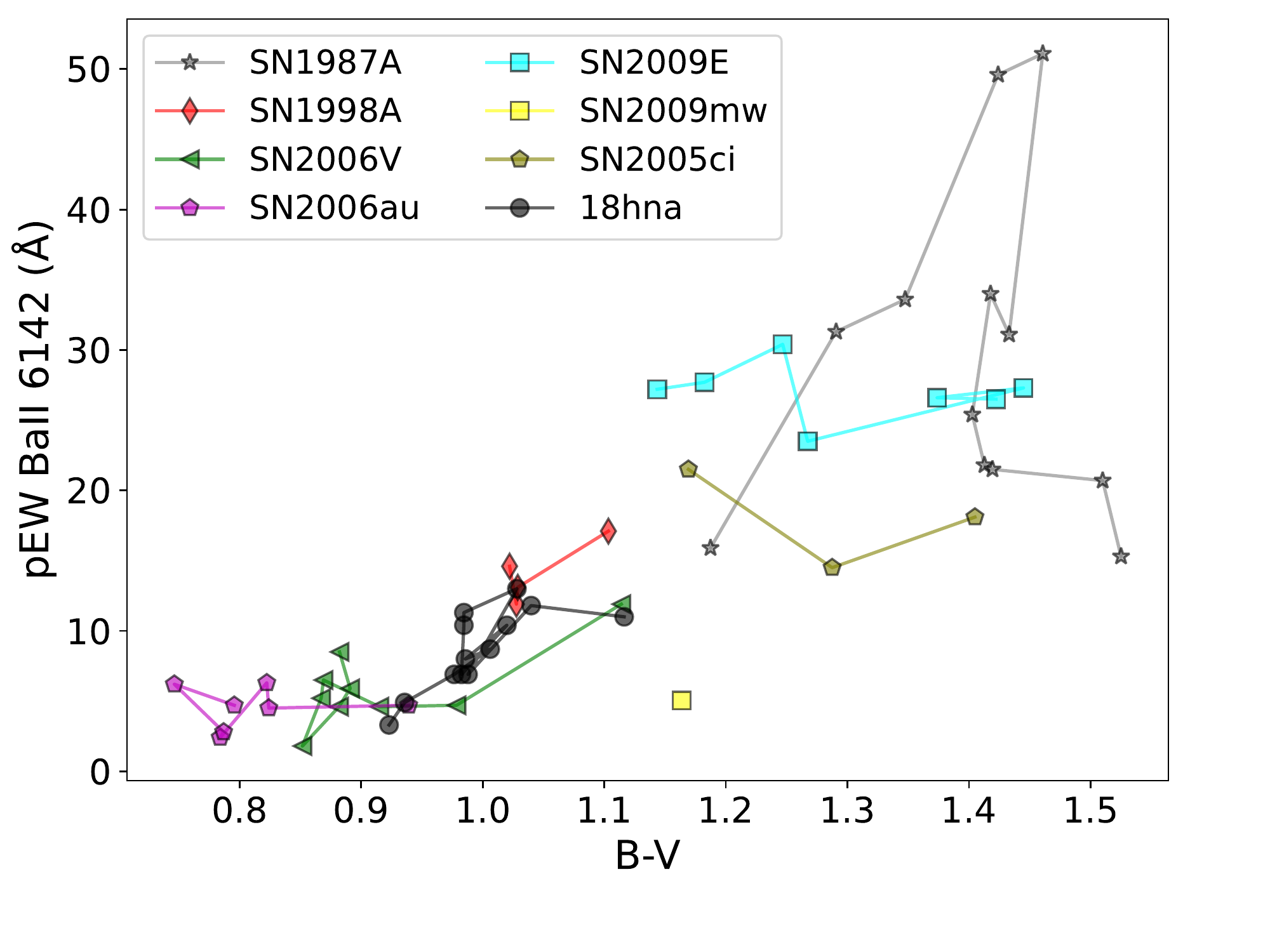}
		\caption{\textit{Upper}: Pseudo-equivalent width (pEW) of the \ion{Ba}{ii}~$\mathrm{\lambda}$6142 line of SN~2018hna compared with other 87A-like SNe. The data for SN~2009mw are taken directly from literature \citep{2016MNRAS.460.3447T}, and data for other SNe are measured from spectra in literature. \textit{Lower}: Correlation of the pEWs of \ion{Ba}{ii}~$\mathrm{\lambda}$6142 and $B-V$ colour of 87A-like SNe.}
		\label{fig:BaII}
	\end{figure}

	The characteristic features of \ion{Ba}{ii} absorption lines in 87A-like SNe are evident in the spectra ($>15$ days) of SN~2018hna \citep{2019ApJ...882L..15S}.
	A previous study by \cite{2016MNRAS.460.3447T} found that the strength (pseudo-equivalent widths; pEWs) of the \ion{Ba}{ii} line in 87A-like SNe appears to split into two groups: the stronger group like SN~1987A and SN~2009E and the weaker one with \ion{Ba}{ii} lines similar to some normal SNe~IIP \citep{2005MNRAS.360..950P}. 
	Spectra of SN~2018hna start to show significant absorption features of \ion{Ba}{ii} and \ion{Sc}{ii} since +20 days (see also \citealp{2019ApJ...882L..15S}). 
	We measured the pEWs of its \ion{Ba}{ii}~$\mathrm{\lambda}$6142 line and compared them with those of other 87A-like SNe. 
	As shown in the upper panel of Fig.~\ref{fig:BaII},  SN~2018hna can be put into the group of weaker \ion{Ba}{ii}~$\mathrm{\lambda}$6142 absorption like SN~1998A and SN~2009mw.
	
	In principle, the strength of \ion{Ba}{ii} lines can be affected by the following factors: an intrinsic difference in Ba abundance of the H-rich ejecta, different physical conditions such as temperature \citep{1995A&A...303..118M} and clumping in SN ejecta \citep{2018A&A...619A..30D}. 
	Among these factors, temperature may play an important role in the strength of \ion{Ba}{ii} lines, as the optical depth of \ion{Ba}{ii} lines increases with decreasing temperature.  
	Thus, a lower mass of \Ni might be responsible for stronger \ion{Ba}{ii} lines, especially after the recombination phase when the SN is mainly powered by \Ni decay.
	
	To further examine quantitatively the temperature effect on \ion{Ba}{ii} absorption, we show the pEWs of \ion{Ba}{ii}~$\mathrm{\lambda}$6142 and the $B-V$ colour in lower panel of Fig.~\ref{fig:BaII}. 
	As it can be seen, these two observables show a strong positive correlation, although there is a large variation at the red side. 
	This suggests that low temperature be at least one of the main factors responsible for strong \ion{Ba}{ii} lines in 87A-like SNe. 
	Note that the temperature of SNe~II is related to their explosion energy and initial radius before recombination ends, and later radioactive decay is heating the ejecta. 
	Thus, we may expect that SNe with lower specific energy ($E/M$) can have stronger \ion{Ba}{ii} lines in their photospheric phases, and those with lower \Ni mass can have stronger \ion{Ba}{ii} lines in the late nebular phases. 
	
	Nevertheless, differences in temperature cannot fully explain the Ba problem. 
	For example, at t$\sim 100$ days when all 87A-like SNe share similar $B - V$ color (and hence temperature), their \ion{Ba}{ii} lines still show large scatter in strength. 
	Thus, Ba may be abundant in some 87A-like SNe.
	Known mechanisms of blue-red-blue evolution of the progenitors of 87A-like SNe are related to He mixing, and therefore to the synthesis of Ba. 
	He mixed into the outer envelope of the progenitor star can be a result of unique convection in the envelope or processes involving binary systems or stellar mergers. 
	So the possible Ba enhanced events (e.g. SN~1987A and SN~2009E) might have more He mixed into the outer H-rich envelope.

\section{Bolometric Light Curve and Hydrodynamic Modeling}\label{sec:bol-prop}
\begin{figure}
		\includegraphics[width=1.0\columnwidth]{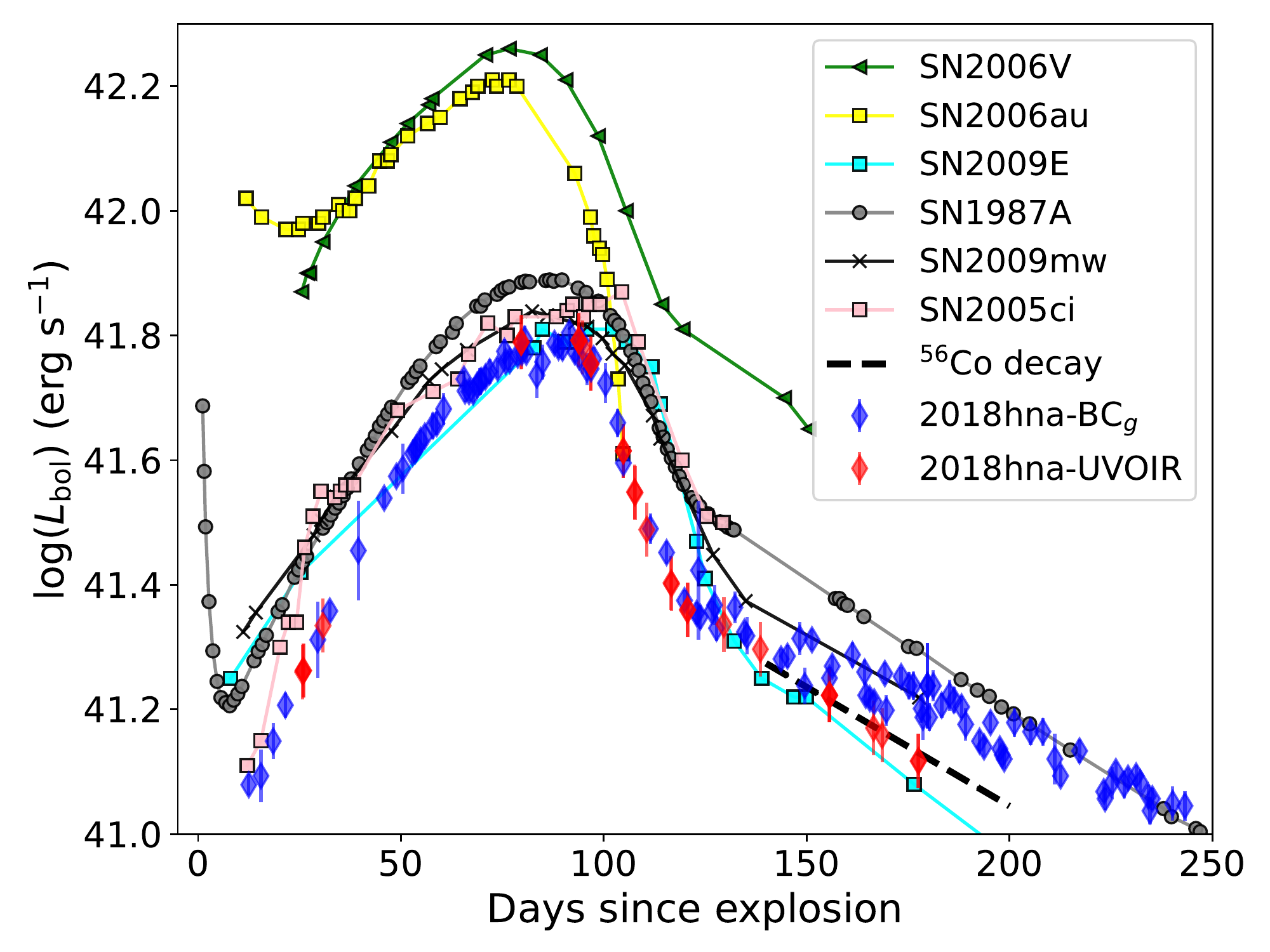}
		\caption{Bolometric light curve of SN~2018hna compared with that of other 87A-like SNe~II. The red diamonds represent the bolometric light curve of SN~2018hna established with the UVOIR light curves, while the blue diamonds indicate the one derived from the ZTF $gr$-band light curves with bolometric corrections (see text for details). The black dashed line shows a \Ni-powered light curve with $M_{\mathrm{Ni}} = 0.05$~\msun.}
		\label{fig:lc_cmp_bol}
	\end{figure}

	The bolometric light curve of SN~2018hna is established via the following two methods: (1) calculating the UV-through-IR (UVOIR) bolometric luminosity by integrating fluxes from optical through NIR wavebands, with the NIR light curves taken from \cite{2019ApJ...882L..15S}, noted as $L_{\mathrm{UVOIR}}$; (2) applying the bolometric correction from \cite{2014MNRAS.437.3848L} to the $g-r$ colour of the SN, noted as $L_{\mathrm{BC_g}}$.
	The results are shown in Fig.~\ref{fig:lc_cmp_bol}. 
	The two methods agree well at <150 days after the explosion. 
	We get a combined bolometric light curve of SN~2018hna using $L_{\mathrm{UVOIR}}$ and $L_{\mathrm{BC_g}}$ at $t<150$~days. 
	Applying an uncertainty in distance of 25\%, we get a peak bolometric luminosity of $L_\mathrm{peak} = (6.2\pm3.1) \times 10^{41}$~erg~s$^{-1}$. 
	The bolometric light curve peaks at MJD = $58497.4 \pm 3.5$, which gives a rise time of $t_{\mathrm{rise}} \approx 86$~days. 
	
\subsection{Estimation of explosion parameters}\label{sec:estimate-exp-par}
	Before performing light curve fitting, we first estimate the explosion parameters of SN~2018hna by quick analytical methods.
	The linear tail (at t $>$ 150 days after explosion) of the bolometric light curve of SNe~II is powered by radioactive decay of \Co, and \Co is produced through the decay of \Ni.
	Assuming a complete $\gamma$-ray trapping, the luminosity of radioactive decay can be calculated from
	\begin{equation}
		L_{\mathrm{Co}} = M(\mathrm{^{56}Ni}) \epsilon_{\mathrm{Co}} e^{-t/\tau_{\mathrm{Co}}}
	\end{equation}
	By fitting this formula to our combined bolometric light curve at t $>$ 150 days, we get an ejected \Ni mass of $\sim$ 0.048($\pm$0.024)~\msun. 
	Referring to \cite{2011ApJ...728...63R}, we estimate the radius of the progenitor by fitting the temperature evolution derived from the early spectral energy distribution (SED) of SN~2018hna.
	Applying $\kappa = 0.34$~cm$^2$~g$^{-1}$ (corresponding to fully ionised H) and $n = 3$, the best-fit parameters are $E_\mathrm{k} = 0.9 \times 10^{51}$~erg, $R_{\star} = 45$~\rsun, and $M_{\mathrm{env}} = 11.7$~\msun.
	Only $R_{\star}$ can be well constrained because the mass and kinetic energy have minor effects on temperature evolution. 
	On the other hand, the diffusion time of radioactively-powered SNe is given as $t_{\mathrm{d}} =  (\kappa M_{\mathrm{ej}}/\beta c v_{\mathrm{ph}})^{1/2}$ \citep{1982ApJ...253..785A,1989ApJ...340..396A}, where $\beta$ is a constant, $v_{\mathrm{ph}}$ is the photospheric velocity inferred from \ion{Fe}{ii} line near the bolometric maximum and $t_{\mathrm{d}}$ is the rise time to peak luminosity in days. 
	Assuming that $\kappa$ is the same for SN~1987A and SN~2018hna, then we can use the parameters of SN~1987A to derive those for SN~2018hna:
	\begin{equation}
		M_{\mathrm{ej,SN}} = M_{\mathrm{ej,87A}}(\frac{t_{\mathrm{d,SN}}}{t_{\mathrm{d,87A}}})^2 \frac{v_{\mathrm{ph,SN}}}{v_{\mathrm{ph,87A}}}
	\end{equation}
	Using $t_\mathrm{d,87A}$ = 84~days, $v_{\mathrm{ph,87A}}$ = 2200~km~s$^{-1}$, $E_{\mathrm{87A}} = 1.1\times10^{51}$~erg, and $M_{\mathrm{ej,87A}}$ = 14~\msun \citep{2000ApJ...532.1132B} for SN~1987A, we get $M_{\mathrm{ej}} \approx 20$~\msun\ and $E_{\mathrm{k}} \propto M_{\mathrm{ej}} v_{\mathrm{ph}}^2 \approx 2.9~\times~10^{51}$~erg for SN 2018hna by inserting its corresponding values of $v_{\mathrm{ph}}$ (i.e., 3000~km~s$^{-1}$) and $t_\mathrm{d}$ (i.e., 86~days). 
	Nevertheless, these estimates are very rough, and we need more detailed model fitting to get more accurate values. 
	In the following subsection, we will use two methods to fit the bolometric light curve of SN~2018hna.
	
\subsection{Light-curve fitting}\label{sec:hydro-modeling}
	The progenitors of 87A-like SNe are thought to be BSGs with small radius ($R < 100$~\rsun) and high effective temperature ($T_{\mathrm{eff}} > 10,000$~K).
	Large density and small radius can result in a slowly-rising SN light curve. 
	The progenitors of 87A-like SNe are thought to have evolved from different channels than those of normal SNe~II.
	For SN~2018hna, \cite{2019ApJ...882L..15S} modeled the early multiband light curves based on the progenitor model proposed by \cite{2000ApJ...532.1132B}, yielding $E_{\mathrm{SN}} \approx 1.7 \times 10^{51}$~erg ($E_{\mathrm{k}}\approx0.97\times10^{51}$~erg) and $M_{\mathrm{ej}} \approx 14$~\msun.
	However, the phase range of their fitting is limited only to the shock-cooling phase (i.e., $t < 30$~days), and the later light curves cannot match the observations (private communication with Petr V. Baklanov). 
	Thus, we revisit this SN by fitting the bolometric light curve to both a semi-analytical model and hydrodynamical models.
	
\begin{figure}
		\includegraphics[width=1.\columnwidth]{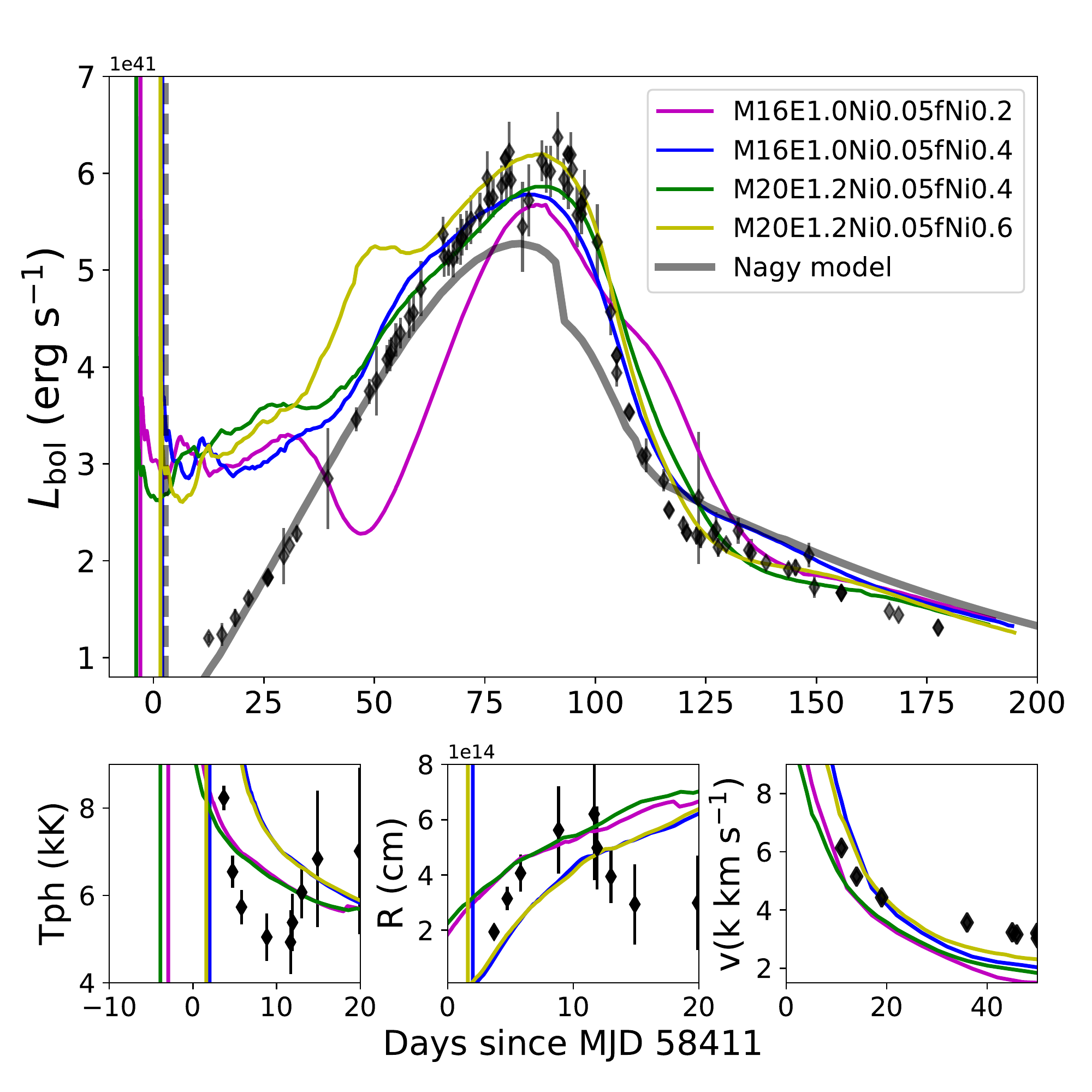}
		\caption{Bolometric light curve models for SN~2018hna overplotted with the observed bolometric light curve (\textit{top}). Light curves of models are shifted horizontally to match the peak date of the observation. The date of discovery (MJD 58413.815) is marked by the dashed vertical line. The photometric temperature, radius, and expansion velocity from STELLA models are compared to the observations (\textit{bottom}). See the text for details.}
		\label{fig:LTR}
	\end{figure}

	A semi-analytical light curve model for SNe~II was published by \cite{2014A&A...571A..77N} and further developed in \cite{2016A&A...589A..53N}. In this model, the SN ejecta are divided into two parts: the inner core and outer shell. 
	As the outer shell component only contributes to the early-time shock-cooling phase, we only consider the core component. 
	Free parameters include $M_{\mathrm{ej}}$, $R_0$, $M$(\Ni), $E_{\mathrm{k}}$, and the initial thermal energy $E_{\mathrm{th,0}}$. 
	We adopt $\kappa = 0.19$~cm$^{2}$~g$^{-1}$, the average opacity of a one-dimensional (1D) model for SNe~II, and the gamma-ray leakage exponent $A_\mathrm{g} = 2.7 \times 10^5$, as \cite{2016A&A...589A..53N} used for SN~1987A. 
	We modified the code by \cite{2016A&A...589A..53N} so that it can be used as subroutines in \texttt{python} and then applied the Markov chain Monte Carlo (MCMC) package \texttt{emcee} \citep{emcee} to do MCMC fitting to the derived bolometric light curve of SN~2018hna in the range 20--200~days after MJD = 58411.
	The resulting best-fit parameters are $M_{\mathrm{ej}} = 13.29^{+0.37}_{-1.82}$~\msun, $R_0 = 55^{+2}_{-9}$~\rsun, $M$(\Ni) = 0.052$^{+0.001}_{-0.002}$~\msun, $E_{\mathrm{k}} = 3.08^{+0.15}_{-0.85} \times 10^{51}$~erg and $E_{\mathrm{th,0}} = 0.16^{+0.02}_{-0.08} \times 10^{51}$~erg.
	The model light curve as well as the observed bolometric light curve of SN~2018hna are shown in Fig.~\ref{fig:LTR}.
	
	We also compare with the hydrodynamical results. 
	First, we use Modules for Experiments in Stellar Astrophysics \cite[MESA;][]{2011ApJS..192....3P,2013ApJS..208....4P,2015ApJS..220...15P,2018ApJS..234...34P} to evolve a massive single star to core-collapse explosion. 
	Then a quantity of energy is deposited into the innermost layer and the shock spread close to the stellar surface. 
	Finally, the 1D radiation hydrodynamics code STELLA \citep{1998ApJ...496..454B,2000ApJ...532.1132B,2006A&A...453..229B} is used to compute the SN evolution.
	
	We tested stars with an initial mass of 12--25~\msun\ and metallicity $Z = 10^{-6}$--$10^{-3}$ using MESA. 
	To obtain a BSG progenitor, we find that very low metallicity is needed. 
	A star with $M = 12$~\msun\ with $Z = 10^{-6}$ ends up as an RSG, and the one with $M = 16$~\msun and $Z = 5 \times 10^{-5}$ produces a hot star with $R > 100$~\rsun. 
	The most massive model with $M = 25$~\msun\ requires a metallicity lower than $10^{-3}$ to make a BSG SN progenitor. 
	All models that collapse as a BSG stop evolving before they become an RSG, similar to that found by \cite{2016A&A...588A...5T}.
	We obtained two pre-SN models, one with $M_{\mathrm{ZAMS}} = 16$~\msun\ and $Z = 10^{-6}$, the other with $M_{\mathrm{ZAMS}} = 20$~\msun\ and $Z = 10^{-6}$. 
	These two models are exploded in MESA with a final energy of $1.0 \times 10^{51}$~erg and $1.2 \times 10^{51}$~erg, respectively.
	The 16~\msun model has a remnant mass of 2.3~\msun\ ($M_{\mathrm{ej}}=13.7$~\msun) and $R = 44$~\rsun; the 20~\msun\ model has a remnant mass of 2.32~\msun\ ($M_{\mathrm{ej}}=17.7$~\msun) and $R = 50$~\rsun. 
	The final masses are almost the same as the initial masses owing to the low metallicity.
	We use these two models to explore the mass range of the SN ejecta.
	
	To match the bolometric light curve of SN~2018hna, we vary the following parameters: total energy of SN explosion, \Ni mass in the ejecta, and mixing length of the \Ni distribution; \Ni is uniformly distributed from the innermost shell to a fraction $f_{\mathrm{Ni}}$ of the total mass. 
	We find that the slow rise and steep decline around the main peak can be explained by a moderate mixing of \Ni ($f_{\mathrm{Ni}}=0.4$). 
	With more \Ni mixed into the outer ejecta, an additional peak would emerge before the main peak. 
	The 16~\msun\ and 20~\msun\ models can both fit the bolometric light curve except for the shock-cooling phase. 
	This is a result of high temperature of the MESA model, leading to high shock-cooling luminosity. 
	The 16~\msun\ model requires a first-light date quite close to the discovery, while the first-light date of the 20~\msun\ model is about 4 days prior to that found by \cite{2019ApJ...882L..15S} with early shock cooling light curve fitting. 
	We would expect that any models with $M_{\mathrm{ej}}\lesssim13.7$~\msun would have rise times shorter than 83.6 days, which is the time from discovery to maximum light, thus unlikely for SN~2018hna. 
	If we consider the estimate of explosion epoch by \cite{2019ApJ...882L..15S} to be relatively accurate, then the ejecta mass of  the SN tends to be less than 17.7~\msun.
	
	The main difference between the explosion parameters yielded by the semi-analytical model and the hydrodynamical models is that the former produces a kinetic energy about twice the latter. 
	The hydrodynamical models may be more plausible, as the ratio of kinetic energy to \Ni mass found by the semi-analytical model seems too high for a core-collapse supernova. 
	According to our STELLA models, SN~2018hna has ejecta mass larger than $\sim$13.7~\msun but less than $\sim$17.7~\msun. 
	Assuming a remnant mass of 1.4~\msun, the pre-explosion star should be more massive than 15~\msun. 
	We further discuss the possible origin of the progenitor of SN~2018hna in Sec.~\ref{sec:progenitor-analysis}. 
	
\section{Discussions}\label{sec:progenitor-analysis}
	In this section, we further discuss the progenitor properties of SN 2018hna and the 87A-like SNe II by comparing the \ion{O}{i}] line features and the deduced oxygen mass in the SN ejecta with those predicted by stellar evolution models. 
	
\subsection{[\ion{O}{i}] line properties}
	In Sec.~\ref{sec:bol-prop}, we derived some explosion parameters for SN~2018hna, helping constrain its progenitor properties.
	Here, we examine the progenitor properties using nebular-phase spectra, especially the [\ion{O}{i}] lines. 
	
	The strength of the [\ion{O}{i}]~$\lambda$5577 and [\ion{O}{i}]~$\lambda\lambda$6300, 6364 lines in nebular spectra of core-collapse SNe can be used to roughly estimate the ejected oxygen mass in the SN explosion \citep{1986ApJ...308..685U,1992ApJ...387..309L,1994ApJ...428L..17C,2014MNRAS.439.3694J}.
	Using the nebular spectra, we measure the flux of the [\ion{O}{i}]~$\lambda$5577 and [\ion{O}{i}]~$\lambda\lambda$6300, 6364 lines of SN~2018hna.
	For comparison, we collected nebular spectra for other 87A-type SNe and performed similar analysis for them.
	Besides SN~1987A, the nebular spectra are only available for SN~1998A and SN~2009E. 
	The mass of oxygen is calculated using the Eqs.~2 and 3 from \cite{2014MNRAS.439.3694J}. 
	We summarise the observations, measurement results of oxygen lines and the inferred mass of oxygen in Tab.~\ref{tab:neb-87Alike}.
	
	The oxygen mass of SN 2018hna is found to be 0.43–-0.73~\msun, close to that of SN 1998A (0.52~\msun), but significantly lower than that of SN~1987A (1.10--1.65~\msun).
	The result derived for SN~1987A is consistent with that inferred in the literature (1.2--1.5~\msun; \citealp{1992ApJ...387..309L,1994ApJ...428L..17C,1998ApJ...497..431K}), but our results for SN~1998A (0.52~\msun) and SN~2009E (0.13--0.46~\msun) are lower than those estimated by \cite{2012A&A...537A.141P} (i.e., 1.18--1.48~\msun\ for SN~1998A and 0.60--0.75~\msun\ for SN~2009E). 
	Differences in the derived oxygen mass are likely related to the methods adopted in the calculations. 
	For example, when adopting Eq.~3 of \cite{2012A&A...537A.141P}, the oxygen mass of SN~2009E is estimated as 0.5--0.6~\msun\, which is slightly larger than our result; while the corresponding estimates are 0.51--0.63~\msun\ for SN~1998A and 0.43--0.65~\msun\ for SN~2018hna, respectively, which are similar to our results.
	
	\begin{table*}
		\centering
		\caption{Nebular spectral properties of 87A-like supernovae.}\label{tab:neb-87Alike}
		\begin{tabular}{lcccccc}
			\hline
			SN  & $M_{\mathrm{Ni}}$ & Phase & $L$([\ion{O}{i}] $\lambda\lambda$6300, 6364) & $L$([\ion{O}{i}] $\lambda$5577)) & $M$([\ion{O}{i}])$^\mathrm{LTE}_{\beta_{\mathrm{ratio}}=1.5,\beta_{6300}=0.5}$ &Data source\\
			&(\msun)           &(day) &(erg~s$^{-1}$)              &(erg~s$^{-1}$)         &(\msun)   &      \\
			\hline
			SN~1987A  &0.075  &198  &3.62$~\times~10^{39}$  &5.74$~\times~10^{38}$  &1.33  &1\\
			&       &339  &2.36$~\times~10^{39}$  &1.81$~\times~10^{38}$  &1.65  &1\\
			&       &360  &2.01$~\times~10^{39}$  &1.44$~\times~10^{38}$  &1.49  &1\\
			&       &375  &1.85$~\times~10^{39}$  &1.21$~\times~10^{38}$  &1.48  &1\\
			&       &408  &1.40$~\times~10^{39}$  &8.38$~\times~10^{37}$  &1.22  &1\\
			&       &428  &1.21$~\times~10^{39}$  &6.84$~\times~10^{37}$  &1.10  &1\\
			&       &526  &3.44$~\times~10^{38}$  &1.64$~\times~10^{37}$  &0.37  &1\\
			\hline
			SN~1998A  &0.11   &161  &2.34$~\times~10^{39}$  &6.57$~\times~10^{38}$  &0.52  &2\\
			&       &343  &1.45$~\times~10^{39}$  &Not covered                   &--    &2\\
			&       &416  &2.82$~\times~10^{38}$  &--          &--    &2\\
			\hline
			SN~2009E  &0.04   &198  &6.44$~\times~10^{38}$  &2.13$~\times~10^{38}$  &0.13  &3\\
			&       &226  &6.73$~\times~10^{38}$  &1.54$~\times~10^{38}$  &0.19  &3\\
			&       &386  &3.69$~\times~10^{38}$  &1.56$~\times~10^{37}$  &0.46  &3\\
			&       &388  &3.07$~\times~10^{38}$  &<3.29$~\times~10^{37}$  &>0.17  &3\\
			\hline
			SN~2018hna &0.05      &247  &1.59$~\times~10^{39}$  &2.10$~\times~10^{38}$  &0.73  &This work\\
			&      &253  &1.13$~\times~10^{39}$  &1.77$~\times~10^{38}$  &0.44  &4\\
			&      &255  &1.16$~\times~10^{39}$  &1.56$~\times~10^{38}$  &0.53  &4\\
			&      &522  &9.61$~\times~10^{37}$  &--                   &--    &This work\\
			\hline
		\end{tabular}
		\\
		{\centering
			\textbf{References}. (1) \cite{1995ApJS...99..223P}; (2) \cite{2005MNRAS.360..950P}; (3) \cite{2012AA...537A.141P}; (4) \cite{2019ApJ...882L..15S}.\par}
	\end{table*}
	
\begin{figure}
		\includegraphics[width=1.1\columnwidth]{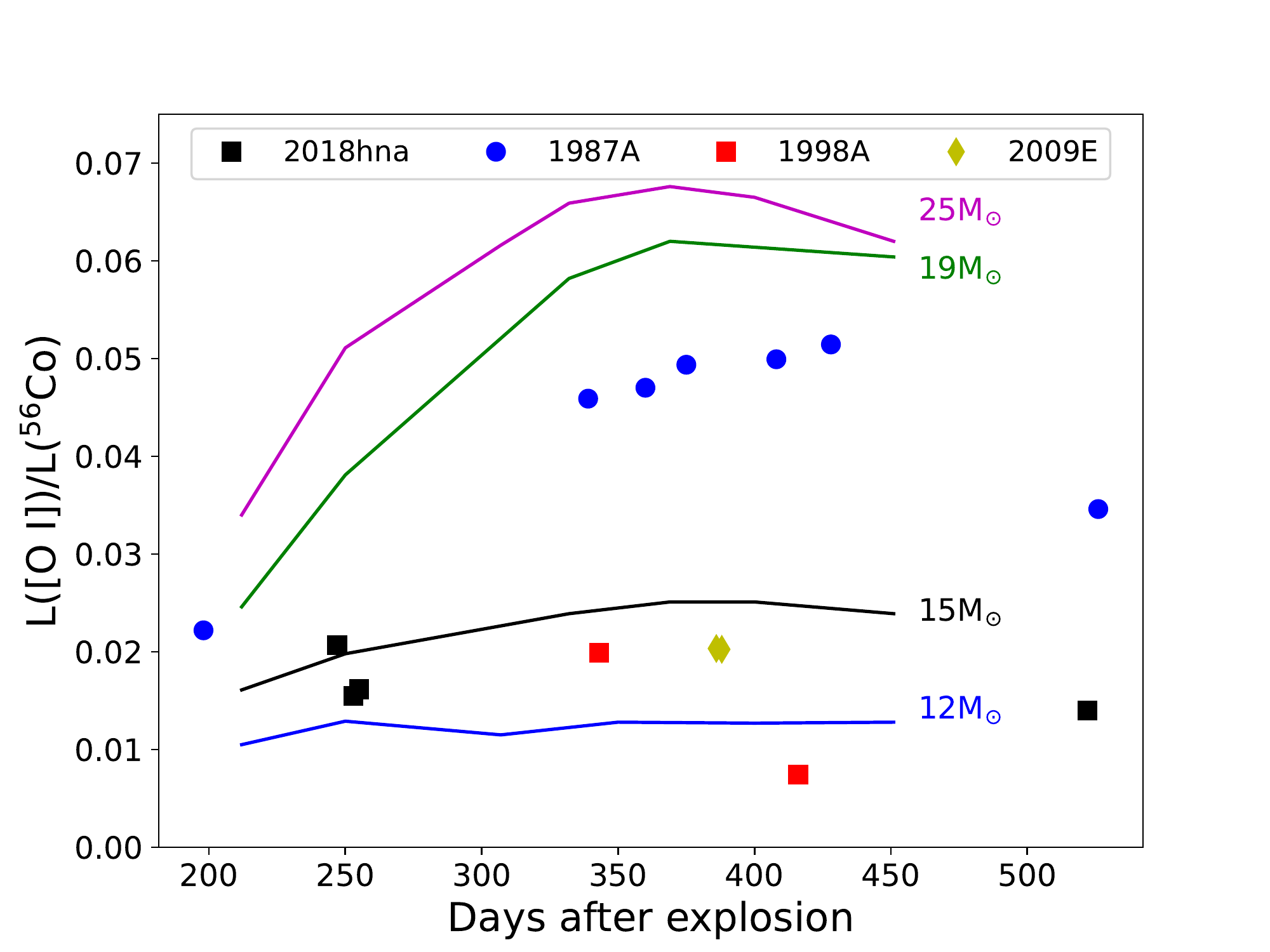}
		\caption{The nebular-phase evolution of [\ion{O}{i}]~$\mathrm{\lambda\lambda}$6300, 6364 line luminosities (normalised to the $^{56}$Co decay power) of four 87A-like SNe, compared with model predictions. The solid lines represent the model predictions with  initial masses of 12 M$_{\odot}$ (blue), 15 M$_{\odot}$ (black), 19 M$_{\odot}$ (green), and 25 M$_{\odot}$ (pink), respectively.}
		\label{fig:LOI-LCo}
	\end{figure}

	Standard stellar evolution theories predict that the oxygen mass in the SN ejecta is positively correlated with the initial mass of the exploding star \citep[e.g.,][]{1995ApJS..101..181W}.
	Metallicity of the host galaxy of SN~2018hna is found to be about 0.3~Z$_{\odot}$ \citep{2019ApJ...882L..15S}. 
	While considering that the SN is located far from the galaxy centre, SN~2018hna could be metal-poorer (i.e. $Z<$0.3~Z$_{\odot}$).
	Comparing the ejected oxygen mass derived for SN~2018hna with that predicted by standard single star evolution models at different metallicities ($Z=$\,Z$_{\odot}$, \citet{2016ApJ...821...38S} and $Z=0.1$\,Z$_{\odot}$, \citet{1995ApJS..101..181W}), the initial mass of SN~2018hna is constrained to have a range of 12--15.5~\msun, while a corresponding range of 16--19~\msun\ is found for SN~1987A. 
	The low initial masses of SNe~2018hna, 1998A and 2009E are in agreement with low mass NLTE spectra models (see Appendix~\ref{sec:neb-SNe} for details). 
	The distinction between SN~1987A and the other three 87A-like SNe can be also seen from the evolution of their [\ion{O}{i}] line luminosities. 
	Like in \cite{2015MNRAS.448.2482J}, we compare the evolution of nebular [\ion{O}{i}]~$\mathrm{\lambda\lambda}$6300, 6364 luminosity with that of the models from \cite{2014MNRAS.439.3694J} in Fig.~\ref{fig:LOI-LCo}. 
	Among the four 87A-like objects, only SN~1987A lies above the 15~\msun\ model, while the other three show the distribution similar to normal SNe~II \cite[see Fig.~9 of][]{2015MNRAS.448.2482J}. 
	
	Through the above analysis of the nebular [\ion{O}{i}] lines, we find that under the assumption of standard single-star evolution, the initial mass of the progenitor star of SN~2018hna will not exceed 16~\msun. 
	However, in standard single-star evolution model, a star with $M_{\mathrm{ZAMS}} \lesssim 16$~\msun\ will evolve to be a BSG prior to core collapse only under extremely low metallicity ($Z \approx 10^{-6}$) condition. 
	This is not the case of any 87A-like SNe, which implies that the progenitor stars of 87A-like SNe are unlikely to have evolved as a single star through standard evolution. 
	Instead, some nonstandard mechanisms are needed, or the progenitor stars are involved in binary or multiple systems. 
	As for SN~1987A, a binary scenario is more favorable. 
	We notice that the large gap of [\ion{O}{i}] line properties between SN~1987A and others may indicate that their progenitors are essentially different, although similar ejected mass and energy can be derived from their light curves. 
	In the next subsection, we will extend our discussions about the possible progenitor origins of SN~2018hna. 
	
\subsection{Possible origins of the progenitors of 87A-like SNe}\label{sec:vsIIP}
	It is now well known that 87A-like SNe are located in lower metallicity galaxies than SNe~IIP, and the partially stripped type IIb SNe have a similar metallicity distribution to that of 87A-like SNe \cite[e.g.,][]{2013A&A...558A.143T}.
	Apparently, differences in metallicity alone cannot make stars evolve in such different ways, since most H-rich SNe arising from similar metal-poor environment tend to become Type IIP \citep[e.g., SN~2015bs arising from $Z \lesssim 0.1$\,Z$_{\odot}$;][]{2018NatAs...2..574A}.
	According to standard stellar evolution theory, stars with initial masses in a range of 8 to 30~\msun\ would end up as RSGs, although no RSG progenitor with mass $>$ 20~\msun\ has been found yet. 
	Archival images of SNe~IIP progenitors indicate an average initial mass of $\sim$10~\msun\ \citep{2015PASA...32...16S}. 
	Progenitors of 87A-like SNe are not settled yet, while there are several possibilities for both single stars and binaries. 
	We discuss these probabilities below.
	
	The ejecta mass of 87A-like SNe can be derived from their light curves using empirical relations --- the relation between diffusion time, SN energy, and mass \citep{1989ApJ...340..396A} --- or analytical/hydrodynamical light-curve models \cite[e.g.,][]{1980ApJ...237..541A,2000ApJ...532.1132B,2003MNRAS.338..711Z,2014A&A...571A..77N,2016A&A...589A..53N}. 
	By examining these SNe, we find that their ejecta masses lie in the range of $\sim 12$--26~\msun, with a typical value of $\sim 17$~\msun. 
	It seems that 87A-like SNe have higher ejected total mass as well as \Ni mass than normal SNe~IIP, while they share similar velocity evolution (see Fig.~\ref{fig:vel-comp}). 
	Thus, we conclude that 87A-like SNe have more massive progenitors and suffer more energetic explosions than normal SNe~II. 
	However, it is shown above that some 87A-like SNe should be relatively oxygen-poor, with SN~1987A itself being an exception. 
	Progenitors of those oxygen-poor 87A-like events are likely to be generated through different channels than SN~1987A.
	
	There must be some mechanisms that reduce the oxygen mass in SN ejecta.
	It is believed that the progenitor star of SN~1987A has undergone blue-red-blue evolution; the abundance ratios of N/C and N/O in the circumstellar gas around the SN are large compared to solar ratios, which was interpreted as an evidence of experiencing mass loss when the star was an RSG. 
	The conditions under which this process occurs were investigated by \cite{1988ApJ...331..388S}, and they concluded that either high envelope mass or enhanced He abundance ($Y$) in the H-rich envelope after core He burning is needed. 
	High envelope mass can be a result of unusually low metallicity and zero mass loss \cite[e.g.,][but it was ruled out for SN~1987A]{1988A&A...197L..11W} or simply higher initial mass. 
	Apparently, the low oxygen mass derived above indicates that SN~2018hna (also SN~1998A and SN~2009E) cannot be produced by a single massive star with standard evolution; we need other mechanisms. 
	Higher $Y$ can be acquired by mixing due to semiconvection and shell convection or rotation. 
	It is easier to reach the acquired $Y$ under lower metallicity. 
	
	Single-star models for a BSG SN progenitor have been discussed extensively in literature  \citep{1987ApJ...319..136A,1987Natur.327..597H,1988ApJ...324..466W,1988A&A...197L..11W,1988ApJ...331..388S,1989ApJ...339..365W,1991A&A...243..155L,2015A&A...584A..54P,2019A&A...624A.116U}.
	\cite{2019A&A...624A.116U} calculated the explosion of several single-star models; their B15-2, W16-3, W18x2, and W20 all have ejected oxygen mass $<0.7$~\msun, with B15 having the least amount of oxygen (i.e., 0.16~\msun). 
	Considering both ejecta mass and oxygen mass, W16-3 is suitable for SN~2018hna. 
	The progenitor of the W16-3 model is evolved by invoking reduced metallicity ($Z=1/3$\,Z$_{\odot}$, close to that of SN~2018hna), restricted semiconvection, and a substantial amount of rotation ($J_{\mathrm{ZAMS}}=2.7 \times 10^{52}$~erg; \citealt{2016ApJ...821...38S}). 
	So, it is still likely that these oxygen-poor events originated from a single-star channel.
	
	If the progenitor star has been in a binary system, accretion from the companion or merging of the stars can enlarge the envelope hence produce BSG SN progenitors.
	Binary models have been presented by \cite{1989Natur.338..401P}, \cite{1989A&A...219L...3H}, \cite{1992ApJ...391..246P} with binary interaction (mass transfer), \cite{2007AIPC..937..125P}, \cite{2017MNRAS.469.4649M}, and \cite{2021ApJ...914....4U} with star mergers. 
	These models successfully match the properties of SN~1987A and its progenitor star, especially the triple-ring nebula around it, which is a direct evidence of a star merger. 
	It is also possible that the progenitors of the oxygen-poor events were evolved from binary interaction, but the ejected oxygen masses are substantially smaller than that of SN~1987A, hence much lower initial mass of the binary system. 
	This may also be related to the lower ratio of oxygen-rich events (so far only 1 out of 4), as the proportion of lower-mass stars is thought to be higher. 
	In addition, the rarity of 87A-like events in SNe~II is also in line with the expectations of binary star evolution \citep{1992ApJ...391..246P,2002RvMP...74.1015W}.
	In the case of SN~2018hna, the similar polarization properties to SN~1987A might also make the binary picture more favourable.
	
\section{Summary}\label{sec:summary}
	The peculiar H-rich SN~2018hna is studied, with new optical photometry and spectra presented in addition to those by \cite{2019ApJ...882L..15S}.
	Like SN~1987A, SN~2018hna shows a slow rise to maximum light in $\sim 86$~days, with a peak bolometric luminosity of $(6.2 \pm 3.1) \times 10^{41}$~erg~s$^{-1}$. 
	A \Ni mass of $(0.048 \pm 0.024)$~\msun\ was derived from the linear declining tail after 150 days. Both the photometric and spectroscopic evolution of SN~2018hna behave like the famous SN 1987A, making it another well-observed peculiar SN~II with observations covering from the early shock-cooling phase to more than 550 days after the explosion. 
	
	In addition, we collected a small sample of 87A-like SNe~II with good photometric and spectroscopic observations in the literature and studied some of their properties in comparison with SN~2018hna. 
	The $B-V$ colour curves of 87A-like SNe all show two peaks, and they share similar evolution before the first peak and after the second peak ($t>100$ days). 
	For them, we find that the earlier the first peak occurs, the shorter time it will take the $B-V$ colour to go redward again. 
	The expansion velocity evolution of SN~2018hna is amont the range of the 87A-like SN family. 
	In the photospheric phase, the spectral evolution of SN~2018hna shows high resemblances to SN~2006V, with weak metal lines and H$\mathrm{\alpha}$ absorption. 
	Metal lines, such as \ion{Fe}{ii} and \ion{Ba}{ii} are unusually strong among this subtype. 
	We measured the pEW of the \ion{Ba}{ii}~$\mathrm{\lambda}$6142 line in the spectra of 87A-like SNe; and we found that it is significantly stronger in SN~1987A and SN~2009E than in other objects.
	The pEW of \ion{Ba}{ii}~$\mathrm{\lambda}$6142 is positively related to the $B-V$ colour of the SNe, implying that low temperature may be the leading factor responsible for strong \ion{Ba}{ii} lines. 
	Nevertheless,  the photospheric temperature cannot fully explain the observed difference.
	Some Ba-strong events are found to have real enhanced Ba abundance in their ejecta. 
	In addition, lower \Ni mass might also be responsible for stronger \ion{Ba}{ii} lines. 
	Further studies are needed for definite conclusions.
	
	We calculated the bolometric light curve of SN~2018hna using its multiband light curves; a stellar radius of $R \approx 45$~\rsun\ is derived from the temperature evolution. 
	Two methods were used to fit the bolometric light curve of SN~2018hna to estimate its explosion parameters. 
	Using the semi-analytical model by \cite{2016A&A...589A..53N}, we found a best-fit model with $M_{\mathrm{ej}} \approx 13.29$~\msun, $R_0 \approx 55$~\rsun, $M$(\Ni) $\approx 0.052$~\msun, $E_{\mathrm{k}} \approx  3.08 \times 10^{51}$~erg and $E_{\mathrm{th,0}} = 0.16 \times 10^{51}$~erg. 
	The ratio of $E_{\mathrm{k}}$ to $M$(\Ni) given by this model seems too high for a core-collapse supernova. 
	On the other hand, the hydrodynamic method (built by MESA and STELLA) gives more plausible fit with $M_{\mathrm{ej}} = 13.7-17.7$~\msun, $E_{\mathrm{k}} = 1.0-1.2\times 10^{51}$~erg, $M_{\mathrm{^{56}Ni}} = 0.05$~\msun, and \Ni mixing length scale $f_{\mathrm{Ni}} = 0.4$.
	
	In order to study the properties of the progenitor star of SN~2018hna, we investigated the luminosities of the [\ion{O}{i}] lines in its nebular phase spectra. 
	As comparison, we also collected nebular spectra of other 87A-like SNe and found two of them (SN~1998A and SN~2009E) in addition to SN~1987A. 
	The nebular [\ion{O}{i}] lines are used to estimate the ejected oxygen masses and hence the initial masses of their progenitors. 
	We estimated the ejected oxygen mass of SN~2018hna to be 0.44--0.73~\msun\ using the [\ion{O}{i}]~$\mathrm{\lambda}$5577 and [\ion{O}{i}]~$\mathrm{\lambda\lambda}$6300, 6364 lines in the nebular spectra. 
	Similar low oxygen masses were found for SN~1998A and SN~2009E. 
	SN~1987A has significantly larger mass of ejected oxygen (1.10--1.65~\msun). 
	The distinction between SN~1987A and SN~2018hna can also be seen from their evolution of the luminosities of neular [\ion{O}{i}]~$\mathrm{\lambda\lambda}$6300, 6364 lines. 
	
	The gap in ejected oxygen mass and behaviour of [\ion{O}{i}] lines in the nebular spectra seen between SN~1987A and SN~2018hna indicate that their progenitors were probably formed through different channels. 
	Some single-star models with reduced metallicity, restricted semiconvection, and a substantial amount of rotation from \cite{2016ApJ...821...38S} might match the low oxygen mass of SN~2018hna. 
	Unlike in the case of SN~1987A where the single-star picture is almost ruled out, it is possible that SN~2018hna was originated from a single massive star. 
	On the other hand, the binary scenario can be plausible. 
	Given the dominance of binary interactions in the evolution of massive stars, and the similar polarization properties to SN~1987A, the binary scenario may be more likely for SN~2018hna. 
	In the binary picture, the low oxygen mass in SN~2018hna may be a result of lower masses of the binary components or less material that was accreted from the companion. 
	More observational and theoretical research is required to uncover the mystery of BSG supernova progenitors.
	The oxygen mass in SN ejecta can be a key point in future studies, and analysis of the explosion rate of 87A-like SNe can be helpful.

\section*{Acknowledgements}
	We acknowledge the support of the staff of the XLT, LJT, TNT, ARC, and Keck-I for assistance with the observations. The operation of XLT is supported by the Open Project Program of the Key Laboratory of Optical Astronomy, National Astronomical Observatories, Chinese Academy of Sciences. Funding for the LJT has been provided by the Chinese Academy of Sciences and the People's Government of Yunnan Province. The LJT is jointly operated and administrated by Yunnan Observatories and Center for Astronomical Mega-Science, CAS. Some of the data presented herein were obtained at the W. M. Keck Observatory, which is operated as a scientific partnership among the California Institute of Technology, the University of California, and NASA; the observatory was made possible by the generous financial support of the W. M. Keck Foundation.
	Distance and Galactic extinction to SN~2018hna are obtained via the NASA/IPAC Extragalactic Database (NED). 
	Photometric and spectral data of referenced SNe were obtained via the the Open Supernova Catalog \cite[OSC;][]{2017ApJ...835...64G} and the Weizmann Interactive Supernova Data Repository \cite[WISeREP;][]{2012PASP..124..668Y}.
		
	This work is supported by the National Natural Science Foundation of China (NSFC grants 12288102, 12033003, and 11633002), the Scholar Program of Beijing Academy of Science and Technology (DZ:BS202002), and the Tencent Xplorer Prize. A.V.F.'s group at U.C. Berkeley received financial support from the Christopher R. Redlich Fund and many individual donors. SB is supported by grant RSF 19-12-00229 for the development of STELLA code.
		
\section*{Data Availability}
		
	Photometric data of SN~2018hna from the TNT are presented in Tab.~\ref{tab:lc_tnt}. 
	Spectroscopic data are available on reasonable request to the corresponding authors.
		
		
		
	\bibliographystyle{mnras}
	\bibliography{ref} 

\begin{thebibliography}{}
\makeatletter
\relax
\def\mn@urlcharsother{\let\do\@makeother \do\$\do\&\do\#\do\^\do\_\do\%\do\~}
\def\mn@doi{\begingroup\mn@urlcharsother \@ifnextchar [ {\mn@doi@}
  {\mn@doi@[]}}
\def\mn@doi@[#1]#2{\def\@tempa{#1}\ifx\@tempa\@empty \href
  {http://dx.doi.org/#2} {doi:#2}\else \href {http://dx.doi.org/#2} {#1}\fi
  \endgroup}
\def\mn@eprint#1#2{\mn@eprint@#1:#2::\@nil}
\def\mn@eprint@arXiv#1{\href {http://arxiv.org/abs/#1} {{\tt arXiv:#1}}}
\def\mn@eprint@dblp#1{\href {http://dblp.uni-trier.de/rec/bibtex/#1.xml}
  {dblp:#1}}
\def\mn@eprint@#1:#2:#3:#4\@nil{\def\@tempa {#1}\def\@tempb {#2}\def\@tempc
  {#3}\ifx \@tempc \@empty \let \@tempc \@tempb \let \@tempb \@tempa \fi \ifx
  \@tempb \@empty \def\@tempb {arXiv}\fi \@ifundefined
  {mn@eprint@\@tempb}{\@tempb:\@tempc}{\expandafter \expandafter \csname
  mn@eprint@\@tempb\endcsname \expandafter{\@tempc}}}

\bibitem[\protect\citeauthoryear{{Anderson} et~al.,}{{Anderson}
  et~al.}{2014}]{2014ApJ...786...67A}
{Anderson} J.~P.,  et~al., 2014, \mn@doi [\apj] {10.1088/0004-637X/786/1/67},
  \href {https://ui.adsabs.harvard.edu/abs/2014ApJ...786...67A} {786, 67}

\bibitem[\protect\citeauthoryear{{Anderson} et~al.,}{{Anderson}
  et~al.}{2018}]{2018NatAs...2..574A}
{Anderson} J.~P.,  et~al., 2018, \mn@doi [Nature Astronomy]
  {10.1038/s41550-018-0458-4}, \href
  {https://ui.adsabs.harvard.edu/abs/2018NatAs...2..574A} {2, 574}

\bibitem[\protect\citeauthoryear{{Arnett}}{{Arnett}}{1980}]{1980ApJ...237..541A}
{Arnett} W.~D.,  1980, \mn@doi [\apj] {10.1086/157898}, \href
  {https://ui.adsabs.harvard.edu/abs/1980ApJ...237..541A} {237, 541}

\bibitem[\protect\citeauthoryear{{Arnett}}{{Arnett}}{1982}]{1982ApJ...253..785A}
{Arnett} W.~D.,  1982, \mn@doi [\apj] {10.1086/159681}, \href
  {https://ui.adsabs.harvard.edu/abs/1982ApJ...253..785A} {253, 785}

\bibitem[\protect\citeauthoryear{{Arnett}}{{Arnett}}{1987}]{1987ApJ...319..136A}
{Arnett} W.~D.,  1987, \mn@doi [\apj] {10.1086/165439}, \href
  {https://ui.adsabs.harvard.edu/abs/1987ApJ...319..136A} {319, 136}

\bibitem[\protect\citeauthoryear{{Arnett} \& {Fu}}{{Arnett} \&
  {Fu}}{1989}]{1989ApJ...340..396A}
{Arnett} W.~D.,  {Fu} A.,  1989, \mn@doi [\apj] {10.1086/167402}, \href
  {https://ui.adsabs.harvard.edu/abs/1989ApJ...340..396A} {340, 396}

\bibitem[\protect\citeauthoryear{{Bellm} et~al.,}{{Bellm}
  et~al.}{2019}]{2019PASP..131a8002B}
{Bellm} E.~C.,  et~al., 2019, \mn@doi [\pasp] {10.1088/1538-3873/aaecbe}, \href
  {https://ui.adsabs.harvard.edu/abs/2019PASP..131a8002B} {131, 018002}

\bibitem[\protect\citeauthoryear{{Bevan}}{{Bevan}}{2018}]{2018MNRAS.480.4659B}
{Bevan} A.,  2018, \mn@doi [\mnras] {10.1093/mnras/sty2094}, \href
  {https://ui.adsabs.harvard.edu/abs/2018MNRAS.480.4659B} {480, 4659}

\bibitem[\protect\citeauthoryear{{Bevan} \& {Barlow}}{{Bevan} \&
  {Barlow}}{2016}]{2016MNRAS.456.1269B}
{Bevan} A.,  {Barlow} M.~J.,  2016, \mn@doi [\mnras] {10.1093/mnras/stv2651},
  \href {https://ui.adsabs.harvard.edu/abs/2016MNRAS.456.1269B} {456, 1269}

\bibitem[\protect\citeauthoryear{{Bevan} et~al.,}{{Bevan}
  et~al.}{2020}]{2020ApJ...894..111B}
{Bevan} A.~M.,  et~al., 2020, \mn@doi [\apj] {10.3847/1538-4357/ab86a2}, \href
  {https://ui.adsabs.harvard.edu/abs/2020ApJ...894..111B} {894, 111}

\bibitem[\protect\citeauthoryear{{Blinnikov}, {Eastman}, {Bartunov},
  {Popolitov}  \& {Woosley}}{{Blinnikov} et~al.}{1998}]{1998ApJ...496..454B}
{Blinnikov} S.~I.,  {Eastman} R.,  {Bartunov} O.~S.,  {Popolitov} V.~A.,
  {Woosley} S.~E.,  1998, \mn@doi [\apj] {10.1086/305375}, \href
  {https://ui.adsabs.harvard.edu/abs/1998ApJ...496..454B} {496, 454}

\bibitem[\protect\citeauthoryear{{Blinnikov}, {Lundqvist}, {Bartunov}, {Nomoto}
   \& {Iwamoto}}{{Blinnikov} et~al.}{2000}]{2000ApJ...532.1132B}
{Blinnikov} S.,  {Lundqvist} P.,  {Bartunov} O.,  {Nomoto} K.,   {Iwamoto} K.,
  2000, \mn@doi [\apj] {10.1086/308588}, \href
  {https://ui.adsabs.harvard.edu/abs/2000ApJ...532.1132B} {532, 1132}

\bibitem[\protect\citeauthoryear{{Blinnikov}, {R{\"o}pke}, {Sorokina},
  {Gieseler}, {Reinecke}, {Travaglio}, {Hillebrandt}  \&
  {Stritzinger}}{{Blinnikov} et~al.}{2006}]{2006A&A...453..229B}
{Blinnikov} S.~I.,  {R{\"o}pke} F.~K.,  {Sorokina} E.~I.,  {Gieseler} M.,
  {Reinecke} M.,  {Travaglio} C.,  {Hillebrandt} W.,   {Stritzinger} M.,  2006,
  \mn@doi [\aap] {10.1051/0004-6361:20054594}, \href
  {https://ui.adsabs.harvard.edu/abs/2006A&A...453..229B} {453, 229}

\bibitem[\protect\citeauthoryear{{Catchpole} et~al.,}{{Catchpole}
  et~al.}{1987}]{1987MNRAS.229P..15C}
{Catchpole} R.~M.,  et~al., 1987, \mn@doi [\mnras] {10.1093/mnras/229.1.15P},
  \href {https://ui.adsabs.harvard.edu/abs/1987MNRAS.229P..15C} {229, 15P}

\bibitem[\protect\citeauthoryear{{Catchpole} et~al.,}{{Catchpole}
  et~al.}{1988}]{1988MNRAS.231P..75C}
{Catchpole} R.~M.,  et~al., 1988, \mn@doi [\mnras] {10.1093/mnras/231.1.75P},
  \href {https://ui.adsabs.harvard.edu/abs/1988MNRAS.231P..75C} {231, 75P}

\bibitem[\protect\citeauthoryear{{Chugai}}{{Chugai}}{1994}]{1994ApJ...428L..17C}
{Chugai} N.~N.,  1994, \mn@doi [\apjl] {10.1086/187382}, \href
  {https://ui.adsabs.harvard.edu/abs/1994ApJ...428L..17C} {428, L17}

\bibitem[\protect\citeauthoryear{{Chugai}}{{Chugai}}{2001}]{2001MNRAS.326.1448C}
{Chugai} N.~N.,  2001, \mn@doi [\mnras] {10.1111/j.1365-2966.2001.04717.x},
  \href {https://ui.adsabs.harvard.edu/abs/2001MNRAS.326.1448C} {326, 1448}

\bibitem[\protect\citeauthoryear{{Dessart} \& {Hillier}}{{Dessart} \&
  {Hillier}}{2005}]{2005A&A...437..667D}
{Dessart} L.,  {Hillier} D.~J.,  2005, \mn@doi [\aap]
  {10.1051/0004-6361:20042525}, \href
  {https://ui.adsabs.harvard.edu/abs/2005A&A...437..667D} {437, 667}

\bibitem[\protect\citeauthoryear{{Dessart} \& {Hillier}}{{Dessart} \&
  {Hillier}}{2008}]{2008MNRAS.383...57D}
{Dessart} L.,  {Hillier} D.~J.,  2008, \mn@doi [\mnras]
  {10.1111/j.1365-2966.2007.12538.x}, \href
  {https://ui.adsabs.harvard.edu/abs/2008MNRAS.383...57D} {383, 57}

\bibitem[\protect\citeauthoryear{{Dessart} \& {Hillier}}{{Dessart} \&
  {Hillier}}{2010}]{2010MNRAS.405.2141D}
{Dessart} L.,  {Hillier} D.~J.,  2010, \mn@doi [\mnras]
  {10.1111/j.1365-2966.2010.16611.x}, \href
  {https://ui.adsabs.harvard.edu/abs/2010MNRAS.405.2141D} {405, 2141}

\bibitem[\protect\citeauthoryear{{Dessart} \& {Hillier}}{{Dessart} \&
  {Hillier}}{2019}]{2019A&A...622A..70D}
{Dessart} L.,  {Hillier} D.~J.,  2019, \mn@doi [\aap]
  {10.1051/0004-6361/201833966}, \href
  {https://ui.adsabs.harvard.edu/abs/2019A&A...622A..70D} {622, A70}

\bibitem[\protect\citeauthoryear{{Dessart} \& {Hillier}}{{Dessart} \&
  {Hillier}}{2022}]{2022A&A...660L...9D}
{Dessart} L.,  {Hillier} D.~J.,  2022, \mn@doi [\aap]
  {10.1051/0004-6361/202243372}, \href
  {https://ui.adsabs.harvard.edu/abs/2022A&A...660L...9D} {660, L9}

\bibitem[\protect\citeauthoryear{{Dessart}, {Hillier}  \& {Wilk}}{{Dessart}
  et~al.}{2018}]{2018A&A...619A..30D}
{Dessart} L.,  {Hillier} D.~J.,   {Wilk} K.~D.,  2018, \mn@doi [\aap]
  {10.1051/0004-6361/201833278}, \href
  {https://ui.adsabs.harvard.edu/abs/2018A&A...619A..30D} {619, A30}

\bibitem[\protect\citeauthoryear{{Dessart}, {Hillier}, {Sukhbold}, {Woosley}
  \& {Janka}}{{Dessart} et~al.}{2021}]{2021A&A...652A..64D}
{Dessart} L.,  {Hillier} D.~J.,  {Sukhbold} T.,  {Woosley} S.~E.,   {Janka}
  H.~T.,  2021, \mn@doi [\aap] {10.1051/0004-6361/202140839}, \href
  {https://ui.adsabs.harvard.edu/abs/2021A&A...652A..64D} {652, A64}

\bibitem[\protect\citeauthoryear{{Falco} et~al.,}{{Falco}
  et~al.}{1999}]{1999PASP..111..438F}
{Falco} E.~E.,  et~al., 1999, \mn@doi [\pasp] {10.1086/316343}, \href
  {https://ui.adsabs.harvard.edu/abs/1999PASP..111..438F} {111, 438}

\bibitem[\protect\citeauthoryear{{Filippenko}}{{Filippenko}}{1997}]{1997ARA&A..35..309F}
{Filippenko} A.~V.,  1997, \mn@doi [\araa] {10.1146/annurev.astro.35.1.309},
  \href {https://ui.adsabs.harvard.edu/abs/1997ARA&A..35..309F} {35, 309}

\bibitem[\protect\citeauthoryear{{Fitzpatrick}}{{Fitzpatrick}}{1999}]{1999PASP..111...63F}
{Fitzpatrick} E.~L.,  1999, \mn@doi [\pasp] {10.1086/316293}, \href
  {https://ui.adsabs.harvard.edu/abs/1999PASP..111...63F} {111, 63}

\bibitem[\protect\citeauthoryear{{Foreman-Mackey}, {Hogg}, {Lang}  \&
  {Goodman}}{{Foreman-Mackey} et~al.}{2013}]{emcee}
{Foreman-Mackey} D.,  {Hogg} D.~W.,  {Lang} D.,   {Goodman} J.,  2013, \mn@doi
  [PASP] {10.1086/670067}, 125, 306

\bibitem[\protect\citeauthoryear{{Gal-Yam}}{{Gal-Yam}}{2017}]{2017hsn..book..195G}
{Gal-Yam} A.,  2017, in {Alsabti} A.~W.,  {Murdin} P.,  eds, , Handbook of
  Supernovae.
p.~195, \mn@doi{10.1007/978-3-319-21846-5_35}

\bibitem[\protect\citeauthoryear{{Gehrels} et~al.,}{{Gehrels}
  et~al.}{2004}]{2004ApJ...611.1005G}
{Gehrels} N.,  et~al., 2004, \mn@doi [\apj] {10.1086/422091}, \href
  {https://ui.adsabs.harvard.edu/abs/2004ApJ...611.1005G} {611, 1005}

\bibitem[\protect\citeauthoryear{{Guillochon}, {Parrent}, {Kelley}  \&
  {Margutti}}{{Guillochon} et~al.}{2017}]{2017ApJ...835...64G}
{Guillochon} J.,  {Parrent} J.,  {Kelley} L.~Z.,   {Margutti} R.,  2017,
  \mn@doi [\apj] {10.3847/1538-4357/835/1/64}, \href
  {https://ui.adsabs.harvard.edu/abs/2017ApJ...835...64G} {835, 64}

\bibitem[\protect\citeauthoryear{{Guti{\'e}rrez} et~al.,}{{Guti{\'e}rrez}
  et~al.}{2017}]{2017ApJ...850...89G}
{Guti{\'e}rrez} C.~P.,  et~al., 2017, \mn@doi [\apj]
  {10.3847/1538-4357/aa8f52}, \href
  {https://ui.adsabs.harvard.edu/abs/2017ApJ...850...89G} {850, 89}

\bibitem[\protect\citeauthoryear{{Hamuy} \& {Pinto}}{{Hamuy} \&
  {Pinto}}{2002}]{2002ApJ...566L..63H}
{Hamuy} M.,  {Pinto} P.~A.,  2002, \mn@doi [\apjl] {10.1086/339676}, \href
  {https://ui.adsabs.harvard.edu/abs/2002ApJ...566L..63H} {566, L63}

\bibitem[\protect\citeauthoryear{{Hillebrandt} \& {Meyer}}{{Hillebrandt} \&
  {Meyer}}{1989}]{1989A&A...219L...3H}
{Hillebrandt} W.,  {Meyer} F.,  1989, \aap, \href
  {https://ui.adsabs.harvard.edu/abs/1989A&A...219L...3H} {219, L3}

\bibitem[\protect\citeauthoryear{{Hillebrandt}, {Hoeflich}, {Weiss}  \&
  {Truran}}{{Hillebrandt} et~al.}{1987}]{1987Natur.327..597H}
{Hillebrandt} W.,  {Hoeflich} P.,  {Weiss} A.,   {Truran} J.~W.,  1987, \mn@doi
  [\nat] {10.1038/327597a0}, \href
  {https://ui.adsabs.harvard.edu/abs/1987Natur.327..597H} {327, 597}

\bibitem[\protect\citeauthoryear{{Hoeflich}}{{Hoeflich}}{1988}]{1988PASA....7..434H}
{Hoeflich} P.,  1988, \mn@doi [\pasa] {10.1017/S1323358000022608}, \href
  {https://ui.adsabs.harvard.edu/abs/1988PASA....7..434H} {7, 434}

\bibitem[\protect\citeauthoryear{{Huang}, {Li}, {Wang}, {Shang}, {Zhang}, {Hu},
  {Qiu}  \& {Jiang}}{{Huang} et~al.}{2012}]{2012RAA....12.1585H}
{Huang} F.,  {Li} J.-Z.,  {Wang} X.-F.,  {Shang} R.-C.,  {Zhang} T.-M.,  {Hu}
  J.-Y.,  {Qiu} Y.-L.,   {Jiang} X.-J.,  2012, \mn@doi [Research in Astronomy
  and Astrophysics] {10.1088/1674-4527/12/11/012}, \href
  {https://ui.adsabs.harvard.edu/abs/2012RAA....12.1585H} {12, 1585}

\bibitem[\protect\citeauthoryear{{Ivezi{\'c}} et~al.,}{{Ivezi{\'c}}
  et~al.}{2007}]{2007AJ....134..973I}
{Ivezi{\'c}} {\v{Z}}.,  et~al., 2007, \mn@doi [\aj] {10.1086/519976}, \href
  {https://ui.adsabs.harvard.edu/abs/2007AJ....134..973I} {134, 973}

\bibitem[\protect\citeauthoryear{{Jerkstrand}, {Smartt}, {Fraser}, {Fransson},
  {Sollerman}, {Taddia}  \& {Kotak}}{{Jerkstrand}
  et~al.}{2014}]{2014MNRAS.439.3694J}
{Jerkstrand} A.,  {Smartt} S.~J.,  {Fraser} M.,  {Fransson} C.,  {Sollerman}
  J.,  {Taddia} F.,   {Kotak} R.,  2014, \mn@doi [\mnras]
  {10.1093/mnras/stu221}, \href
  {https://ui.adsabs.harvard.edu/abs/2014MNRAS.439.3694J} {439, 3694}

\bibitem[\protect\citeauthoryear{{Jerkstrand} et~al.,}{{Jerkstrand}
  et~al.}{2015}]{2015MNRAS.448.2482J}
{Jerkstrand} A.,  et~al., 2015, \mn@doi [\mnras] {10.1093/mnras/stv087}, \href
  {https://ui.adsabs.harvard.edu/abs/2015MNRAS.448.2482J} {448, 2482}

\bibitem[\protect\citeauthoryear{{Jordi}, {Grebel}  \& {Ammon}}{{Jordi}
  et~al.}{2006}]{2006A&A...460..339J}
{Jordi} K.,  {Grebel} E.~K.,   {Ammon} K.,  2006, \mn@doi [\aap]
  {10.1051/0004-6361:20066082}, \href
  {https://ui.adsabs.harvard.edu/abs/2006A&A...460..339J} {460, 339}

\bibitem[\protect\citeauthoryear{{Karachentsev}, {Makarov}  \&
  {Kaisina}}{{Karachentsev} et~al.}{2013}]{2013AJ....145..101K}
{Karachentsev} I.~D.,  {Makarov} D.~I.,   {Kaisina} E.~I.,  2013, \mn@doi [\aj]
  {10.1088/0004-6256/145/4/101}, \href
  {https://ui.adsabs.harvard.edu/abs/2013AJ....145..101K} {145, 101}

\bibitem[\protect\citeauthoryear{{Kelly} et~al.,}{{Kelly}
  et~al.}{2016}]{2016ApJ...831..205K}
{Kelly} P.~L.,  et~al., 2016, \mn@doi [\apj] {10.3847/0004-637X/831/2/205},
  \href {https://ui.adsabs.harvard.edu/abs/2016ApJ...831..205K} {831, 205}

\bibitem[\protect\citeauthoryear{{Kleiser} et~al.,}{{Kleiser}
  et~al.}{2011}]{2011MNRAS.415..372K}
{Kleiser} I. K.~W.,  et~al., 2011, \mn@doi [\mnras]
  {10.1111/j.1365-2966.2011.18708.x}, \href
  {https://ui.adsabs.harvard.edu/abs/2011MNRAS.415..372K} {415, 372}

\bibitem[\protect\citeauthoryear{{Kozma} \& {Fransson}}{{Kozma} \&
  {Fransson}}{1998}]{1998ApJ...497..431K}
{Kozma} C.,  {Fransson} C.,  1998, \mn@doi [\apj] {10.1086/305452}, \href
  {https://ui.adsabs.harvard.edu/abs/1998ApJ...497..431K} {497, 431}

\bibitem[\protect\citeauthoryear{{Langer}}{{Langer}}{1991}]{1991A&A...243..155L}
{Langer} N.,  1991, \aap, \href
  {https://ui.adsabs.harvard.edu/abs/1991A&A...243..155L} {243, 155}

\bibitem[\protect\citeauthoryear{{Li} \& {McCray}}{{Li} \&
  {McCray}}{1992}]{1992ApJ...387..309L}
{Li} H.,  {McCray} R.,  1992, \mn@doi [\apj] {10.1086/171082}, \href
  {https://ui.adsabs.harvard.edu/abs/1992ApJ...387..309L} {387, 309}

\bibitem[\protect\citeauthoryear{{Lyman}, {Bersier}  \& {James}}{{Lyman}
  et~al.}{2014}]{2014MNRAS.437.3848L}
{Lyman} J.~D.,  {Bersier} D.,   {James} P.~A.,  2014, \mn@doi [\mnras]
  {10.1093/mnras/stt2187}, \href
  {https://ui.adsabs.harvard.edu/abs/2014MNRAS.437.3848L} {437, 3848}

\bibitem[\protect\citeauthoryear{{Maund} et~al.,}{{Maund}
  et~al.}{2021}]{2021MNRAS.503..312M}
{Maund} J.~R.,  et~al., 2021, \mn@doi [\mnras] {10.1093/mnras/stab391}, \href
  {https://ui.adsabs.harvard.edu/abs/2021MNRAS.503..312M} {503, 312}

\bibitem[\protect\citeauthoryear{{Mazzali} \& {Chugai}}{{Mazzali} \&
  {Chugai}}{1995}]{1995A&A...303..118M}
{Mazzali} P.~A.,  {Chugai} N.~N.,  1995, \aap, \href
  {https://ui.adsabs.harvard.edu/abs/1995A&A...303..118M} {303, 118}

\bibitem[\protect\citeauthoryear{{Mazzali}, {Lucy}  \& {Butler}}{{Mazzali}
  et~al.}{1992}]{1992A&A...258..399M}
{Mazzali} P.~A.,  {Lucy} L.~B.,   {Butler} K.,  1992, \aap, \href
  {https://ui.adsabs.harvard.edu/abs/1992A&A...258..399M} {258, 399}

\bibitem[\protect\citeauthoryear{{Menon} \& {Heger}}{{Menon} \&
  {Heger}}{2017}]{2017MNRAS.469.4649M}
{Menon} A.,  {Heger} A.,  2017, \mnras, \href
  {https://ui.adsabs.harvard.edu/abs/2017MNRAS.469.4649M} {469, 4649}

\bibitem[\protect\citeauthoryear{{Menzies} et~al.,}{{Menzies}
  et~al.}{1987}]{1987MNRAS.227P..39M}
{Menzies} J.~W.,  et~al., 1987, \mn@doi [\mnras] {10.1093/mnras/227.1.39P},
  \href {https://ui.adsabs.harvard.edu/abs/1987MNRAS.227P..39M} {227, 39P}

\bibitem[\protect\citeauthoryear{{Nagy} \& {Vink{\'o}}}{{Nagy} \&
  {Vink{\'o}}}{2016}]{2016A&A...589A..53N}
{Nagy} A.~P.,  {Vink{\'o}} J.,  2016, \mn@doi [\aap]
  {10.1051/0004-6361/201527931}, \href
  {https://ui.adsabs.harvard.edu/abs/2016A&A...589A..53N} {589, A53}

\bibitem[\protect\citeauthoryear{{Nagy}, {Ordasi}, {Vink{\'o}}  \&
  {Wheeler}}{{Nagy} et~al.}{2014}]{2014A&A...571A..77N}
{Nagy} A.~P.,  {Ordasi} A.,  {Vink{\'o}} J.,   {Wheeler} J.~C.,  2014, \mn@doi
  [\aap] {10.1051/0004-6361/201424237}, \href
  {https://ui.adsabs.harvard.edu/abs/2014A&A...571A..77N} {571, A77}

\bibitem[\protect\citeauthoryear{{Niculescu-Duvaz} et~al.,}{{Niculescu-Duvaz}
  et~al.}{2022}]{2022MNRAS.515.4302N}
{Niculescu-Duvaz} M.,  et~al., 2022, \mn@doi [\mnras] {10.1093/mnras/stac1626},
  \href {https://ui.adsabs.harvard.edu/abs/2022MNRAS.515.4302N} {515, 4302}

\bibitem[\protect\citeauthoryear{{Oke} et~al.,}{{Oke}
  et~al.}{1995}]{1995PASP..107..375O}
{Oke} J.~B.,  et~al., 1995, \mn@doi [\pasp] {10.1086/133562}, \href
  {https://ui.adsabs.harvard.edu/abs/1995PASP..107..375O} {107, 375}

\bibitem[\protect\citeauthoryear{{Pastorello} et~al.,}{{Pastorello}
  et~al.}{2005}]{2005MNRAS.360..950P}
{Pastorello} A.,  et~al., 2005, \mn@doi [\mnras]
  {10.1111/j.1365-2966.2005.09079.x}, \href
  {https://ui.adsabs.harvard.edu/abs/2005MNRAS.360..950P} {360, 950}

\bibitem[\protect\citeauthoryear{{Pastorello} et~al.,}{{Pastorello}
  et~al.}{2012a}]{2012A&A...537A.141P}
{Pastorello} A.,  et~al., 2012a, \mn@doi [\aap] {10.1051/0004-6361/201118112},
  \href {https://ui.adsabs.harvard.edu/abs/2012A&A...537A.141P} {537, A141}

\bibitem[\protect\citeauthoryear{{Pastorello} et~al.,}{{Pastorello}
  et~al.}{2012b}]{2012AA...537A.141P}
{Pastorello} A.,  et~al., 2012b, \mn@doi [\aap] {10.1051/0004-6361/201118112},
  \href {https://ui.adsabs.harvard.edu/abs/2012A&A...537A.141P} {537, A141}

\bibitem[\protect\citeauthoryear{{Paxton}, {Bildsten}, {Dotter}, {Herwig},
  {Lesaffre}  \& {Timmes}}{{Paxton} et~al.}{2011}]{2011ApJS..192....3P}
{Paxton} B.,  {Bildsten} L.,  {Dotter} A.,  {Herwig} F.,  {Lesaffre} P.,
  {Timmes} F.,  2011, \mn@doi [\apjs] {10.1088/0067-0049/192/1/3}, \href
  {https://ui.adsabs.harvard.edu/abs/2011ApJS..192....3P} {192, 3}

\bibitem[\protect\citeauthoryear{{Paxton} et~al.,}{{Paxton}
  et~al.}{2013}]{2013ApJS..208....4P}
{Paxton} B.,  et~al., 2013, \mn@doi [\apjs] {10.1088/0067-0049/208/1/4}, \href
  {https://ui.adsabs.harvard.edu/abs/2013ApJS..208....4P} {208, 4}

\bibitem[\protect\citeauthoryear{{Paxton} et~al.,}{{Paxton}
  et~al.}{2015}]{2015ApJS..220...15P}
{Paxton} B.,  et~al., 2015, \mn@doi [\apjs] {10.1088/0067-0049/220/1/15}, \href
  {https://ui.adsabs.harvard.edu/abs/2015ApJS..220...15P} {220, 15}

\bibitem[\protect\citeauthoryear{{Paxton} et~al.,}{{Paxton}
  et~al.}{2018}]{2018ApJS..234...34P}
{Paxton} B.,  et~al., 2018, \mn@doi [\apjs] {10.3847/1538-4365/aaa5a8}, \href
  {https://ui.adsabs.harvard.edu/abs/2018ApJS..234...34P} {234, 34}

\bibitem[\protect\citeauthoryear{{Perley} et~al.,}{{Perley}
  et~al.}{2020}]{2020ApJ...904...35P}
{Perley} D.~A.,  et~al., 2020, \mn@doi [\apj] {10.3847/1538-4357/abbd98}, \href
  {https://ui.adsabs.harvard.edu/abs/2020ApJ...904...35P} {904, 35}

\bibitem[\protect\citeauthoryear{{Petermann}, {Langer}, {Castro}  \&
  {Fossati}}{{Petermann} et~al.}{2015}]{2015A&A...584A..54P}
{Petermann} I.,  {Langer} N.,  {Castro} N.,   {Fossati} L.,  2015, \mn@doi
  [\aap] {10.1051/0004-6361/201526302}, \href
  {https://ui.adsabs.harvard.edu/abs/2015A&A...584A..54P} {584, A54}

\bibitem[\protect\citeauthoryear{{Podsiadlowski}}{{Podsiadlowski}}{1992}]{1992PASP..104..717P}
{Podsiadlowski} P.,  1992, \mn@doi [\pasp] {10.1086/133043}, \href
  {https://ui.adsabs.harvard.edu/abs/1992PASP..104..717P} {104, 717}

\bibitem[\protect\citeauthoryear{{Podsiadlowski} \& {Joss}}{{Podsiadlowski} \&
  {Joss}}{1989}]{1989Natur.338..401P}
{Podsiadlowski} P.,  {Joss} P.~C.,  1989, \mn@doi [\nat] {10.1038/338401a0},
  \href {https://ui.adsabs.harvard.edu/abs/1989Natur.338..401P} {338, 401}

\bibitem[\protect\citeauthoryear{{Podsiadlowski}, {Joss}  \&
  {Hsu}}{{Podsiadlowski} et~al.}{1992}]{1992ApJ...391..246P}
{Podsiadlowski} P.,  {Joss} P.~C.,   {Hsu} J.~J.~L.,  1992, \mn@doi [\apj]
  {10.1086/171341}, \href
  {https://ui.adsabs.harvard.edu/abs/1992ApJ...391..246P} {391, 246}

\bibitem[\protect\citeauthoryear{{Podsiadlowski}, {Morris}  \&
  {Ivanova}}{{Podsiadlowski} et~al.}{2007}]{2007AIPC..937..125P}
{Podsiadlowski} P.,  {Morris} T.~S.,   {Ivanova} N.,  2007, in {Immler} S.,
  {Weiler} K.,   {McCray} R.,  eds,  American Institute of Physics Conference
  Series Vol. 937, Supernova 1987A: 20 Years After: Supernovae and Gamma-Ray
  Bursters. pp 125--133, \mn@doi{10.1063/1.3682893}

\bibitem[\protect\citeauthoryear{{Pun} et~al.,}{{Pun}
  et~al.}{1995}]{1995ApJS...99..223P}
{Pun} C. S.~J.,  et~al., 1995, \mn@doi [\apjs] {10.1086/192185}, \href
  {https://ui.adsabs.harvard.edu/abs/1995ApJS...99..223P} {99, 223}

\bibitem[\protect\citeauthoryear{{Rabinak} \& {Waxman}}{{Rabinak} \&
  {Waxman}}{2011}]{2011ApJ...728...63R}
{Rabinak} I.,  {Waxman} E.,  2011, \mn@doi [\apj] {10.1088/0004-637X/728/1/63},
  \href {https://ui.adsabs.harvard.edu/abs/2011ApJ...728...63R} {728, 63}

\bibitem[\protect\citeauthoryear{{Rho} et~al.,}{{Rho}
  et~al.}{2019}]{2019ATel12897....1R}
{Rho} J.,  et~al., 2019, The Astronomer's Telegram, \href
  {https://ui.adsabs.harvard.edu/abs/2019ATel12897....1R} {12897, 1}

\bibitem[\protect\citeauthoryear{{Roming} et~al.,}{{Roming}
  et~al.}{2005}]{2005SSRv..120...95R}
{Roming} P. W.~A.,  et~al., 2005, \mn@doi [\ssr] {10.1007/s11214-005-5095-4},
  \href {https://ui.adsabs.harvard.edu/abs/2005SSRv..120...95R} {120, 95}

\bibitem[\protect\citeauthoryear{{Saio}, {Kato}  \& {Nomoto}}{{Saio}
  et~al.}{1988}]{1988ApJ...331..388S}
{Saio} H.,  {Kato} M.,   {Nomoto} K.,  1988, \mn@doi [\apj] {10.1086/166565},
  \href {https://ui.adsabs.harvard.edu/abs/1988ApJ...331..388S} {331, 388}

\bibitem[\protect\citeauthoryear{{Schlafly} \& {Finkbeiner}}{{Schlafly} \&
  {Finkbeiner}}{2011}]{2011ApJ...737..103S}
{Schlafly} E.~F.,  {Finkbeiner} D.~P.,  2011, \mn@doi [\apj]
  {10.1088/0004-637X/737/2/103}, \href
  {https://ui.adsabs.harvard.edu/abs/2011ApJ...737..103S} {737, 103}

\bibitem[\protect\citeauthoryear{{Singh} et~al.,}{{Singh}
  et~al.}{2019}]{2019ApJ...882L..15S}
{Singh} A.,  et~al., 2019, \mn@doi [\apjl] {10.3847/2041-8213/ab3d44}, \href
  {https://ui.adsabs.harvard.edu/abs/2019ApJ...882L..15S} {882, L15}

\bibitem[\protect\citeauthoryear{{Smartt}}{{Smartt}}{2015}]{2015PASA...32...16S}
{Smartt} S.~J.,  2015, \mn@doi [\pasa] {10.1017/pasa.2015.17}, \href
  {https://ui.adsabs.harvard.edu/abs/2015PASA...32...16S} {32, e016}

\bibitem[\protect\citeauthoryear{{Sukhbold}, {Ertl}, {Woosley}, {Brown}  \&
  {Janka}}{{Sukhbold} et~al.}{2016}]{2016ApJ...821...38S}
{Sukhbold} T.,  {Ertl} T.,  {Woosley} S.~E.,  {Brown} J.~M.,   {Janka} H.~T.,
  2016, \mn@doi [\apj] {10.3847/0004-637X/821/1/38}, \href
  {https://ui.adsabs.harvard.edu/abs/2016ApJ...821...38S} {821, 38}

\bibitem[\protect\citeauthoryear{{Suntzeff}, {Hamuy}, {Martin}, {Gomez}  \&
  {Gonzalez}}{{Suntzeff} et~al.}{1988}]{1988AJ.....96.1864S}
{Suntzeff} N.~B.,  {Hamuy} M.,  {Martin} G.,  {Gomez} A.,   {Gonzalez} R.,
  1988, \mn@doi [\aj] {10.1086/114933}, \href
  {https://ui.adsabs.harvard.edu/abs/1988AJ.....96.1864S} {96, 1864}

\bibitem[\protect\citeauthoryear{{Taddia} et~al.,}{{Taddia}
  et~al.}{2012}]{2012A&A...537A.140T}
{Taddia} F.,  et~al., 2012, \mn@doi [\aap] {10.1051/0004-6361/201118091}, \href
  {https://ui.adsabs.harvard.edu/abs/2012A&A...537A.140T} {537, A140}

\bibitem[\protect\citeauthoryear{{Taddia} et~al.,}{{Taddia}
  et~al.}{2013}]{2013A&A...558A.143T}
{Taddia} F.,  et~al., 2013, \mn@doi [\aap] {10.1051/0004-6361/201322276}, \href
  {https://ui.adsabs.harvard.edu/abs/2013A&A...558A.143T} {558, A143}

\bibitem[\protect\citeauthoryear{{Taddia} et~al.,}{{Taddia}
  et~al.}{2016}]{2016A&A...588A...5T}
{Taddia} F.,  et~al., 2016, \mn@doi [\aap] {10.1051/0004-6361/201527811}, \href
  {https://ui.adsabs.harvard.edu/abs/2016A&A...588A...5T} {588, A5}

\bibitem[\protect\citeauthoryear{{Tak{\'a}ts} et~al.,}{{Tak{\'a}ts}
  et~al.}{2016}]{2016MNRAS.460.3447T}
{Tak{\'a}ts} K.,  et~al., 2016, \mn@doi [\mnras] {10.1093/mnras/stw1122}, \href
  {https://ui.adsabs.harvard.edu/abs/2016MNRAS.460.3447T} {460, 3447}

\bibitem[\protect\citeauthoryear{{Tinyanont} et~al.,}{{Tinyanont}
  et~al.}{2021}]{2021NatAs...5..544T}
{Tinyanont} S.,  et~al., 2021, \mn@doi [Nature Astronomy]
  {10.1038/s41550-021-01320-4}, \href
  {https://ui.adsabs.harvard.edu/abs/2021NatAs...5..544T} {5, 544}

\bibitem[\protect\citeauthoryear{{Uomoto} \& {Kirshner}}{{Uomoto} \&
  {Kirshner}}{1986}]{1986ApJ...308..685U}
{Uomoto} A.,  {Kirshner} R.~P.,  1986, \mn@doi [\apj] {10.1086/164540}, \href
  {https://ui.adsabs.harvard.edu/abs/1986ApJ...308..685U} {308, 685}

\bibitem[\protect\citeauthoryear{{Utrobin}, {Wongwathanarat}, {Janka},
  {M{\"u}ller}, {Ertl}  \& {Woosley}}{{Utrobin}
  et~al.}{2019}]{2019A&A...624A.116U}
{Utrobin} V.~P.,  {Wongwathanarat} A.,  {Janka} H.~T.,  {M{\"u}ller} E.,
  {Ertl} T.,   {Woosley} S.~E.,  2019, \mn@doi [\aap]
  {10.1051/0004-6361/201834976}, \href
  {https://ui.adsabs.harvard.edu/abs/2019A&A...624A.116U} {624, A116}

\bibitem[\protect\citeauthoryear{{Utrobin}, {Wongwathanarat}, {Janka},
  {M{\"u}ller}, {Ertl}, {Menon}  \& {Heger}}{{Utrobin}
  et~al.}{2021}]{2021ApJ...914....4U}
{Utrobin} V.~P.,  {Wongwathanarat} A.,  {Janka} H.~T.,  {M{\"u}ller} E.,
  {Ertl} T.,  {Menon} A.,   {Heger} A.,  2021, \mn@doi [\apj]
  {10.3847/1538-4357/abf4c5}, \href
  {https://ui.adsabs.harvard.edu/abs/2021ApJ...914....4U} {914, 4}

\bibitem[\protect\citeauthoryear{{Wampler}, {Wang}, {Baade}, {Banse},
  {D'Odorico}, {Gouiffes}  \& {Tarenghi}}{{Wampler}
  et~al.}{1990}]{1990ApJ...362L..13W}
{Wampler} E.~J.,  {Wang} L.,  {Baade} D.,  {Banse} K.,  {D'Odorico} S.,
  {Gouiffes} C.,   {Tarenghi} M.,  1990, \mn@doi [\apjl] {10.1086/185836},
  \href {https://ui.adsabs.harvard.edu/abs/1990ApJ...362L..13W} {362, L13}

\bibitem[\protect\citeauthoryear{{Wang} et~al.,}{{Wang}
  et~al.}{2008}]{2008ApJ...675..626W}
{Wang} X.,  et~al., 2008, \mn@doi [\apj] {10.1086/526413}, \href
  {https://ui.adsabs.harvard.edu/abs/2008ApJ...675..626W} {675, 626}

\bibitem[\protect\citeauthoryear{{Weiss}}{{Weiss}}{1989}]{1989ApJ...339..365W}
{Weiss} A.,  1989, \mn@doi [\apj] {10.1086/167302}, \href
  {https://ui.adsabs.harvard.edu/abs/1989ApJ...339..365W} {339, 365}

\bibitem[\protect\citeauthoryear{{Weiss}, {Hillebrandt}  \& {Truran}}{{Weiss}
  et~al.}{1988}]{1988A&A...197L..11W}
{Weiss} A.,  {Hillebrandt} W.,   {Truran} J.~W.,  1988, \aap, \href
  {https://ui.adsabs.harvard.edu/abs/1988A&A...197L..11W} {197, L11}

\bibitem[\protect\citeauthoryear{{Whitelock} et~al.,}{{Whitelock}
  et~al.}{1988}]{1988MNRAS.234P...5W}
{Whitelock} P.~A.,  et~al., 1988, \mn@doi [\mnras] {10.1093/mnras/234.1.5P},
  \href {https://ui.adsabs.harvard.edu/abs/1988MNRAS.234P...5W} {234, 5P}

\bibitem[\protect\citeauthoryear{{Woosley} \& {Weaver}}{{Woosley} \&
  {Weaver}}{1995}]{1995ApJS..101..181W}
{Woosley} S.~E.,  {Weaver} T.~A.,  1995, \mn@doi [\apjs] {10.1086/192237},
  \href {https://ui.adsabs.harvard.edu/abs/1995ApJS..101..181W} {101, 181}

\bibitem[\protect\citeauthoryear{{Woosley}, {Pinto}  \& {Ensman}}{{Woosley}
  et~al.}{1988}]{1988ApJ...324..466W}
{Woosley} S.~E.,  {Pinto} P.~A.,   {Ensman} L.,  1988, \mn@doi [\apj]
  {10.1086/165908}, \href
  {https://ui.adsabs.harvard.edu/abs/1988ApJ...324..466W} {324, 466}

\bibitem[\protect\citeauthoryear{{Woosley}, {Heger}  \& {Weaver}}{{Woosley}
  et~al.}{2002}]{2002RvMP...74.1015W}
{Woosley} S.~E.,  {Heger} A.,   {Weaver} T.~A.,  2002, \mn@doi [Reviews of
  Modern Physics] {10.1103/RevModPhys.74.1015}, \href
  {https://ui.adsabs.harvard.edu/abs/2002RvMP...74.1015W} {74, 1015}

\bibitem[\protect\citeauthoryear{{Yaron} \& {Gal-Yam}}{{Yaron} \&
  {Gal-Yam}}{2012}]{2012PASP..124..668Y}
{Yaron} O.,  {Gal-Yam} A.,  2012, \mn@doi [\pasp] {10.1086/666656}, \href
  {https://ui.adsabs.harvard.edu/abs/2012PASP..124..668Y} {124, 668}

\bibitem[\protect\citeauthoryear{{Zampieri}, {Pastorello}, {Turatto},
  {Cappellaro}, {Benetti}, {Altavilla}, {Mazzali}  \& {Hamuy}}{{Zampieri}
  et~al.}{2003}]{2003MNRAS.338..711Z}
{Zampieri} L.,  {Pastorello} A.,  {Turatto} M.,  {Cappellaro} E.,  {Benetti}
  S.,  {Altavilla} G.,  {Mazzali} P.,   {Hamuy} M.,  2003, \mn@doi [\mnras]
  {10.1046/j.1365-8711.2003.06082.x}, \href
  {https://ui.adsabs.harvard.edu/abs/2003MNRAS.338..711Z} {338, 711}

\bibitem[\protect\citeauthoryear{{Zhang} et~al.,}{{Zhang}
  et~al.}{2022}]{2022MNRAS.509.2013Z}
{Zhang} X.,  et~al., 2022, \mn@doi [\mnras] {10.1093/mnras/stab3007}, \href
  {https://ui.adsabs.harvard.edu/abs/2022MNRAS.509.2013Z} {509, 2013}

\bibitem[\protect\citeauthoryear{{de Vaucouleurs}, {de Vaucouleurs}, {Corwin},
  {Buta}, {Paturel}  \& {Fouque}}{{de Vaucouleurs}
  et~al.}{1991}]{1991rc3..book.....D}
{de Vaucouleurs} G.,  {de Vaucouleurs} A.,  {Corwin} Herold~G. J.,  {Buta}
  R.~J.,  {Paturel} G.,   {Fouque} P.,  1991, {Third Reference Catalogue of
  Bright Galaxies}

\makeatother
\end{thebibliography}
		
		
		
\appendix
\section{Nebular Spectra of 87A-like Peculiar SNe~II}\label{sec:neb-SNe}
\begin{figure*}
		\centering
		\includegraphics[width=1.8\columnwidth]{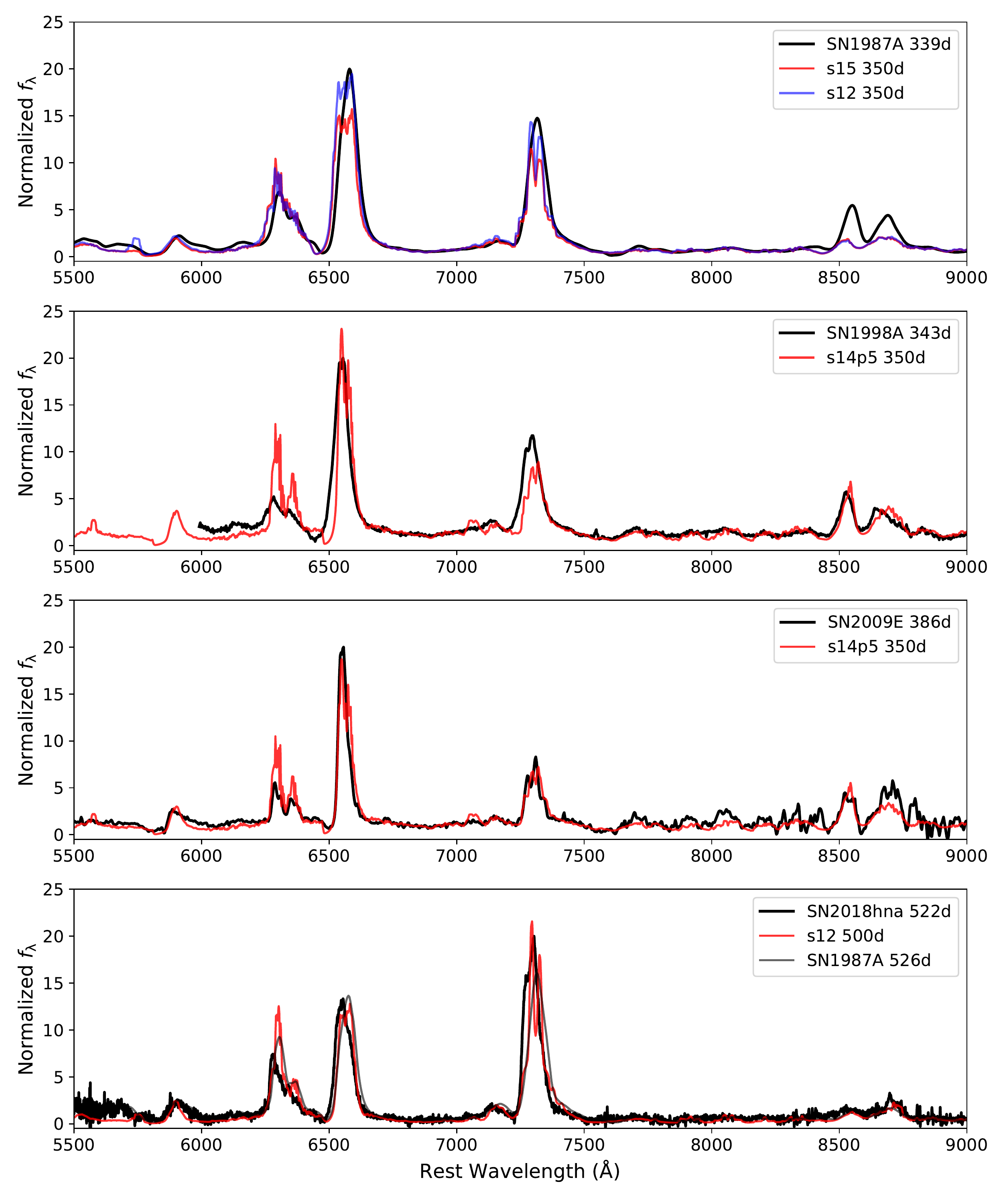}
		\caption{Nebular spectra of four 87A-like SNe compared with their best-fit models, showing the range of [\ion{O}{i}], H$\mathrm{\alpha}$, [\ion{Ca}{ii}], and \ion{Ca}{ii} lines. For SN~2018hna, we also compare it with SN~1987A at the same phase. The spectrum flux in each panel is scaled by normalising the maximum flux of the observed spectrum to be 20.}
		\label{fig:neb-SNe}
	\end{figure*}

	In this section, we compare the nebular-phase spectra of several SNe with those predicted by NLTE models of \cite{2021A&A...652A..64D}. 
	These SNe~II models are extended to about 500 days after the explosion. 
	The observed spectra of SNe~1987A, 1998A, and 2009E taken at about 1~yr after explosion and the t $\sim$ 522 day spectrum of SN~2018hna are used to compare with these NLTE models in Fig.~\ref{fig:neb-SNe}. 
	For each spectrum, the model spectra are normalised to match the flux of the observed spectrum in the range 6780--7000~\AA, and the model with the least squares is selected as the best fit. 
	The initial masses of the best-fit models are 15, 14.5, 14.5, and 12~\msun\ for SN~1987A, SN~1998A, SN~2009E, and SN~2018hna, respectively.
		
	Compared with model s12 at day 500, the H$\mathrm{\alpha}$, [\ion{Ca}{ii}], \ion{Ca}{ii} NIR triplet, and [\ion{O}{i}]~$\lambda$5577 lines are of comparable strength in the t$\sim$522 day spectrum of SN~2018hna, but the observation has weaker [\ion{O}{i}]~$\lambda\lambda$6300, 6364. 
	Moreover, the profile of [\ion{Ca}{ii}] doublet near 7300~\AA\ is quite different from the model: it is a broad and flat line rather than two spikes as in the model.
	We caution that there may be contamination from emission of [\ion{Ni}{ii}] and [\ion{Fe}{ii}] in the observed spectrum, so this difference is hard to determine. 
	Also, it is widely known that dust formation is common in core-collapse SNe, and dust emission will affect the line profiles of their nebular spectra. 
	The H$\mathrm{\alpha}$ line profile shows a clear flux excess in blue and deficit in red, which is a sign of dust emission (see Appendix.~\ref{sec:dust}).
	Compared with the spectrum of the same epoch in SN~1987A, the [\ion{O}{i}]~$\lambda\lambda$6300, 6364 line of SN~2018hna is weaker, but the [\ion{Ca}{ii}] lines are stronger, and the line profiles are also different.
		
	The [\ion{O}{i}] line of SN~1987A is consistent with the s15A model, but has stronger H$\mathrm{\alpha}$, [\ion{Ca}{ii}] and \ion{Ca}{ii} lines. 
	If only the wavelength range shown in the figure is considered in fitting, the best match is the s12 model, whose initial mass is 12~\msun, but the metal lines of the model at the blue end are significantly stronger than those in SN~1987A.
	The H$\mathrm{\alpha}$ line of SN~1998A is consistent with the s14p5 model, but the [\ion{Ca}{ii}] lines are stronger, and the [\ion{O}{i}] lines are much weaker than in the model.
	The H$\mathrm{\alpha}$ line and [\ion{Ca}{ii}] lines of SN~2009E are consistent with model s14p5, while the [\ion{O}{i}]~$\lambda\lambda$6300, 6364 line is weaker than the model, although [\ion{O}{i}]~$\lambda$5577 is comparable to the model. 
		
	Because the models of \cite{2021A&A...652A..64D} make use of pre-explosion profiles of stars with $Z =$ Z$_{\odot}$, which differs from the environments of the observed SNe, the above fitting results do not mean that their progenitor stars have the same initial masses as the best-fit models. 
	But the takeaway is that 87A-like SNe~II are all oxygen-poor, with SN~1998A having the lowest oxygen abundance. 
	Although the host galaxy of SN~1998A has approximately the solar metallicity \citep{2013A&A...558A.143T}, the intensity of its [\ion{O}{i}] lines is substantially lower than that of the model, which also has a relatively small initial mass (14.5~\msun).
	Note that stars of this low mass are likely to explode as RSGs rather than BSGs. 
	Combining the analysis of oxygen mass in the SN ejecta (Sec.~\ref{sec:progenitor-analysis}), the evolution mechanism of the progenitor stars of SNe~1998A, 2009E, and 2018hna may be different from that of SN~1987A.
		
\section{Dust Formation}\label{sec:dust}
	Dust formation is common in SNe~IIP \citep[e.g.,][]{2016MNRAS.456.1269B,2022MNRAS.515.4302N,2022MNRAS.509.2013Z}.
	Newly formed dust within the ejecta of SNe can cause a deficit in the red side of the emission lines and produce blueshifted peaks. 
	Also, emission of CO ($\sim 2.3$~$\mu$m) was detected in SN~2018hna in the NIR spectrum on day 153 \citep{2019ATel12897....1R}.
	To estimate the dust formation for SN~2018hna, we use the code DAMOCLES \citep{2016MNRAS.456.1269B,2018MNRAS.480.4659B} to model the H$\alpha$ line of SN~2018hna in its $t \approx 522$~day spectrum. 
	DAMOCLES is a Monte Carlo line radiative transfer code which models the extent and shape of the dust-affected line profiles.
	We follow the method described by \cite{2018MNRAS.480.4659B}, which applied the affine-invariant ensemble sampler for \texttt{emcee} to DAMOCLES.
\begin{figure}
	\includegraphics[width=1.0\columnwidth]{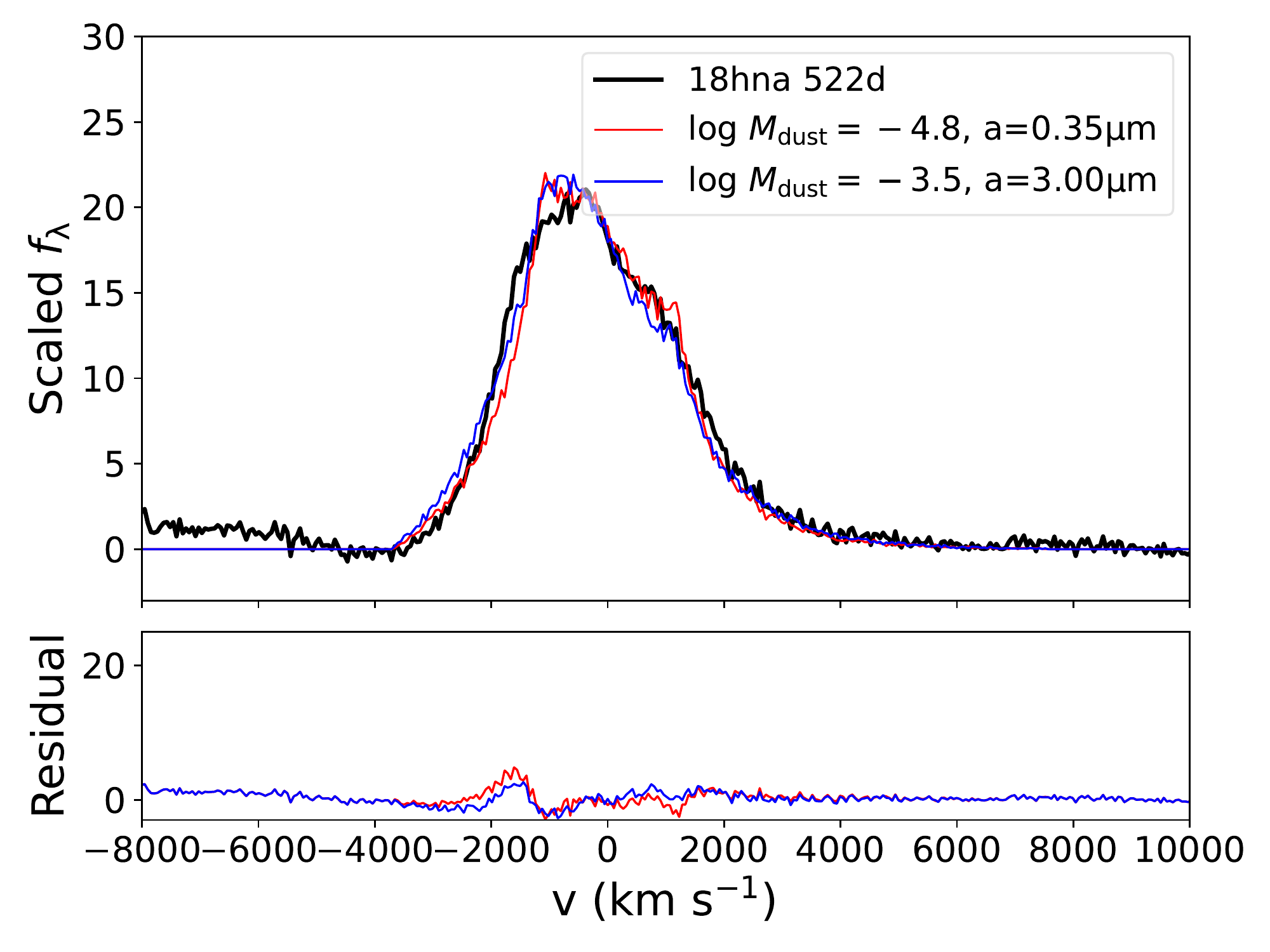}
	\caption{DAMOCLES dust model (\textit{upper}) and residual (\textit{lower}) for the H$\mathrm{\alpha}$ line of SN~2018hna at 522 days after explosion.}
	\label{fig:dust-Ha}
\end{figure}
	The code has 10 free parameters, while for simplicity and saving computing resources, we vary only the dust mass and fix the remaining 9 parameters. 
	The values of the fixed parameters and dust properties are referenced from the results for SN~1987A by \cite{2016MNRAS.456.1269B} as follows. 
	We assume a smooth dust distribution without clumping. 
	The dust and gas are coupled with the same density profile, $\rho\propto r^{-\beta}$ with $\beta = 2$. 
	The values of $V_{\mathrm{max}}$ and $V_{\mathrm{min}}$ are measured from the observed spectrum. 
	The dust is assumed to be 100\% amorphous carbon with a grain size of $a = 0.35$~$\mu$m. 
	The best-fit model (see Fig.~\ref{fig:dust-Ha}) shows that the total dust formed up to $t \approx 522$ days after explosion is log\,$M_{\mathrm{dust}} = -4.80^{+0.04}_{-0.09}$~\msun. 
	If the newly formed dust follows the prediction of the relationship suggested by \cite{2020ApJ...894..111B}, we get log\,$M_{\mathrm{dust}}\approx -3.5$~\msun, which is over one order of magnitude larger than the value we derived for SN~2018hna. 
	If log\,$M_{\mathrm{dust}}= -3.5$~\msun\ is applied, a larger dust grain size of $\sim$3~$\mu$m is required to match the observation. 
	This may imply that the dust formed in SN~2018hna has different properties than that in SN~1987A.
		
\section{Observational Data}\label{sec:appendixA}
	Magnitudes of the reference stars in the field of \mbox{SN~2018hna} are listed in Tab.~\ref{tab:refstars}.
	Photometric results from the TNT images are displayed in Tab.~\ref{tab:lc_tnt}.
	The details of TNS spectra and our optical spectroscopic observations for SN~2018hna are presented in Tab.~\ref{tab:TNS-spec} and Tab.~\ref{tab:spec-log}, respectively.
\begin{table*}
	\caption{Magnitudes and $1\sigma$ uncertainties of reference stars in field of \mbox{SN~2018hna}.}
	\label{tab:refstars}
	\begin{tabular}{cccccccc}
		\hline
		Star No. &RA (J2000) & Dec. (J2000) &\textit{B} &\textit{V} &\textit{g}  &\textit{r} &\textit{i}\\
		\hline
		1   &12h26m38s.16   &+58\degree18$^{\prime}$17$^{\prime\prime}$.48   &13.96(02)  &13.12(08)  &13.52(01)  &12.96(02)  &12.81(01)\\
		2   &12h26m05s.78   &+58\degree20$^{\prime}$33$^{\prime\prime}$.33   &16.72(03)  &15.21(09)  &16.04(02)  &14.69(02)  &14.18(02)\\
		3   &12h26m01s.30   &+58\degree20$^{\prime}$02$^{\prime\prime}$.59   &14.56(02)  &13.34(08)  &13.99(02)  &12.96(03)  &12.64(01)\\
		4   &12h26m28s.42   &+58\degree15$^{\prime}$00$^{\prime\prime}$.04   &15.29(03)  &14.58(06)  &14.90(01)  &14.47(01)  &14.32(02)\\
		5   &12h25m58s.62   &+58\degree17$^{\prime}$05$^{\prime\prime}$.73   &14.74(01)  &13.84(09)  &14.28(02)  &13.66(01)  &13.48(02)\\
		6   &12h26m04s.51   &+58\degree21$^{\prime}$39$^{\prime\prime}$.44   &16.40(01)  &15.75(05)  &16.03(01)  &15.73(01)  &15.73(03)\\
		7   &12h25m47s.17   &+58\degree15$^{\prime}$00$^{\prime\prime}$.19   &16.72(07)  &16.01(06)  &16.33(01)  &15.80(02)  &15.64(03)\\
		8   &12h26m00s.69   &+58\degree18$^{\prime}$33$^{\prime\prime}$.11   &16.83(06)  &16.00(05)  &16.40(02)  &15.74(02)  &15.51(02)\\
		9   &12h25m35s.11   &+58\degree13$^{\prime}$49$^{\prime\prime}$.34   &17.11(07)  &16.03(09)  &16.59(02)  &15.72(03)  &15.51(02)\\
		10  &12h25m49s.71   &+58\degree17$^{\prime}$06$^{\prime\prime}$.76   &17.83(06)  &16.80(06)  &17.33(01)  &16.44(02)  &16.13(03)\\
		\hline
	\end{tabular}
\end{table*}

\begin{table*}
\caption{$BVgri$ photometric observations (mag, $1\sigma$) of \mbox{SN~2018hna} from TNT.}
\label{tab:lc_tnt}
\begin{tabular}{cccccc}
\hline
MJD  &\textit{B} &\textit{V} &\textit{g}  &\textit{r} &	\textit{i}\\ 
\hline
58456.8818  &15.768(023)  &14.775(030)  &15.198(007)  &14.653(007)  &14.713(012)\\ 
58459.8841  &15.683(021)  &14.695(028)  &15.109(006)  &14.565(004)  &14.631(004)\\ 
58463.8584  &15.583(022)  &14.596(028)  &15.021(008)  &14.475(009)  &14.551(031)\\ 
58464.8363  &15.553(020)  &14.575(029)  &14.998(007)  &14.452(006)  &14.526(005)\\ 
58465.8534  &15.532(022)  &14.550(029)  &14.967(015)  &14.422(017)  &14.489(010)\\ 
58466.8849  &15.517(021)  &14.530(028)  &14.949(008)  &14.405(010)  &14.479(005)\\ 
58468.9026  &15.518(073)  &14.496(042)  &14.900(023)  &14.364(030)  &14.437(038)\\ 
58469.8456  &15.438(030)  &14.471(048)  &14.893(017)  &14.353(020)  &14.419(013)\\ 
58476.8029  &15.305(022)  &14.356(028)  &14.743(022)  &14.228(008)  &14.308(007)\\ 
58477.7920  &15.308(024)  &14.351(030)  &14.741(008)  &14.230(011)  &14.299(013)\\ 
58478.8260  &15.306(028)  &14.349(029)  &14.728(008)  &14.237(016)  &14.291(009)\\ 
58479.7832  &15.311(030)  &14.329(028)  &14.735(008)  &14.202(007)  &14.263(006)\\ 
58480.7888  &15.255(021)  &14.318(029)  &14.721(009)  &14.182(006)  &14.258(005)\\ 
58481.8202  &15.243(022)  &14.292(030)  &14.699(010)  &14.170(013)  &...\\ 
58482.8812  &15.225(084)  &14.281(100)  &14.701(008)  &14.144(032)  &...\\ 
58484.8716  &15.201(024)  &14.253(030)  &14.660(010)  &14.135(014)  &14.199(035)\\ 
58486.8623  &15.157(026)  &14.241(029)  &14.640(005)  &14.107(004)  &14.191(016)\\ 
58487.8341  &15.172(020)  &14.234(028)  &14.632(008)  &14.104(005)  &14.184(016)\\ 
58489.8028  &15.150(021)  &14.219(031)  &14.610(009)  &14.081(006)  &14.148(009)\\ 
58490.8340  &15.148(021)  &14.206(029)  &14.580(006)  &14.075(006)  &14.138(014)\\ 
58491.9155  &15.130(039)  &14.207(034)  &14.602(012)  &14.071(009)  &14.126(028)\\ 
58495.9295  &...  &...  &14.601(039)  &14.117(045)  &...\\ 
58498.8507  &15.134(022)  &14.186(028)  &14.582(005)  &14.031(005)  &14.098(018)\\ 
58499.8604  &15.143(021)  &14.188(031)  &14.598(009)  &14.047(021)  &14.094(016)\\ 
58500.8375  &15.168(031)  &14.195(029)  &14.615(025)  &14.049(017)  &14.099(016)\\ 
58503.8620  &15.224(031)  &14.242(029)  &14.640(011)  &14.063(006)  &14.118(007)\\ 
58504.8879  &15.247(025)  &14.250(031)  &14.684(011)  &14.080(008)  &14.121(017)\\ 
58506.8683  &15.360(045)  &14.308(038)  &14.780(039)  &14.131(036)  &14.160(039)\\ 
58507.8622  &15.373(031)  &14.316(044)  &14.781(010)  &14.130(013)  &14.146(011)\\ 
58515.8312  &16.055(022)  &14.826(028)  &15.375(010)  &14.538(007)  &14.497(004)\\ 
58530.8982  &16.855(026)  &15.530(029)  &16.132(011)  &15.149(011)  &15.047(015)\\ 
58533.8840  &16.935(067)  &15.601(039)  &16.172(021)  &15.199(013)  &15.090(041)\\ 
58534.7725  &16.913(048)  &15.594(034)  &16.167(017)  &15.205(025)  &15.102(024)\\ 
58537.6673  &17.184(033)  &15.624(030)  &16.192(012)  &15.199(007)  &15.120(013)\\ 
58538.7867  &16.954(035)  &15.671(036)  &16.222(024)  &15.255(025)  &15.133(019)\\ 
58543.6943  &16.967(026)  &...  &...  &...  &...\\ 
58545.6754  &17.011(046)  &15.735(038)  &16.297(008)  &15.291(010)  &15.224(019)\\ 
58554.6898  &17.167(027)  &15.853(030)  &16.437(011)  &15.410(012)  &15.331(019)\\ 
58559.7581  &17.088(132)  &15.899(050)  &...  &15.459(026)  &...\\ 
58560.5376  &17.132(131)  &15.938(039)  &16.410(045)  &...  &...\\ 
58566.6336  &17.259(030)  &15.973(032)  &16.549(020)  &15.499(017)  &15.432(018)\\ 
58575.5722  &17.350(025)  &16.095(029)  &16.653(010)  &15.584(009)  &15.526(011)\\ 
58576.5732  &17.336(029)  &16.102(030)  &16.657(011)  &15.593(011)  &15.542(010)\\ 
58577.6453  &17.398(032)  &16.135(030)  &16.679(011)  &15.608(006)  &15.556(006)\\ 
58580.6479  &17.451(058)  &16.263(060)  &16.719(034)  &15.644(036)  &15.656(057)\\ 
58589.6703  &17.555(157)  &16.229(076)  &16.874(079)  &15.726(026)  &15.698(031)\\ 
58590.7650  &...  &...  &16.652(132)  &15.558(103)  &15.687(074)\\ 
58599.6243  &17.665(025)  &...  &...  &...  &...\\ 
58603.6161  &17.659(030)  &16.429(029)  &16.970(010)  &15.822(011)  &15.789(017)\\ 
58604.7262  &17.684(027)  &16.454(030)  &16.978(008)  &15.838(008)  &15.796(017)\\ 
58608.6191  &17.749(041)  &16.511(030)  &17.049(009)  &15.874(012)  &15.841(011)\\ 
58609.7037  &17.750(038)  &16.535(030)  &17.067(010)  &15.902(009)  &15.876(017)\\ 
58623.6030  &17.911(045)  &16.694(032)  &17.219(015)  &16.012(012)  &15.999(012)\\ 
58631.6280  &17.970(038)  &16.784(039)  &17.294(043)  &...  &...\\ 
58634.5618  &17.998(034)  &16.818(029)  &17.337(023)  &16.114(011)  &16.096(017)\\ 
58645.6379  &18.074(080)  &16.934(050)  &17.425(037)  &16.183(013)  &16.191(018)\\ 
58649.5714  &18.111(225)  &17.082(079)  &17.549(085)  &16.233(032)  &16.277(036)\\ 
58663.6125  &18.208(040)  &17.146(037)  &17.627(025)  &...  &...\\ 
\hline
\end{tabular}
\end{table*}
	\begin{table*}
	\caption{Details of TNS spectra of SN~2018hna.}
	\label{tab:TNS-spec}
	\begin{tabular}{cccccc}
		\hline
		ID     &UT             &MJD       &Telescope(Instrument)  &Exp. time (s)   &Observer(s)\\
		\hline
		2933   &2018/10/24.80  &58415.80  &Kanata / HOWPol        &900             &K. Takagi, T. Nakaoka, M. Kawabata \\
		2930   &2018/10/26.72  &58417.72  &ALPY                   &3006            &R. Leadbeater \\
		2974   &2018/11/02.53  &58424.53  &P60/SEDM               &1600            &ZTF \\
		\hline
	\end{tabular}		
\end{table*}

\begin{table*}
\caption{Spectrascopic observations of SN~2018hna.}
\label{tab:spec-log}
\begin{tabular}{ccccc}
\hline
UT  &MJD  &Telescope &Instrument & Exp. time (s)\\ 
\hline
2018/12/04.89  &58456.89  &BAO 2.16~m  &BFOSC  &3300\\
2018/12/08.88  &58460.88  &BAO 2.16~m  &BFOSC  &3300\\
2018/12/09.50  &58462.50  &APO 3.5~m   &DIS    &2400\\
2018/12/11.87  &58463.87  &BAO 2.16~m  &BFOSC  &3300\\
2018/12/14.87  &58466.87  &BAO 2.16~m  &BFOSC  &3300\\
2018/12/16.88  &58468.88  &BAO 2.16~m  &BFOSC  &3000\\
2018/12/22.88  &58474.88  &BAO 2.16~m  &BFOSC  &3300\\
2019/01/01.90  &58484.90  &BAO 2.16~m  &BFOSC  &2400\\
2019/01/06.90  &58489.90  &BAO 2.16~m  &BFOSC  &2400\\
2019/01/25.91  &58508.91  &LJO 2.4-m   &YFOSC  &1800\\
2019/01/29.80  &58512.80  &BAO 2.16~m  &BFOSC  &3000\\
2019/01/31.84  &58514.84  &BAO 2.16~m  &BFOSC  &3000\\
2019/02/11.66  &58525.66  &BAO 2.16~m  &BFOSC  &3000\\
2019/02/15.70  &58529.70  &LJO 2.4-m   &YFOSC  &1000\\
2019/02/20.75  &58534.75  &LJO 2.4-m   &YFOSC  &1000\\
2019/02/23.84  &58554.84  &BAO 2.16~m  &BFOSC  &3000\\
2019/03/12.85  &58565.85  &BAO 2.16~m  &BFOSC  &3000\\
2019/03/26.77  &58568.77  &BAO 2.16~m  &BFOSC  &3300\\
2019/04/02.75  &58575.75  &LJO 2.4-m   &YFOSC  &1500\\
2019/05/06.61  &58609.61  &BAO 2.16~m  &BFOSC  &2700\\
2019/06/11.55  &58645.55  &BAO 2.16~m  &BFOSC  &2700\\
2019/06/23.54  &58657.54  &BAO 2.16~m  &BFOSC  &2700\\
2020/03/24.63  &58932.63  &Keck-I 10~m &LRIS   &610\\
\hline
\end{tabular}
\end{table*}
		
	\bsp	
	\label{lastpage}
\end{document}